\documentclass[nofootinbib,superscriptaddress,twocolumn]{revtex4-2}

\usepackage[T1]{fontenc}
\usepackage[english]{babel}

\usepackage{amsmath,amsfonts,amssymb,amsthm}
\usepackage{graphicx}
\usepackage{hyperref}
\usepackage{footnotebackref}
\usepackage{braket}
\usepackage{mathtools}
\usepackage[dvipsnames]{xcolor}
\usepackage{tcolorbox}
\usepackage{enumerate}
\usepackage{bm}
\usepackage{multirow}
\usepackage{booktabs}
\usepackage{quantikz}
\usepackage{transparent}
\input{llangle_rrangle}

\usepackage[capitalize]{cleveref}
\crefname{section}{\text{Sec.}}{\text{Secs.}}
\crefname{appendix}{\text{App.}}{\text{Apps.}}
\crefname{table}{\text{Tab.}}{\text{Tabs.}}

\definecolor{quantum}{RGB}{83, 37, 128}
\definecolor{burntorange}{HTML}{C04E01}
\definecolor{fig_2_red}{HTML}{A70000}

\colorlet{equiv_diagram_left_colback}{Mulberry!25!white}
\colorlet{equiv_diagram_left_colframe}{Mulberry}
\colorlet{equiv_diagram_right_colback}{orange!15!white}
\colorlet{equiv_diagram_right_colframe}{orange!75!black}

\hypersetup{
    colorlinks=true,
    linkcolor=blue,
    filecolor=burntorange,
    urlcolor=burntorange,
    citecolor=burntorange,
    pdfpagemode=FullScreen,
}

\newtheorem{theorem}{Theorem}
\newtheorem{definition}{Definition}[section]
\newtheorem{lemma}[theorem]{Lemma}

\newenvironment{redbox}{\begin{tcolorbox}[colback=red!15!white,colframe=red!50!black,boxsep=0pt,right=10pt,left=10pt]}{\end{tcolorbox}}
\newenvironment{highlight}{\begin{tcolorbox}[colback=green!15!white,colframe=green!50!black,boxsep=0pt,right=7pt,left=7pt]}{\end{tcolorbox}}
\newenvironment{convention}{\begin{tcolorbox}[colback=Cerulean!15!white,colframe=Cerulean!50!black,boxsep=0pt,right=7pt,left=7pt]}{\end{tcolorbox}}

\newcommand{\C}{\mathbb{C}}
\newcommand{\R}{\mathbb{R}}

\newcommand{\Span}{\operatorname{span}}

\newcommand{\btheta}{\vec{\theta}}
\newcommand{\act}[1]{\cdot_{#1}}
\newcommand{\utcless}{\mf u^{\mf t}_\circ}
\newcommand{\tcless}{\mf t_\circ}

\renewcommand{\vec}{\boldsymbol}
\newcommand{\orth}[1]{\overline{#1}} 
\newcommand{\uperp}{\mf u^\perp} 
\newcommand{\real}[1]{\mf{Re}\left\{#1\right\}}
\newcommand{\imag}[1]{\mf{Im}\left\{#1\right\}}
\newcommand{\tr}[1]{\operatorname{tr}\De{#1}}
\newcommand{\de}[1]{\left(#1\right)}
\newcommand{\De}[1]{\left[#1\right]}
\newcommand{\DE}[1]{\left\{#1\right\}}
\newcommand{\dE}[1]{\left\llangle#1\right\rrangle}

\newcommand\mc[1]{\mathcal{#1}}
\newcommand\mf[1]{\mathfrak{#1}}
\newcommand\mbb[1]{\mathbb{#1}}
\newcommand\mlnode[1]{\begin{tabular}{@{}c@{}}#1\end{tabular}}

\makeatletter 
    
\renewcommand\onecolumngrid{
    \do@columngrid{one}{\@ne}%
    \def\set@footnotewidth{\onecolumngrid}
    \def\footnoterule{\kern-6pt\hrule width 1.5in\kern6pt}%
}

\renewcommand\twocolumngrid{
    \def\footnoterule{
    \dimen@\skip\footins\divide\dimen@\thr@@
    \kern-\dimen@\hrule width.5in\kern\dimen@}
    \do@columngrid{mlt}{\tw@}
}%

\makeatother    

\begin{document}

\title{Symmetric derivatives of parametrized quantum circuits}

\author{David Wierichs}
\affiliation{Xanadu, Toronto, ON, M5G 2C8, Canada}

\author{Richard D. P. East}
\affiliation{Xanadu, Toronto, ON, M5G 2C8, Canada}

\author{Mart{\'i}n Larocca}
\affiliation{Theoretical Division, Los Alamos National Laboratory, Los Alamos, New Mexico 87545, USA}
\affiliation{Center for Nonlinear Studies, Los Alamos National Laboratory, Los Alamos, New Mexico 87545, USA}

\author{M. Cerezo}
\affiliation{Information Sciences, Los Alamos National Laboratory, Los Alamos, NM 87545, USA}

\author{Nathan Killoran}
\affiliation{Xanadu, Toronto, ON, M5G 2C8, Canada}

\begin{abstract}
    Symmetries are crucial for tailoring parametrized quantum circuits to applications, due to their capability to capture the essence of physical systems.
    In this work, we shift the focus away from incorporating symmetries in the circuit design and towards symmetry-aware training of variational quantum algorithms.
    For this, we introduce the concept of projected derivatives of parametrized quantum circuits, in particular the equivariant and covariant derivatives.
    We show that the covariant derivative gives rise to the quantum Fisher information and quantum natural gradient.
    This provides an operational meaning for the covariant derivative, and allows us to extend the quantum natural gradient to all continuous symmetry groups.
    Connecting to traditional particle physics, we confirm that our covariant derivative is the same as the one introduced in physical gauge theory.
    This work provides tools for tailoring variational quantum algorithms to symmetries by incorporating them locally in derivatives, rather than into the design of the circuit.
\end{abstract}

\maketitle

\section{Introduction}
Symmetries are a central concept throughout modern physics.
Recently, they also have become increasingly important as a tool in the field of machine learning, in a discipline coined \emph{geometric deep learning}~\cite{bronstein2021geometric}.
Effective machine learning architectures typically encode information about the symmetries of the task at hand as inductive biases.
Geometric deep learning takes problems with a symmetry and creates architectures using maps that in some way respect this symmetry.
A particularly effective way of doing this is by using group equivariant maps, as is famously seen in the convolutional layers of convolutional neural networks (CNNs).
These layers (approximately) implement translation group equivariant maps, which have a natural use in image analysis~\cite{cohen2016group,kondor2018generalization}.
A local formulation of equivariance has been applied to finding symmetries~\cite{dehmamy2021automatic}.

This development in geometric deep learning soon after carried over to \emph{quantum machine learning (QML)}, including \emph{variational quantum algorithms (VQAs)}.
This opened the research direction of \emph{geometric quantum machine learning (GQML)}, which has had success in improving trainability and generalization~\cite{bowles2023backpropagation,sauvage2022building,west2022reflection,zheng2023sncqa,larocca2022group,zheng2022super,schatzki2022theoretical,nguyen2022theory,skolik2023equivariant,meyer2023exploiting,heredge2023permutation,zheng2023speeding}.
Depending on the groups of interest, different techniques exist for constructing equivariant gates in practice.
Twirling for example, as proposed in Ref.~\cite{meyer2023exploiting}, allows one to build equivariant gates, but for many-qubit gates it is highly non-trivial to perform the required twirling formula.
For a special unitary group as symmetry, one can turn to the generalised permutations of Refs.~\cite{zheng2022super,zheng2023speeding}, but this also leads to difficulties in constructing the ans\"{a}tze for large systems.
The work done in Ref.~\cite{east2023all} offers a concrete unitary construction built from so-called spin networks, but the technique is presently limited to the special unitary group on spins alone.
This highlights that encoding symmetry at the architecture level is presently not, and in general may never be, practical for all groups of interest.

A number of reasons make QML applications natural candidates to incorporate symmetries and geometry in both model development and algorithm analysis.
First, many problems of interest come with symmetries, and it is important to include such task-specific information in the model design \cite{ragone2022representation,nguyen2022theory,meyer2023exploiting,larocca2022group,seki2020symmetry_open,gard2020efficient,wierichs2020avoiding,anselmetti2021local,lyu2023symmetry}.
Second, these symmetries typically can be represented as subgroups of the unitary group.
QML models are built using \emph{parametrized quantum circuits (PQCs)}, which map to the unitary group, so that they can naturally accommodate unitary symmetries.
In addition, the vast mathematical fields of differential geometry and Lie theory offer the right tools to analyze PQCs.
Third, geometric methods are frequently used to understand or improve VQAs, and PQCs as subroutines more generally, even without the context of symmetries.

A central tool for this is the \emph{dynamical Lie algebra} of the circuit ansatz \cite{bohm1988dynamical,schirmer2002degrees,albertini2001notions,anand2022exploring}.
It can be used to analyze the complexity of PQC-based tasks \cite{somma2005quantum}, to classify spin chain models \cite{wiersema2023classification}, and to understand untrainable regions of the parameter space, called barren plateaus \cite{larocca2022diagnosing,ragone2023unified,fontana2023adjoint}.
Productively, this knowledge can be leveraged to create efficient simulation tools for PQC expectation values \cite{goh2023lie} and for time evolution \cite{kokcu2022fixed}.
Understanding the geometry of the Hilbert space, the global phase symmetry and quantum information enables the development of dedicated optimization methods for QML \cite{facchi2010classical,wiersema2023optimizing}.
Finally, a geometric description of PQCs makes it possible to put them into context with the mathematical literature on quantum physics and optimization \cite{hackl2020geometry,mahony1994optimization}.

Thus far, the main approach to incorporate symmetries into PQC models uses circuit ans{\"a}tze that commute with the symmetry action, i.e.,~\emph{equivariant} quantum circuits.
This has been done for QML applications \cite{meyer2023exploiting,larocca2022group,nguyen2022theory,skolik2023equivariant,heredge2023permutation,east2023all,schatzki2022theoretical} as well as for quantum chemistry tasks where the ansatz preserves \emph{quantum numbers}, the conserved quantities within symmetry sectors \cite{ogorman2019generalized,anselmetti2021local,arrazola2022universalquantum}.
In this approach, the considered symmetries only determine the allowed circuit ans{\"a}tze, but do not impact their training.

In other modern disciplines of physics, symmetries frequently enter in the form of \emph{gauge symmetries}, which capture essential features of the described system and power the mathematical formulation of fundamental theories, e.g.,~for high-energy particle physics.
In these theories, mathematical gauge theory manifests itself in the form of intricate group actions on unphysical fields, which imprint symmetries on physical fields.
In stark contrast, PQCs allow to implement gauge theory directly.
We can thus exploit the machinery of gauge theories to better understand and incorporate symmetries into quantum computations\footnote{Interestingly, in particle physics, gauges are often ``unphysical'' quantities, whereas here we can treat them as physical but unwanted degrees of freedom.}.

\subsection*{Summary of main results}
\begin{figure}
    \centering
    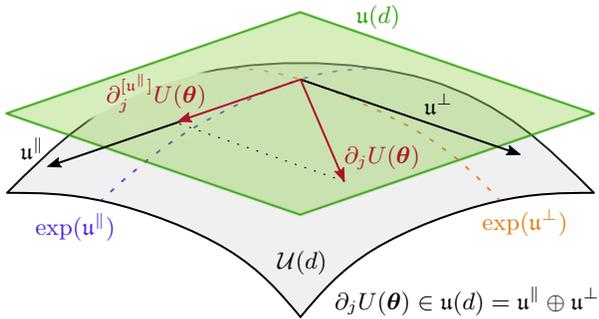
    \caption{%
    \textbf{The unitary group $\mc U(d)$ and its tangent space,} which is isomorphic to the unitary algebra $\mf u(d)$ (green).
    The algebra is split into the subspace $\mf u^\|$ of desired directions of change and its orthogonal complement $\uperp$, the restricted directions.
    A parametrized unitary $U(\btheta)$ has partial derivatives $\partial_j U(\btheta)$.
    A central tool of this work is the projected derivative $\partial_j^{[u^\|]}U(\btheta)$ that captures the component of $\partial_j U(\btheta)$ in the desired directions of change.
    Typically, one of $\mf u^\|$ and $\uperp$ also is a subalgebra, in which case it gives rise to a subgroup of $\mc U(d)$.
    \label{fig:general_decomp}}
\end{figure}
In this work we will look beyond the prevailing approach of equivariant quantum circuits and discuss how else symmetries can be taken into account, for example during training.
In particular, we introduce the \emph{covariant derivative} and the \emph{equivariant derivative} for parametrized unitaries, parametrized quantum states and PQC cost functions.
Both derivatives are special cases of an essential tool for our work, which we sketch in \cref{fig:general_decomp}:
the \emph{projected derivative} based on some orthogonal decomposition of the unitary Lie algebra $\mf u$.

While equivariance is the natural concept to apply at the circuit design level, the covariant derivative seems the obvious choice for implementing symmetries in the training of PQCs.
However, we show that mixing the concepts works as well: the concept of equivariance can be realized during training with the \emph{equivariant derivative}, and there is a notion of \emph{covariant quantum circuits} for special combinations of total groups and symmetry groups.
In essence, equivariant and covariant quantum circuits implement symmetries at the \emph{global} level via subgroups of the unitary group.
The covariant and equivariant derivatives instead are realized in the tangent space of the unitary group or of Hilbert space, and thus act \emph{locally}.
For the latter, we show that equivariance and covariance split the tangent space into orthogonal subspaces in two different ways, which turn out to be particularly compatible at the level of the unitary Lie algebra, resulting in the decomposition
\begin{align}
    \mf u =
    \underset%
    {\text{covariant }\de{\orth{\mf t}}\hspace{-7pt}}%
    {\underbrace{%
        \mf r \hspace{3pt}\oplus \hspace{3pt}\lefteqn{\hspace{-10pt}%
            \overset%
            {\text{equivariant } \de{\mf u^{\mf t}}}%
            {\overbrace%
                {\phantom{\utcless \hspace{3pt}\oplus \mf z(\mf t)}}%
            }
        }%
        \utcless}%
    }%
    \oplus \underset{\text{vertical } (\mf t)}{\underbrace{\mf z(\mf t) \hspace{3pt}\oplus \hspace{3pt}\tcless}}.
\end{align}
This preview into \cref{sec:equi_and_co:connect}, where these spaces are discussed in detail, shows the relationship of covariance and equivariance at the algebraic level.

If the orthogonal subspace onto which the derivative is projected is a subalgebra in addition, our local construction gives rise to a subgroup and thus provides a way to realize the projection globally instead, akin to equivariant quantum circuits.
While a global restriction allows us to reduce the parameter space itself, the local approach based on a projected derivative maintains the full parameter space but traverses it only in a subset of desired optimization directions at each point.
The projected derivative therefore offers an alternative for incorporating the symmetry into the VQA whenever the cost function itself can not be made symmetry-aware, e.g.,~if modifying the circuit ansatz is not an option due to hardware requirements or challenges in compiling the necessary gates.
It replaces reducing the parameter count during the ansatz design stage by symmetry-informed training, and promises to move in the right directions within a bigger parameter space.
This way, the equivariant derivative performs updates that are unchanged by symmetries, and the covariant derivative excludes symmetry directions from the training.

After exploring equivariance and covariance in depth, we consider the \emph{quantum natural gradient (QNG)} \cite{stokes2020quantum,hackl2020geometry}, which incorporates the geometry of Hilbert space.
QNG is based on the global phase symmetry and we show how the covariant derivative with respect to this symmetry gives rise to the quantum Fisher information, the core ingredient of QNG.
We then use this relationship to extend QNG to arbitrary continuous symmetries.

\subsection*{Structure}
This work is structured as follows.
The preliminaries in \cref{sec:preliminaries} introduce parametrized quantum circuits (\cref{sec:preliminaries:pqc}) as well as group representations and actions (\cref{sec:preliminaries:symmetries}).
In \cref{sec:equi_and_co} we then discuss equivariance and covariance at an abstract level in depth.
For this we introduce the \emph{equivariant} and \emph{covariant} derivatives and associated subgroups, and discuss their relationship.
This is complemented by \cref{sec:circuitry_app}, which introduces the same concepts from the perspective of quantum circuits without relying on higher mathematical abstractions.
These concepts are applied to gates, quantum states, and PQC cost functions in \cref{sec:circuitry}, where we provide explicit computational recipes and examples.
In \cref{sec:qng} we discuss the relationship between the covariant derivative and quantum natural gradient and we extend the latter to symmetry groups beyond global phases.
A connection to physical gauge theory is made in \cref{sec:gauge_theory} and we conclude in \cref{sec:conclusion}.
We make use of a number of lemmas in \cref{sec:lie_lemmas} and defer a series of calculations to \cref{sec:covariant_derivative_is_equivariant,app:example_calculations,sec:hadamard_test_F_c,sec:qng_comp,sec:gauge_theory_comp}.

\begin{convention}
    We will highlight main results in green, and conventions (such as this one) in blue.
\end{convention}

\section{Preliminaries}\label{sec:preliminaries}

\subsection{Parametrized quantum circuits}\label{sec:preliminaries:pqc}

We will denote a parametrized unitary acting on a $d$-dimensional Hilbert space as $U:\R^p\to\mc U(d)$%
%
\begin{align}
     U(\btheta) = \prod_{j=p}^1 U_j(\theta_j),
    \quad U_j(\theta_j)=\exp(-i\theta_j H_j)\,,
\end{align}
where $\btheta=(\theta_1,\ldots,\theta_p)$ are trainable parameters and $\mc G = \DE{-iH_j}_j$ is a set of gate generators.
Note that we absorbed the factor $-i$ into the generators, making them skew-Hermitian.
This is for notational clarity and useful when thinking about Lie algebras, which we intend to do quite a bit.
We have assumed that each gate depends on a single parameter and that no parameters are reused between gates.
This can be generalized to multivariate and perturbed gates as well as circuits that reuse parameters by applying the description via effective generators below directly.

Parametrized quantum circuits have seen widespread use in variational quantum computing, where the parameters $\btheta$ are optimized to minimize a given cost function
\begin{align}\label{eq:cost_function}
    C
    : \R^p &\to \R
    : \vec\theta \mapsto C(\btheta)\equiv C(U(\btheta))\,,
\end{align}
which quantifies the degree to which a target problem has been solved. 
The parameters are commonly trained via gradient-based methods, which require the evaluation of the derivatives $\partial_j C(\btheta)$, implicitly using $\partial_jU(\btheta)$ through the chain rule.

The derivative of a single gate is 
\begin{align}
    \frac{\partial}{\partial \theta_j} U_j(\theta_j)
    \eqqcolon\partial_j U_j(\theta_j)
    &=U_j(\theta_j) (-iH_j)
    = -i H_j U_j(\theta_j)\nonumber.
\end{align}
We will also need the generators that capture the directions of change of the \emph{entire} unitary $U(\btheta)$.
In contrast to the single gate, they will generally depend on the side to which we pull $-iH_j$ out of the circuit.
The \emph{right-effective generator} is defined via $\partial_jU(\btheta)=U(\btheta)\Omega^R_j(\btheta)$.
We rewrite this to get
\begin{align}
    \Omega_j^R(\btheta)
    &= U^\dagger(\btheta) \partial_j U(\btheta)
    = U_{[:j]}^\dagger(\btheta) (-iH_j) U_{[:j]}(\btheta).
\end{align}
Analogously, we can define the \emph{left-effective generator}%
%
\begin{align}
    \Omega_j^L(\btheta)
    &= \partial_j U(\btheta)U^\dagger(\btheta)
    = U_{[j:]}(\btheta) (-iH_j) U^\dagger_{[j:]}(\btheta).
\end{align}
Put simply, increasing the parameter $\theta_j$ in $U(\btheta)$ is equivalent to appending an incremental gate with the generator $\Omega^{R(L)}_j(\btheta)$ to the right (left) of $U$.

The fact that $H_j$ need not commute with the gates $U_{k\neq j}$ has two important consequences.
On one hand, it makes $\Omega^{R(L)}$ depend on all circuit parameters before (after) $U_j$.
On the other hand, the structure of the effective generators usually is more complex than that of the gate generators $H_j$, because they are transformed by parts of the circuit.
Accordingly, the circuit determines how complex the $\Omega^{R(L)}$ can get.
More precisely, this complexity is captured by an algebra associated to the circuit.
It is a Lie algebra, defined by the set of gate generators $\mc G=\{H_j\}_j$~\cite{zeier2011symmetry,larocca2022diagnosing}, and all circuit transformations lie in the Lie group corresponding to the algebra.
\begin{definition}[Dynamical Lie algebra and group]
    Given a parametrized quantum circuit's set of generators $\mc G$, its \emph{dynamical Lie algebra (DLA)} is $\mf{g} =\Span_{\mathbb{R}}\langle \mc G\rangle_{{\rm Lie}}$, and its \emph{dynamical Lie group (DLG)} is $\exp(\mf g)$.
\end{definition}
\noindent Here, $\langle\cdot\rangle_{{\rm Lie}}$  is the Lie closure, i.e.,~the set of nested commutators of operators in $\mc G$.

As the DLA is closed (under commutators) and $U(\btheta)$ lies in the DLG, the effective generators are guaranteed to be part of the DLA.
This also means we can bound the maximum number of independent directions the circuit can explore through the dimension of its DLA.
As a consequence, simulating parametrized quantum circuits is efficient in $\mathrm{dim}(\mf g)$~\cite{somma2005quantum,goh2023lie} and Hamiltonian simulation can be performed efficiently in the dimension of the DLA associated to the Hamiltonian~\cite{kokcu2022fixed}.

At a mathematical level, the partial derivatives of $U(\btheta)$provide us with local coordinates in $\mf{u}(d)$, the unitary Lie algebra, which also is the tangent plane at the identity on the Lie group $\mc U(d)$. 
That is, $\Omega_j(\btheta)$ corresponds to a vector in the DLA, within the vector space $\mf{u}(d)$ (see \cref{fig:general_decomp}).
It is worth remarking that these coordinates may contain redundancies and usually will not form an orthonormal basis with respect to any canonical metric. 

The previous geometric perspective indicates that performing an optimization step $\btheta\rightarrow \btheta+\vec{\delta}$ translates into changing the unitary along some direction in the DLA $\mf g\subset \mf u(d)$.
In standard gradient-based methods, the direction of change is determined by that of steepest descent in the cost function landscape.
In contrast, here we consider the general question: \emph{how can we modify the standard derivative in such a way that we optimize the cost while respecting a given set of constraints?}
Specifically, we will consider a general partition of the tangent plane of the form (see \cref{fig:general_decomp})
\begin{equation}\label{eq:general_algebra_split}
    \mf{u}=\mf{u}^\|\oplus \uperp\,,
\end{equation}
and assume that our goal is to only allow parameter updates that move along directions in $\mf{u}^\|$, but not along $\uperp$.
At an infinitesimal level, this can be achieved by projecting the direction of change in $\mf u$ onto $\mf u^\|$ and modifying the parameter update direction accordingly.
We call this strategy \emph{projected derivative} and formalize it as
\begin{align}\label{eq:projected_derivative}
    \partial_j^{[\mf u^\|]} U(\btheta) = \mbb P_{\mf u^\|} \partial_j U(\btheta),
\end{align}
where $\mbb P_{\mf u^\|}$ is the projector onto $\mf u^\|$ and is implicitly defined to act at a specific point during the parametrized quantum circuit.
As an example, $\mbb P_{\mf u^\|}$ could act after the circuit, so that we project the left-effective generator $\Omega_j^L(\btheta)$.
We will discuss examples for such a decomposition of $\mf u(d)$ based on continuous symmetry groups and their Lie algebras, resulting in the equivariant and covariant derivatives.
There, the action of the symmetry group will tell us where in the circuit the projector is applied.

An interesting open question is to characterize the result of training quantum gates or parametrized states via the projected derivative.
This can be expected to depend on the initial position, the projection, and whether infinitesimal or finite-stepsize learning is considered.

\subsection{Group representations and actions}\label{sec:preliminaries:symmetries}
Here we discuss symmetry groups and how they interact with unitaries, observables, and parametrized quantum circuits.
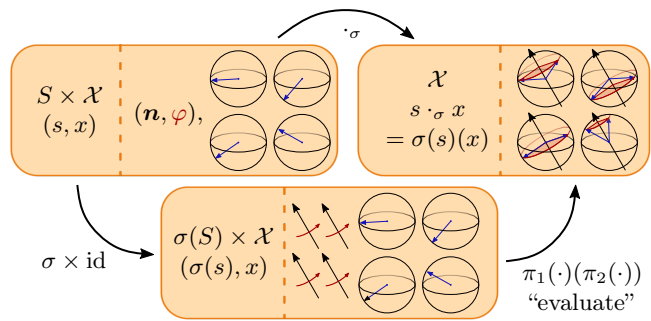
\begin{figure}
    \centering
\begingroup%
  \makeatletter%
  \providecommand\color[2][]{%
    \errmessage{(Inkscape) Color is used for the text in Inkscape, but the package 'color.sty' is not loaded}%
    \renewcommand\color[2][]{}%
  }%
  \providecommand\transparent[1]{%
    \errmessage{(Inkscape) Transparency is used (non-zero) for the text in Inkscape, but the package 'transparent.sty' is not loaded}%
    \renewcommand\transparent[1]{}%
  }%
  \providecommand\rotatebox[2]{#2}%
  \newcommand*\fsize{\dimexpr\f@size pt\relax}%
  \newcommand*\lineheight[1]{\fontsize{\fsize}{#1\fsize}\selectfont}%
  \ifx\svgwidth\undefined%
    \setlength{\unitlength}{241.9234869bp}%
    \ifx\svgscale\undefined%
      \relax%
    \else%
      \setlength{\unitlength}{\unitlength * \real{\svgscale}}%
    \fi%
  \else%
    \setlength{\unitlength}{\svgwidth}%
  \fi%
  \global\let\svgwidth\undefined%
  \global\let\svgscale\undefined%
  \makeatother%
  \begin{picture}(1,0.49227529)%
    \lineheight{1}%
    \setlength\tabcolsep{0pt}%
    \put(0,0){\includegraphics[width=\unitlength,page=1]{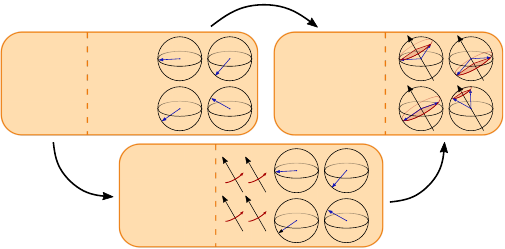}}%
    \put(0.66711471,0.36338591){\makebox(0,0)[t]{\lineheight{1.25}\smash{\begin{tabular}[t]{c}$\mathcal{X}$\\$s\cdot_\sigma x$\\$=\sigma(s)(x)$\end{tabular}}}}%
    \put(0.24770786,0.31096568){\makebox(0,0)[t]{\lineheight{1.25}\smash{\begin{tabular}[t]{c}$(\vec{n}, \textcolor{fig_2_red}{\varphi}),$\end{tabular}}}}%
    \put(0.09424039,0.33854234){\makebox(0,0)[t]{\lineheight{1.25}\smash{\begin{tabular}[t]{c}$S\times \mathcal{X}$\\$(s, x)$\end{tabular}}}}%
    \put(0.33426578,0.11793964){\makebox(0,0)[t]{\lineheight{1.25}\smash{\begin{tabular}[t]{c}$\sigma(S)\times \mathcal{X}$\\$(\sigma(s), x)$\end{tabular}}}}%
    \put(0.89266795,0.06220901){\makebox(0,0)[t]{\lineheight{1.25}\smash{\begin{tabular}[t]{c}$\pi_1(\cdot)(\pi_2(\cdot))$\\``evaluate''\end{tabular}}}}%
    \put(0.51869801,0.44322014){\makebox(0,0)[lt]{\lineheight{1.25}\smash{\begin{tabular}[t]{l}$\cdot_\sigma$\end{tabular}}}}%
    \put(0.04994793,0.08065769){\makebox(0,0)[lt]{\lineheight{1.25}\smash{\begin{tabular}[t]{l}$\sigma\times \mathrm{id}$\end{tabular}}}}%
  \end{picture}%
\endgroup%

    \caption{\textbf{A group action of a symmetry group $S$ on a $d$-dimensional vector space $\mc X$} based on a linear representation $\sigma$ of $S$ on $\mc X$.
    $\sigma$ maps group elements to invertible linear transformations on $\mc X$---elements of $GL(\mc X)$---and the corresponding action consists of applying the result to a vector space element $x$.
    As an example, we depict a rotation $(\vec{n},\varphi)=s\in S=\mc{SU}(2)$ that acts on four spins by applying the same rotation to each spin individually.
    Accordingly, $\sigma$ is the tensor power representation on four qubits.
    Any representation we will consider consists of a mapping to a complex-valued matrix and a rule of how to apply this matrix to $x$ in order to realize the transformation in $GL(\mc X)$.
    This could be simple matrix multiplication from the left, or could be less generic, as in the third example in the main text.
    $\pi_{1,2}$ are the canonical projections onto the first and second elements used here to formalize the function ``evaluate''.
    }
    \label{fig:group_rep_action}
\end{figure}
%
We start with the standard definition of a group representation, limited to unitary (sub)groups:
\begin{definition}[Group representation]
    A representation $\sigma$ of $S<\mc U$ on a vector space $\mc X$ maps from $S$ to the group $GL(\mc X)$ of general linear transformations on $\mc X$
    and preserves the group structure, i.e.,~is a group homomorphism. In short:
    \begin{align}
        \sigma
        : S &\to GL(\mc X)\\
        s &\mapsto \sigma(s),
        \quad \text{with}\quad
        \sigma(st)=\sigma(s)\sigma(t).\nonumber
    \end{align}
\end{definition}
A group element $s$ acting on a vector $x\in\mc X$ is denoted as $\sigma(s)(x)$.
We abbreviate this as $s\act{\sigma}x$, using the \emph{group action} $\act{\sigma}$ induced by the representation $\sigma$, which we depict in \cref{fig:group_rep_action}.
Not all group actions stem from representations, but we often can make actions explicit via a representation on a vector space.

We give two common examples and one less common example for representations.
First, there is the standard representation of the group $\mc U(d)$ itself as $d$-dimensional complex matrices.
Lie groups often are understood---or even defined---via this representation, identifying the abstract group elements with these matrices.
We sometimes denote group elements next to vectors directly, implying that this representation has been applied.
Note that even if we fix the matrices for each group element, different representations can be associated to this.

Second, consider the adjoint action of a subgroup $S$ on $\mc U$, given by conjugation\footnote{An example of an action not induced by a representation.}: 
\begin{align}
    \Psi_s : G\to G : g\mapsto sgs^{-1}.
\end{align}
The adjoint representation $\rm{Ad}$ of $S$ on the Lie algebra $\mf u$ of $\mc U$ is given by the derivative of $\Psi$ at the group identity $e$.
Using the standard representation of $\mc U$ (and of $\mf u$) from above, this can be written as $\mathrm{Ad}_s(x)=s\act{\mathrm{Ad}}x=s x s^{-1}$.
If we extend the vector space from $\mf u$ to all matrices\footnote{This is a complexification, as we include $i\mf u$ with this step.} of the given size, the adjoint representation captures how $s$ acts on a Hamiltonian $H$ in the Heisenberg picture, or on a quantum state $\rho$ in the Schr\"odinger picture.

Our third example is less generic.
Consider cost functions given by the expectation value of any observable $M$ and a fixed PQC with initial state $\rho_0$:
\begin{align}
    C_M(\btheta) = \tr{M U(\btheta) \rho_0 U^\dagger(\btheta)},\  M^\dagger=M.
\end{align}
This is a vector space over $\mbb R$ if we define addition and scalar multiplication via the observable, $(C_M+\alpha C_N)=C_{M+\alpha N}$.
A representation of $S<\mc U$ on this vector space is given by
\begin{align}
    (s\act{}C)(\btheta) = C_{s^\dagger\act{\mathrm{Ad}}M}(\btheta) = \tr{M s U(\btheta)\rho_0 U^\dagger(\btheta)s^\dagger}\nonumber.
\end{align}
Clearly, we simply inherited the vector space structure and the adjoint representation from the space of observables above to this new vector space.
Nonetheless, this is a useful construction because it showcases how to apply symmetry-aware derivatives to cost functions directly.

\section{Equivariance and Covariance}\label{sec:equi_and_co}
Depending on the field of research and on context, the words \emph{covariance}, \emph{equivariance}, and \emph{invariance} with respect to a symmetry are used for different concepts, and sometimes some of them are used interchangeably, which can cause confusion.
We thus want to be clear:
\begin{redbox}
  Equivariance and covariance in general are not the same throughout this work.
  Invariance is a special case of equivariance.
\end{redbox}

In \cref{sec:equi_and_co:equivariance_groups,sec:equi_and_co:equivariance_algebras}, we will treat equivariance and apply it to groups and algebras, corresponding to a global and a local perspective, respectively.
This is followed by \cref{sec:equi_and_co:covariance_algebras,sec:equi_and_co:covariance_groups}, which discuss covariance, and \cref{sec:equi_and_co:connect}, which interrelates equivariance and covariance.
While we aim to give a comprehensive overview, this section keeps explanations brief and readers curious about deeper aspects of differential geometry and Lie theory may appreciate Refs.~\cite{hall2015lie_1,fulton2004representation_1,hamilton2017mathematical,hackl2021principal}.
For a review oriented to QML, consider~\cite{ragone2022representation}.
To complement this, we present a less abstract approach to covariant derivatives in \cref{sec:circuitry_app}, focusing on quantum circuits.

\subsection{Equivariance--groups}\label{sec:equi_and_co:equivariance_groups}

\begin{definition}[Equivariance]\label{def:equivariance}
Consider a group $S$ with actions $\act{\mc X}$ and $\act{\mc Y}$ on two spaces $\mc{X}$ and $\mc{Y}$. A map $\phi:\mc{X}\rightarrow \mc{Y}$ is called equivariant if it satisfies
\begin{equation}
    \phi(s\act{\mc X}x) = s\act{\mc Y}\phi(x)
\end{equation}
for all $s\in S$ and $x\in \mc{X}$.
\end{definition}

That is, applying a symmetry transformation $s$ to $x\in\mc X$ and mapping to $\mc Y$ via $\phi$ is the same as using $\mathrm{id}\times\phi$ and applying $s$ to $\phi(x)\in\mc Y$.
As a commutative diagram:
\[
\begin{tikzcd}[column sep=large,row sep=large]
    \mlnode{$S\times \mc X$\\$(s, x)$} \arrow{r}{\act{\mc X}} \arrow[swap]{d}{\mathrm{id} \times \phi } & \mlnode{$\mc X$\\$s\act{\mc X}x$} \arrow{d}{\phi} \\
    \mlnode{$S\times \mc Y$\\$(s,\phi(x))$} \arrow{r}{\act{\mc Y}} & \mlnode{$\mc Y$\\$s\act{\mc Y}\phi(x)$\\$=\phi(s\act{\mc X}x)$}
\end{tikzcd}\,.
\]
In particular, if $\mc X=\mc Y$ and $\act{\mc X}=\act{\mc Y}$, $\phi$ commutes with the group action.

We apply this general definition of equivariance to a number of relevant maps.
First, a unitary operator $U$, as a map from $\mc U$ to $\mc U$, is usually considered equivariant if it commutes with the left(right)-regular action $\act{L(R)}$, which is equivalent to the commutation relation $Us=sU$ on the group level.
For a representation $\sigma$ of $S$, the relation reads $[U, \sigma(s)]=0$ on the matrix level.
Parametrized unitaries are equivariant if they are equivariant for all input parameters $\btheta$.

Second, take a data encoding $W:\R^p\to\mc U$, as commonly used in QML to encode classical data in a quantum state.
It is considered equivariant if it is compatible with a symmetry action $\act{\rm data}$ on the data points themselves and the adjoint action $\act{\mathrm{Ad}}$ on $\mc U$, in short $W(s\act{\rm data} \vec{x})=s W(\vec{x}) s^\dagger$.
This way, symmetry-related data points are encoded as symmetry-related state vectors, which is captured by the commutative diagram
\[
    \begin{tikzcd}[column sep=1.75cm]
        \mlnode{$S\times \R^p$\\$(s, \vec{x})$} \arrow{r}{\act{\mathrm{data}}} \arrow[swap]{d}{\mathrm{id} \times W } & \mlnode{$\R^p$\\$s\act{\mathrm{data}}\vec{x}$} \arrow[shorten >=-10pt]{d}{W} \\
        \mlnode{$S\times \mc U$\\$(s,W(\vec{x}))$} \arrow{r}{\act{\mathrm{Ad}}} & \mlnode{\\ $\mc U$\\$W(s\act{\mathrm{data}}\vec{x})$\\$=s\act{\mathrm{Ad}}W(\vec{x})$}
    \end{tikzcd}\,.
\]

While both the data encoding and the parametrized unitary discussed above are maps from real vectors to $\mc U$, the notion of equivariance for the two differs significantly; the encoding relates a symmetry action on the input vector to an action on quantum states, whereas the parametrized unitary simply commutes with the action on quantum states.

Third, \emph{invariance} of a map $\phi$, i.e.,~$\phi(s\act{\mc X}x)=\phi(x)$ for all $s \in S$, is a special case of equivariance:
\begin{highlight}
    If the action on the target space $\mc{Y}$ is the trivial ``do nothing'' action, we find
    \begin{align}
        \phi \text{ invariant} \ \ \Leftrightarrow\ \ \phi(s\act{\mc{X}} x)=s\act{\mc{Y}} \phi(x) = \phi(x).\nonumber
    \end{align}
    Conversely, an invariant map implies that the action on $\phi(\mc X)\subset\mc Y$ must be trivial.
\end{highlight}

For statements about equivariance and invariance, it is important to consider the symmetry group actions explicitly.
For example, the matrix-level condition $[\rho, \sigma(s)]=0$ may or may not imply invariance, depending on the context, as we discuss in \cref{sec:invariance_or_not}. 

The following definition allows us to rewrite equivariance, but also will be useful for us more generally.
\begin{definition}[Commutant]
Given two sets $A$, $B$ of (square) matrices with $A\subset B$ we define the commutant of $A$ (in $B$) by
\begin{equation}
    B^A=\{ X\in B | [X, Y]=0\ \forall Y\in A\}.
\end{equation}
\end{definition}
Note that we require $A\subset B$, not because it is necessary to make the definition sound, but because we will only care about this case and it provides us with a nice property:
if $B$ is a group (algebra), $B^A$ is a subgroup (subalgebra); see \cref{lemma:commutant_algebra} for the algebraic level.
We could define the commutant for a subgroup of a Lie group on a more abstract level without reference to the matrix commutator.
However, we usually care about the commutant with respect to some representation of a group as matrices, so the above will be sufficient.

Writing the represented symmetry group as $\sigma(S)$, the condition for $U$ being equivariant can now be restated as $U \in \mc{U}^{\sigma(S)}$, and similarly for parametrized unitaries with $U(\btheta)\in \mc{U}^{\sigma(S)}\ \forall\btheta$.
A more abstract definition of the commutant allows to rephrase the equivariance of general maps (like the data encoding $W$ or the density matrix $\rho$ above) between distinct vector spaces.

\subsection{Equivariance--algebras}\label{sec:equi_and_co:equivariance_algebras}
So far, we focused on equivariance at the group level.
We now look at the algebraic level, paving the way for a \emph{local} implementation of equivariance.
From here on, we restrict ourselves to Lie groups.
Concretely, our playing field is $\mf u$, the Lie algebra of the unitary group $\mc U$.
\begin{definition}[Algebra representation]
    A representation $\sigma$ of a subalgebra $\mf s \subset \mf u$ on a vector space $\mc X$ maps from $\mf s$ to the tangent space of $GL(\mc X)$, which is isomorphic to $GL(\mc X)$ itself. In addition, it preserves the Lie bracket so that
    \begin{align}
        \sigma :\mf s &\to GL(\mc X)\\
        x&\mapsto \sigma(x),\quad \text{with} \quad \sigma([x, y])=[\sigma(x), \sigma(y)].\nonumber
    \end{align}
\end{definition}
The differential of a Lie group representation defines a representation on its Lie algebra, which is guaranteed to exist due to the Lie group-Lie algebra correspondence.
This holds vice versa iff the group is simply connected: a Lie algebra representation induces a representation of $\exp(\sigma(\mf s))$ in this case~\cite{lee2012smooth}.
Note that this makes the following less interesting for discrete groups, which are not simply connected.
\begin{convention}
    We denote a group representation and the associated algebra representation with the same symbol.
    From here onwards, we write the represented symmetry algebra as $\mf t=\sigma(\mf s)$ and assume that $S$ is connected and compact.
\end{convention}

The commutants of a represented Lie subgroup and its algebra $\mf t\subset \mf u$ play together nicely:
the commutant $\mf u^{\sigma(S)}$ of $\sigma(S)<\mc U$ in $\mf u$ is itself a subalgebra, and equal to the commutant $\mf u^{\mf t}$ of the represented algebra $\mf t$ in $\mf u$ (\cref{lemma:group_and_algebra_commutants}).
It is equal to the Lie algebra of $\mc U^{\sigma(S)}$, the group of equivariant quantum gates (\cref{lemma:algebra_of_commutant}).
This fact makes equivariant PQCs well-defined in the first place, and allows us to construct them using gate generators from $\mf u^{\mf t}$~\cite{meyer2023exploiting};
for any $x\in\mf u^{\mf t}$, we obtain an equivariant parametrized gate in $\mc U^{\sigma(S)}$ via $\exp(\theta x)$.
We thus call $\mf u^{\mf t}$ the equivariant subspace of $\mf u$.

\begin{convention}
    We denote orthogonal complements as $\overline{\,\cdot\,}$, while $\perp$ is reserved as part of the name $\uperp$ for the space of constrained directions.    
\end{convention}
The commutant and its orthogonal complement $\uperp=\orth{\mf u^{\mf t}}$ with respect to some inner product\footnote{Inner products need to be non-degenerate by definition. This requires the algebra representation $\sigma$ to be faithful, i.e.,~$\sigma(x)\neq 0\ \forall x\neq 0$.} provides an orthogonal split of $\mf u$ as discussed in \cref{sec:preliminaries:pqc}:
\begin{align}\label{eq:equivariant_decomp_algebra}
    \mf u = \mf u^{\mf t} \oplus \uperp.
\end{align}
We now have all the ingredients to obtain the projected derivative (see \cref{eq:projected_derivative}) onto $\mf u^{\mf t}$.
\begin{highlight}
    The \emph{equivariant derivative} is the projected derivative onto the equivariant subspace $\mf u^{\mf t}$:
    \begin{align}\label{eq:equivariant_derivative_abstract}
        E_j f(\btheta) \coloneqq \partial_j^{[\mf u^{\mf t}]}f(\btheta)= \mbb P_{\mf u^{\mf t}} \partial_j f(\btheta),
    \end{align}
    where the projection acts on the derivative of a parametrized unitary $U(\btheta)$ that is used within some function $f$.
\end{highlight}
That is, the equivariant derivative projects the common derivative to the equivariant subspace that commutes with the symmetry group.
To the best of our knowledge, this is a new tool to incorporate symmetries into PQCs at a local level.

While we required a representation for the above decomposition, it is independent of the point in the vector space $\mc X$ on which the represented group acts.
Crucially, the covariant derivative introduced in \cref{sec:equi_and_co:covariance_algebras} will not have this property.
We will discuss the equivariant derivative in detail for gates (\cref{sec:circuitry:gates}), quantum states (\cref{sec:circuitry:states}), and PQC cost functions (\cref{sec:circuitry:cost_functions}).

\paragraph*{Twirling.}
A useful tool to create and test for equivariant algebra elements is twirling.
The twirling (super)operator with respect to a represented symmetry group $\sigma(S)$ acts on unitary algebra elements and is defined as
\begin{align}
    \mc T_{\sigma (S)} : \mf u &\to \mf u\\
    x&\mapsto \mc T_{\sigma(S)}[x] = \int_S \mathrm{d}\mu(s) \sigma(s) x \sigma(s)^\dagger\nonumber,
\end{align}
where $\mu$ is the (normalized) Haar measure of the symmetry group, and the group and algebra elements interact via matrix multiplication.

Twirled vectors are equivariant, i.e.,~they lie in $\mf u^{\mf t}$, because shifts within $S$ do not change the Haar measure\footnote{As we assumed $S$ to be connected, $\mf u^{\mf t} = \mf u^{\sigma(S)}$; see \cref{lemma:group_and_algebra_commutants}.}:
\begin{align}
    \sigma(t) \mc T_{\sigma(S)}[x] \sigma(t)^\dagger 
    &= \int_S \mathrm{d}\mu(s) \sigma(ts) x \sigma(ts)^\dagger\nonumber\\
    &= \mc T_{\sigma(S)}[x].
\end{align}
At the same time, any $x\in\mf u^{\mf t}$ is unaffected by $\mc T_{\sigma(s)}$ due to $\sigma(s)x\sigma(s)^\dagger=x$.
Overall, this makes the range of $\mc T_{\sigma(S)}$ equal to $\mf u^{\mf t}$.
This equivalence between twirled and equivariant operators led to the suggestion to construct equivariant quantum circuits using twirled gate generators \cite{meyer2023exploiting,nguyen2022theory}.

Note that $\mc T_{\sigma(S)}$ is idempotent, i.e.,~applying it a second time does not have any effect,
\begin{align}
    \mc T_{\sigma(S)}^2[x]
    &= \int_S \mathrm{d}\mu(t)\int_S \mathrm{d}\mu(s) \sigma(t)\sigma(s) x \sigma(s)^\dagger\sigma(t)^\dagger\nonumber\\
    &= \int_S \mathrm{d}\mu(s') \sigma(s') x \sigma(s')^\dagger\nonumber\\
    &= \mc T_{\sigma(S)}[x].
\end{align}
Here we used the shift-invariance of $\mu$ again, and that $\sigma$ is a group homomorphism.
We conclude that $\mc T_{\sigma(S)}$ is an idempotent linear operator onto $\mf u^{\mf t}$, making it the projector onto the equivariant subalgebra, i.e.,~the operator $\mbb P_{\mf u^{\mf t}}$ in \cref{eq:equivariant_derivative_abstract}.

\subsection{Covariance--algebras}\label{sec:equi_and_co:covariance_algebras}
\begin{figure}
    \centering
    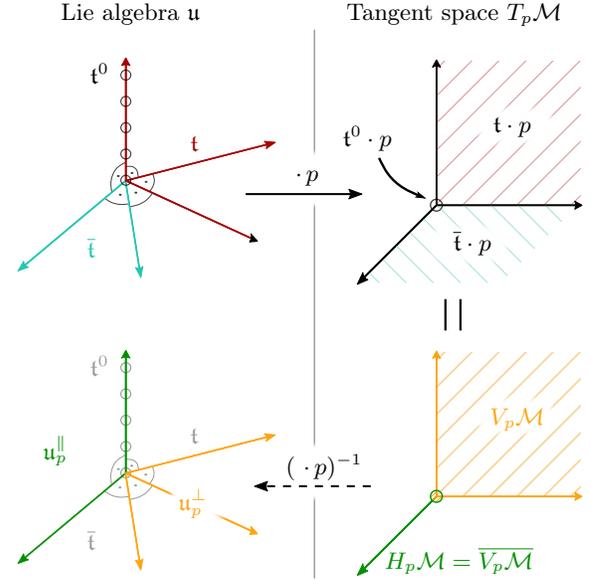
    \caption{\textbf{Local notion of covariance.} Relevant subspaces of the Lie algebra $\mf u(d)$ and the tangent space $T_p\mc M$ of a manifold $\mc M$ on which $\mc U(d)$ acts, with five (three) orthogonal directions in the left (right) column.
    (\emph{top left}) the unitary Lie algebra decomposes as $\mf u =\mf t\oplus \orth{\mf t}$, which is an example for \cref{eq:general_algebra_split}.
    Algebra elements act on a point $p\in\mc M$ via the group action, sending them to $T_p\mc M$ via the map $\act{}p$.
    (\emph{top right}) this map leads to (potentially overlapping) subspaces $\mf t\act{} p$ and $\orth{\mf t}\act{}p$ of the tangent space $T_p\mc M$.
    Some subspace $\mf t^0$ of $\mf t$ may produce the zero tangent only.
    (\emph{bottom right}) the vertical subspace of $T_p\mc M$ is $V_p\mc M=\mf t\act{} p$, the covariant/horizontal subspace $H_p\mc M$ is the orthogonal complement (\cref{def:equi_and_co:vertical_tangents}).
    The zero tangents, including $\mf t^0\act{} p$, are part of both the vertical and the covariant subspace.
    (\emph{bottom left}) a new decomposition of $\mf u$ can be defined by considering the preimage $\mf u^\|_p$ of $H_p\mc M$ with respect to $\act{}p$ as covariant subspace (\cref{eq:pull_back_horizontal_space}).
    The vertical subspace $\uperp_p$ is its orthogonal complement.
    As the map $x\mapsto x\act{} p$ is not injective in general, the obtained decomposition of $\mf u$ need not coincide with that in the top left.
    }
    \label{fig:covariance_algebras}
\end{figure}
For equivariance, we first discussed groups from a global perspective and then the local, algebraic perspective.
For covariance we start local.
Covariance is a concept used and defined rather differently in various contexts.
Physicists tend to identify objects by their transformation behaviour (``A vector is something that transforms like a vector''); also see \cref{sec:gauge_theory}.
From a mathematical perspective, the covariant derivative is induced on physical fields (``associated vector bundle'') by that on unphysical gauge fields (``principal bundle'').
Often, if not always, these definitions can be shown to be equivalent.

Our starting point is another decomposition of $\mf u$ like that in \cref{eq:general_algebra_split}. It is again based on a symmetry group $S$, specifically on the represented Lie algebra $\mf t=\sigma(\mf s)$:
\begin{align}\label{eq:covariant_canonical_decomp}
    \mf u(d) = \mf t \oplus \orth{\mf t}.
\end{align}
Here, $\mf t$ itself forms the constraining space $\uperp$, and its orthogonal complement takes the role of $\mf u^\|$.
That is, for covariance we want to move in all directions but those caused by symmetry transformations, as the latter do not impact the system we are describing.

In addition to the unitary group and its algebra, we usually also care about the action of $\mc U$ on some manifold $\mc M$.
To connect the decomposition of $\mf u$ above with this action, the \emph{tangent space} of $\mc M$ and its \emph{vertical} and \emph{covariant} subspaces will play a central role.
In short, the tangent space consists of the changes caused by an infinitesimal unitary group action on $\mc M$\footnote{Technically, tangents at a point $p\in \mc M$ are equivalence classes of curves that pass through $p$.
Essentially, here we will work with representatives of these classes directly.}, and the vertical space restricts to changes caused by actions of a represented symmetry group $\sigma(S)$.
\begin{convention}
    In this work we only ever consider tangents arising from unitary transformations, independent of the full tangent space of $\mc M$.    
\end{convention}
We will mostly be concerned with $\mc M$ being the unitary group $\mc U(d)$ or the Hilbert space $\mc H$, with the group actions discussed in \cref{sec:equi_and_co:equivariance_groups}. 
We sketch the tangent space and its relevant subspaces in \cref{fig:covariance_algebras}.

\begin{definition}[Tangent space \& its subspaces]\label{def:equi_and_co:vertical_tangents}
    Consider an action $\act{}$ of the unitary group $\mc U(d)$ on a smooth manifold $\mc M$, and a symmetry Lie group $S$ with Lie algebra $\mf s$ and a unitary representation $\sigma$.
    The \emph{tangent space} of $\mc M$ at a point $p\in\mc M$ and its \emph{vertical subspace} are
    \begin{align}
        T_p\mc M &= \left \{\frac{d}{dt} \left(\exp(xt) \act{} p\right)\big|_{t=0} | x\in \mf u\right\}\\
        V_p\mc M &= \left \{\frac{d}{dt} \left(\exp(xt) \act{\sigma} p\right)\big|_{t=0} | x\in \mf s\right\},
    \end{align}
    where $s\act{\sigma}p = \sigma(s)\act{} p$.
    The \emph{covariant}, or \emph{horizontal}, subspace is the orthogonal complement of $V_p\mc M$ within $T_p\mc M$, for some inner product\footnote{Geometrically, a \emph{connection} provides the covariant subspaces at each $p\in\mc M$. Here we restrict ourselves to the case where this connection stems from an inner product.}:
    \begin{align}
        H_p\mc M = \orth{V_p \mc M}.
    \end{align}
\end{definition}

\begin{convention}
    We will stick to the name ``covariant'' for horizontal objects.
\end{convention}
The vertical space is called vertical because it is projected away when dividing out symmetry transformations, which is traditionally drawn in a vertical direction.
While this convention is rather arbitrary, it is used consistently in the literature.

\begin{convention}
Extending the notation for the group actions to the algebra, we may rewrite the tangent space and its subspaces as
\begin{alignat}{3}
    T_p \mc M &= \mf u \act{} p \quad
    V_p \mc M &&= \mf s \act{\sigma} p \quad 
    H_p \mc M &&= \orth{\mf s \act{\sigma} p}\\
    &
    &&=\mf t\act{} p
    &&=\orth{\mf t \act{} p}\ .
\end{alignat}
\end{convention}
In mathematics, Lie algebra elements are set in relation to tangents of $\mc M$ via so-called action fields~\cite{ziller2013group}.
Because we only ever consider tangents arising from unitary transformations, the map $\act{}p:\mf u\to T_p\mc M$ is surjective.

We want to make the above more concrete with two examples.
First, consider the right-regular action $\act{R}$ of $\mc U(d)$ on itself, i.e.,~$\mc M=\mc U(d)$ and $g\act{R} g'=g' g^\dagger$.
In addition, take a $d$-dimensional matrix representation $\sigma$ of $S<\mc U(m)$\footnote{$m$ and $d$ do not need to match.}, so that
\begin{align}
    s\act{R,\sigma} U = U\sigma(s) \ \forall s\in S \text{ and } x\act{R,\sigma} U = U\sigma(x)\ \forall x\in \mf s\nonumber.
\end{align}
The tangent space at a point $U\in \mc U(d)$ then is $U\mf u(d)$ and its vertical subspace is $U\mf t$.
If we choose a unitarily invariant inner product on $U\mf u$, the covariant subspace satisfies $H_U\mc U(d)=\orth{U\mf t}=U \orth{\mf t}$.
That is, the decompositions of the tangent spaces $U\mf u(d)$ and $V\mf u(d)$ at different points $U, V\in \mc U(d)$ are compatible with each other.
In particular, they are consistent with the one at $U=\mbb I$, which is the canonical decomposition in \cref{eq:covariant_canonical_decomp}.
For a parametrized unitary acting on the unitary group itself, we recognize that the derivatives $U(\btheta)\Omega_j^R(\btheta)$ match the description $U\mf u(d)$.

The second example is the standard action $\act{\mc H}$ of $\mc U(d)$ on the Hilbert space $\mc M=\mc H$, i.e.,~$g\act{\mc H}\ket{\psi}=g\ket{\psi}$, again with a representation $\sigma$ of a unitary subgroup $S<\mc U(m)$ represented on $\mc U(d)$:
\begin{align}
    s\act{\mc H, \sigma} \ket{\psi} &= \sigma(s)\ket{\psi}\ \forall s\in S\\
    \text{ and }\ x\act{\mc H, \sigma} \ket{\psi}&= \sigma(x)\ket{\psi}\ \forall x\in \mf s.
\end{align}
The tangent space is given by $T_\psi \mc H=\mf u(d)\ket{\psi}$ and is a real vector space, because $\mf u$ is.
Its vertical subspace is $V_\psi\mc H=\mf t\ket{\psi}$ and the covariant subspace $H_\psi \mc H=\orth{\mf t\ket{\psi}}$ is defined via an inner product on $T_\psi \mc H$.
In contrast to the first example, we have $\orth{\mf t\ket{\psi}}\neq \orth{\mf t}\ket{\psi}$, because the inner product on $T_\psi \mc H$ is not equivalent to that on $\mf u$.

From a gauge theoretical point of view, the first example can be formalized as a principal bundle, which is the object to describe gauge fields, whereas the second example is an associated vector bundle, typically used to describe physical quantities.
It is noteworthy that in our setup, the principal bundle becomes tangible in measurements and optimization tasks, while the gauge fields in particle physics are not.

\begin{highlight}
    The covariant derivative is the derivative projected onto the covariant subspace $H_p\mc M$:
    \begin{align}\label{eq:covariant_derivative_abstract}
        D_j f(\btheta) = \mbb H_p \partial_j f(\btheta)=(\mbb I - \mbb V_p) \partial_j f(\btheta),
    \end{align}
    where we imply $f$ to depend on $\btheta$ via a parametrized unitary $U(\btheta)$ that acts on $\mc M$.
\end{highlight}

The covariant (vertical) projector $\mbb H_p$ ($\mbb V_p$) is defined on the tangent space of $\mc M$, which is specified by the action of $U(\btheta)$.
In contrast, the projector $\mbb P_{\mf u^{\mf t}}$ in \cref{eq:equivariant_derivative_abstract} is defined on the tangent space of the unitary group itself, which is $\mf u$.
In the next section, we show how to define a projector on $\mf u$ for the covariant derivative as well, making it a projected derivative as introduced in \cref{eq:projected_derivative}.
\begin{highlight}
    Note that $\mbb H_p$ depends on $p\in\mc M$, while the equivariant projector does not, because the tangent space decomposition changes with $p$.    
\end{highlight}

We stressed before that equivariance and covariance, as defined in this work, are different concepts.
There is a good reason to call the covariant derivative equivariant, though: as a map on functions, $D_j$ is linear and commutes with local symmetry transformations.
We show this property on an abstract level in \cref{sec:covariant_derivative_is_equivariant}, and in the context of gauge theory in \cref{sec:gauge_theory,sec:gauge_theory_comp}.
For the latter, covariance is identified via the symmetry transformation behavior, so that equivariance becomes the \emph{defining} property to call an object covariant, making the confusion perfect.

\subsubsection{Splitting \texorpdfstring{$\mf u$}{u} via \texorpdfstring{$T_p\mc M$}{TpM}}\label{sec:induced_split}
In \cref{eq:covariant_canonical_decomp} we showed a decomposition of $\mf u$ into $\mf t$ and its complement.
Based on the decomposition of the tangent space $T_p\mc M$ caused by a group action, we can define yet another split of $\mf u$, as shown in \cref{fig:covariance_algebras}.
For this, we ``pull back'' the covariant space to $\mf u$:
fix a point $p\in \mc M$ and consider the preimage 
\begin{align}\label{eq:pull_back_horizontal_space}
    \mf u^\|_p\coloneqq \DE{x\in \mf u| x\act{}p \in H_p\mc M}.
\end{align}
This is a vector space and could be called a covariant subspace of $\mf u$, as it is derived from $H_p\mc M$.
We then construct the desired split using an inner product of our choice on $\mf u$, producing the ``new'' vertical subspace $\uperp_p$ as the orthogonal complement of $\mf u^\|_p$.
This induced decomposition of $\mf u$ lets us write the covariant derivative as a special case of \cref{eq:projected_derivative}:
\begin{align}
    D_j f(\btheta) = \partial_j^{[\mf u^\|_p]}f(\btheta).
\end{align}

Note that elements in $\mf t^0\subset \mf t$ end up in $\mf u^\|_p$, because they produce the zero tangent $0\in H_p\mc M$.
Conversely, a vector in $\orth{\mf t}$ may produce a tangent in $V_p\mc M$, so that it is not in the preimage $\mf u^\|_p$.
As a consequence, the decomposition into $\mf u^\|_p$ and $\uperp_p$ does differ from that into $\mf t$ and $\orth{\mf t}$.
What is more, neither of $\mf u^\|_p$ and $\uperp_p$ need be a subalgebra.

This mismatch between the decompositions is corrected if the group action is free, i.e.,~if only the group identity can preserve any point $p$: $g\act{}p=p\Rightarrow g=e$.
Equivalently, the stabilizer group of all points is trivial.
In this case we have $\uperp_p=\mf t$ and $\mf u^\|_p=\orth{\mf t}$ for all $p\in \mc M$, as we show in \cref{lemma:free_action}.

We observe that the group action on $\mc U$ in the first example from above is free, because only $\mbb I\in \mc U(d)$ preserves any unitary matrix, so that $\uperp_U=\mf t$ and $\mf u_U^\|=\orth{\mf t}$.
In contrast, the group action on $\mc H$ from the second example is not free, and thus induces a split that is incompatible with the orthogonal complement $\orth{\mf t}\subset\mf u(d)$.
That is, $\orth{\mf t\ket{\psi}}\neq \orth{\mf t}\ket{\psi}$ and thus $\mf u^\|\neq \orth{\mf t}$.
We give additional explicit examples in \cref{sec:equi_and_co:connect}.

\subsection{Covariance--groups}\label{sec:equi_and_co:covariance_groups}
The definition of the covariant subspace is inherently local, and it is not clear (and in fact not true) that we can always derive a global notion of covariance from this.
As mentioned above, we require a free group action to align the split $T_p\mc M=V_p\mc M\oplus H_p\mc M$ with $\mf u(d)=\mf t\oplus\orth{\mf t}$.
Even then, a global notion of covariance is not guaranteed because unlike the (represented) algebra $\mf t$, the orthogonal complement need not be a subalgebra\footnote{An alternative to requiring a free group action and $\orth{\mf t}$ to be a subalgebra is that the induced covariant space $\mf u^\|_{p}$ is a subalgebra itself.}.
This can be fixed by requiring $\mf t$ to be an \emph{ideal}, or equivalently $\sigma(S)$ to be a normal subgroup, so that the quotient space $\mc U(d)/\sigma(S)$ becomes a group and the covariant subspace of $\mf u(d)$ its Lie algebra.
In this case, we call $\mc U/ \sigma(S)$ the \emph{covariant subgroup}.

The only normal subgroup of $\mc U(d)$ is that of global phases.
Usually, global phases are not of interest, making the covariant subgroup less relevant if $\mc U(d)$ is the total group.
However, hardware constraints or algorithmic design decisions may lead to another (simply connected, compact) total group. 
This can lead to nontrivial covariant subgroups and thus to a new class of symmetry-constrained circuit ans\"atze.

\subsection{Connecting equivariance and covariance}\label{sec:equi_and_co:connect}
In this section we investigate connections between equivariance and covariance, in particular at the algebraic level.
This is possible because the decompositions of the tangent space $\mf u$ in \cref{sec:equi_and_co:equivariance_algebras,sec:equi_and_co:covariance_algebras} are compatible with each other, resulting in a purely equivariant, a purely covariant and a mixed\footnote{Which shall not be called, perish the thought, \emph{equovariant}.} component, as well as a remainder subspace.
This algebraic decomposition will allow us to construct subgroups of $\mc U$ that contain the gates with the corresponding symmetry properties.
In this section we will treat actions of the symmetry group on unitaries and on quantum states separately, as they differ significantly\footnote{As mentioned before, this difference is a consequence of $S\act{}\mc U$ inducing an $S$-principle bundle but $S\act{}\mc H$ inducing an associated vector bundle.}.

\subsubsection{Equivariant and covariant tangents of \texorpdfstring{$\ \mc U$}{U}}\label{sec:equi_and_co:tangent_decomp_U}

\begin{figure}
    \centering
    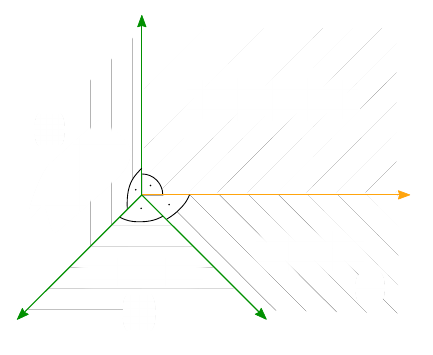
    \caption{\textbf{Decomposition of the unitary algebra $\mf u$ with respect to a symmetry subalgebra $\mf s$,} represented as $\mf t=\sigma(\mf s)$, into four orthogonal subspaces; see \cref{eq:tangent_decomp_U}.
    The (represented) symmetry subalgebra decomposes into $\mf t=\mf z(\mf t)\oplus \tcless$ and the equivariant subalgebra into $\mf u^{\mf t}=\mf z(\mf t)\oplus \utcless$, and all three subspaces are subalgebras (green).
    The remaining orthogonal complement $\mf r$ (orange) is not a subalgebra in general.
    Also see \cref{sec:algebra_4_split_derivation} for details.
    }
    \label{fig:algebra_4_split}
\end{figure}
We start with the comparison of equivariance and covariance for unitaries.
Consider a linear representation of a unitary subgroup $S$ with Lie algebra $\mf s$ on a $d$-dimensional vector space, as well as the tangent space of $\mc U(d)$ at the identity, $T_{\mbb I}\mc U(d)=\mf u(d)$, and pick the trace inner product on $\mf u(d)$.
We will split $\mf u$ into four subspaces based on the relation to symmetry tangents in $\mf s$, and sketch them in \cref{fig:algebra_4_split}.

We start with an intermediate subspace, the commutant $\mf u^{\mf t}$ of $\mf t$ within $\mf u$.
Recall that it consists of all $x\in\mf u$ that commute with all of $\mf t$ and thus is the equivariant subspace of $\mf u$.
The first subspace in our decomposition of $\mf u$ is the \emph{center} $\mf z(\mf t)=\mf u^{\mf t}\cap \mf t$ of $\mf t$.
Note that $\mf z(\mf t)\equiv\mf z(\sigma(\mf s))=\sigma(\mf z(\mf s))$ (\cref{eq:centers_coincide}).
The second subspace is the orthogonal complement of $\mf z(\mf t)$ within $\mf t$, i.e.,~the centerless part of $\mf t$, which we denote as $\tcless$.
Similarly, we may remove the center $\mf z(\mf t)$ from $\mf u^{\mf t}$ to obtain the third subspace $\utcless$.
Note that $\mf z(\mf u^{\mf t})\neq \mf z(\mf t)$ in general.
Finally, the fourth subspace is the orthogonal complement of all we considered so far: $\mf r = \orth{\mf u^{\mf t}\cup \mf t}$.
Overall, we obtain the decomposition
\begin{highlight}
\begin{align}\label{eq:tangent_decomp_U}
    \mf u =
    \underset%
    {\text{covariant }\de{\orth{\mf t}}\hspace{-7pt}}%
    {\underbrace{%
        \mf r \hspace{3pt}\oplus \hspace{3pt}\lefteqn{\hspace{-10pt}%
            \overset%
            {\text{equivariant } \de{\mf u^{\mf t}}}%
            {\overbrace%
                {\phantom{\utcless \hspace{3pt}\oplus \mf z(\mf t)}}%
            }
        }%
        \utcless}%
    }%
    \oplus \underset{\text{vertical } (\mf t)}{\underbrace{\mf z(\mf t) \hspace{3pt}\oplus \hspace{3pt}\tcless}}.
\end{align}
\end{highlight}
%




In \cref{fig:algebra_4_split} we sketch this decomposition.
In \cref{sec:algebra_4_split_derivation} we show that it is well-defined and that all subspaces but $\mf r$ are not only vector spaces but also subalgebras.
A similar decomposition, where the symmetry algebra is replaced by the bond algebra of a spin Hamiltonian has been used to explain Hilbert space fragmentation~\cite{moudgalya2022hilbert}.
As mentioned above, $\mf u^{\mf t}$ is the set of equivariant tangents, or generators, and $\mf t$ contains the symmetry tangents and thus forms the vertical subspace of $\mf u$, so that $\orth{\mf t}$ is the space of covariant, or horizontal, tangents.
Put differently, $\mf z(\mf t)$ contains the purely equivariant and $\mf r$ the purely covariant operators, whereas operators in $\utcless$ are both covariant and equivariant and those in $\tcless$ are neither.

\paragraph*{Examples.}
A generic yet instructive example is $S=\mc U(1)<\mc U(2)$ as a non-normal subgroup, e.g.,~via the representation $\sigma(S)=\DE{\exp(-i\alpha Z)\,|\,\alpha\in[0,2\pi)}$, which we also discuss in \cref{sec:circuitry_app:gates}.
$S$ and $\mf s$ are Abelian so that $\mf z(\mf t)=\mf t=i\R Z$.
Other single-qubit operations that commute with $\sigma(S)$ are global phases, that is $\utcless=i\R\mbb I$.
With the ordering from \cref{eq:tangent_decomp_U}, we obtain the decomposition
\begin{align}\label{eq:u1_in_u2_tangent_decomp}
    \mf u(2) = \Span_{i\R}\DE{X, Y} \oplus i\R\mbb I \oplus i\R Z \oplus \DE{0}.
\end{align}
We find the purely equivariant direction $iZ$, the purely covariant directions $iX$ and $iY$ as well as the direction $i\mbb I$ which is both equivariant and covariant.
This example demonstrates that $\mf r$ is not an algebra in general: $[iX, iY]=2iZ\not\in\mf r$.

We now move from $\mc U(1)<\mc U(2)$ to the normal subgroup of global phases, denoted as $\mc U(1)\triangleleft\mc U(2)$.
Its representation reads $\sigma(S)=\DE{\exp(-i\alpha \mbb I)|\alpha\in[0,2\pi)}$ and for the algebra decomposition, we obtain
\begin{align}
    \mf u(2)=\DE{0}\oplus \Span_{i\R}\DE{X, Y, Z}\oplus i\R\mbb I \oplus \DE{0}.
\end{align}
We find all but $i\mbb I$ to be covariant and all directions to be equivariant.
The latter is not surprising, as all quantum circuits commute with global phases.

A more interesting example may be $S=\mc{SU}(2)$ with its tensor power representation $\sigma(S)=\DE{s\otimes s| s\in \mc{SU}(2)}$ as subgroup of $\mc U(4)$.
Its center is trivial and we have $\utcless=\mf u^{\mf t}=\Span_{i\R}\DE{\mbb I, \operatorname{SWAP}}$ as well as
\begin{align}
    \tcless= \Span_{i\R}\DE{(X_1+X_2),(Y_1+Y_2),(Z_1+Z_2)},
\end{align}
so that
\begin{align}\label{eq:su2_tensorpower_tangent_decomp}
    \mf u(4) =&\mf r\oplus\Span_{i\R}\DE{\mbb I, \operatorname{SWAP}} \oplus \DE{0} \oplus \tcless\\
    \mf r =& \Span_{i\R}\{(XX-YY), (XX-ZZ), \nonumber\\
    &\phantom{\Span_{i\R}\{}\hspace{2pt}(X_1-X_2), (Y_1-Y_2), (Z_1-Z_2),\nonumber\\
    &\phantom{\Span_{i\R}\{}\hspace{2pt}XY, YX, XZ, ZX, YZ, ZY\}\nonumber.
\end{align}
We see that the $13$-dimensional covariant tangent space makes up most of $\mf u(4)$, whereas the two equivariant directions $i\mbb I$ and $i\operatorname{SWAP}$, which are covariant as well, represent a rather small portion of the tangent space.
As a consequence, moving orthogonally to symmetry directions allows for many more degrees of freedom than an equivariant circuit.
The latter can only be compiled from global phase and (partial) $\operatorname{SWAP}$ gates because all other operations are affected by symmetry operations.
Considering that global phases are not of importance in most applications, this leaves us with a single update direction for equivariant training.
In contrast, covariant training only avoids moving into symmetry directions, which poses a weaker constraint.

\subsubsection{Equivariant and covariant tangents of \texorpdfstring{$\mc H$}{H}}\label{sec:equi_and_co:tangent_decomp_H}
\begin{figure}
    \centering
    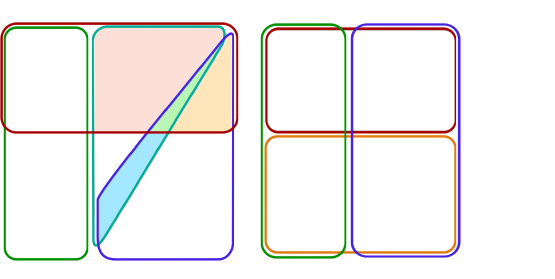
    \caption{%
    \textbf{Decompositions of the tangent space $T_\psi\mc H$ of a Hilbert space $\mc H$} at the point $\ket{\psi}$ based on a (represented) symmetry subalgebra $\mf t$.
    (\emph{left}) the orthogonal decomposition of $\mf u$ in \cref{eq:tangent_decomp_U} leads to non-orthogonal subspaces of $T_\psi\mc H$, \cref{eq:state_space_decomp_nonorth}.
    This is due to redundancies when describing changes of a state with unitary transformations.
    Shaded regions indicate overlaps caused by these redundancies, which do not exist on the level of unitaries.
    (\emph{right}) An orthogonal decomposition of $T_\psi\mc H$, based on the vertical subspace $\mf t\ket{\psi}$ and the equivariant space $\mf u^{\mf t}\ket{\psi}$, \cref{eq:state_space_decomp_orth}.
    }
    \label{fig:tangent_decomps}
\end{figure}

We now move to the tangents of Hilbert space $\mc H$.
\begin{convention}
    We will use the inner product
    \begin{align}\label{eq:inner_product_H}
        (x\ket{\psi}, y\ket{\psi})\mapsto \dE{\psi|x^\dagger y|\psi}=\real{\braket{\psi|x^\dagger y|\psi}}
    \end{align}
    for the (real) vector space $T_{\ket{\psi}}\mc H$.
\end{convention}
As for the ``bra-ket'' notation, here is a certain ambiguity in $\dE{X|O|Y}$ for non-Hermitian operators $O$.
We always mean $\real{\bra{X}(O\ket{Y})}$, i.e.,~any operators between the states are read as denoted, not as part of $\bra{X}$.

When translating the concepts from the previous section to $T_{\ket{\psi}}\mc H$, we need to consider redundancies in its description via tangents of the unitary group.
In particular, we need to keep in mind that covariant tangents are defined indirectly, as the (orthogonal) complement of vertical tangents.

Thus, the covariant tangents of $\mc H$ are \emph{not} given by applying the covariant tangents $\orth{\mf s}\subset\mf u$ of $\mc U$ to $\ket{\psi}$, but by the orthogonal complement of applying the vertical tangents (\cref{sec:equi_and_co:covariance_algebras} and \cref{eq:vert_hor_states}): $H_\psi\mc H=\orth{\mf s\ket{\psi}}\neq \orth{\mf s}\ket{\psi}$.
As a consequence, the decomposition into vertical and covariant subspaces depends on the state $\ket{\psi}$.

In contrast to this, the concept of equivariance usually is fixed to the level of gates, and this will be our choice in the following.
This means that equivariant tangents on the state level are defined as $\mf u^{\mf t}\ket{\psi}$.
We may be tempted to define a decomposition of $T_\psi\mc H$ into orthogonal subspaces based on \cref{eq:tangent_decomp_U}, but need to notice that expressing the space via tangents of $\mc U$ applied to $\ket{\psi}$ disrupts the orthogonality between the resulting subspaces.
Instead we get the decomposition
\begin{align}\label{eq:state_space_decomp_nonorth}
    T_\psi\mc H &= 
    \overset{\text{equivariant}}{\overbrace{\utcless\ket{\psi} + \big[\,\lefteqn{\underset{\text{vertical}}{\underbrace{\phantom{[\mf z(\mf t)\ket{\psi} + \tcless\ket{\psi}]}}}}[\mf z(\mf t)\ket{\psi}}} + \tcless\ket{\psi}]
    \oplus \underset{\text{covariant}}{\underbrace{H_\psi\mc H}}\big],
\end{align}
where $+$ denotes vector space sums that are not orthogonal in general.
In particular, the overlap of the equivariant and covariant subspaces may or may not be trivial, also see \cref{fig:tangent_decomps}

While the orthogonal structure does not carry over from $\mf u$ to $T_\psi\mc H$, we may still define projectors onto the above subspaces, for example via an explicit diagonalization procedure as in \cref{sec:circuitry:states}.
We need to keep in mind though that these projectors are not complementary, i.e.,~do not sum to the identity.
For convenience, we may define an alternative decomposition that is orthogonal by design:
first, define purely equivariant tangents to be $T_\psi^\text{equi}\mc H=\mf u^{\mf t} \ket{\psi} \cap \orth{H_\psi\mc H}$.
Similarly, purely covariant tangents are $T^\text{cov}_\psi\mc H = H_\psi\mc H \cap \orth{\mf u^{\mf t} \ket{\psi}}$.
Tangents that are both covariant and equivariant are denoted as $T^{\diamond}_\psi\mc H = H_\psi\mc H \cap T^\text{equi}_\psi\mc H$.
The fourth and last subspace contains vertical tangents that are not equivariant, $T^\text{vert}_\psi\mc H = V_\psi\mc H \cap \orth{T^\text{equi}_\psi\mc H}$.
In summary, we obtain the decomposition
\begin{alignat}{3}\label{eq:state_space_decomp_orth}
    T_\psi\mc H &= \underset{\text{covariant}}{\underbrace{T^\text{cov}_\psi\mc H \oplus \lefteqn{\overset{\text{equivariant}}{\overbrace{\phantom{T^{\diamond}_\psi\mc H\oplus T^\text{equi}_\psi\mc H}}}}T^{\diamond}_\psi\mc H}}\oplus \lefteqn{\underset{\text{vertical}}{\underbrace{\phantom{T^\text{equi}_\psi\mc H\oplus T^\text{vert}_\psi\mc H}}}}&&T^\text{equi}_\psi\mc H&&\oplus  T^\text{vert}_\psi\mc H,
\end{alignat}
which we show alongside the non-orthogonal decomposition in \cref{fig:tangent_decomps}.
This decomposition, while not directly related to the decomposition of $\mf u$, allows us to define complementary projectors and simplifies our understanding of the sectors of the total tangent space in the following sections.
As discussed in \cref{sec:equi_and_co:covariance_algebras}, it also allows us to categorize unitary generators $x\in\mf u$ via the state tangent $x\ket{\psi}$ they produce, but this categorization will not lead to subalgebras of $\mf u$ in general.

\paragraph*{Examples.}
We continue with the first example from the previous section, $\sigma(S)=\mc U(1)<\mc U(2)$, and choose the position $\ket{+}\in\mc H$.
The vertical subspace then is $\mf t\ket{+} = i\R\ket{-}$ and we obtain the decompositions\footnote{Recall that the tangent state space is a real vector space so that, e.g.,~$\ket{+}$ and $i\ket{+}$ are linearly independent vectors.}
\begin{align}
    T_{\ket{+}}\mc H
    &= 
    \underset{\utcless\ket{+}}{\underbrace{i\R\ket{+}}}
    +\bigg[\big[\underset{\mf z(\mf t)\ket{+}}{\underbrace{i\R\ket{-}}}
    \oplus\hspace{-5pt}\underset{\tcless\ket{+}}{\underbrace{\DE{0}}}\hspace{-5pt}\big]
    \oplus \underset{\orth{\mf t\ket{+}}}{\underbrace{[\R\ket{-}\oplus i\R\ket{+}]}}\bigg]\label{eq:states_ex1_nonorth}\\
    &= \underset{\text{covariant}}{\underbrace{\R\ket{-} \oplus \lefteqn{\overset{\text{equivariant}}{\overbrace{\phantom{i\R\ket{+}\oplus i\R\ket{-}}}}}i\R\ket{+}}}\oplus \lefteqn{\underset{\text{vertical}}{\underbrace{\phantom{i\R\ket{-}\oplus \DE{0}}}}}i\R\ket{-}\oplus  \DE{0}\label{eq:states_ex1_orth},
\end{align}
where the direct sums between $\mf z(\mf t)\ket{+}$ and $\tcless\ket{+}$ as well as between $\utcless\ket{+}$ and $\mf t\ket{+}$ (not denoted) in \cref{eq:states_ex1_nonorth} are coincidental.
From \cref{eq:states_ex1_orth} we observe that, of the two equivariant tangent directions, one is vertical ($i\R\ket{-}$) and one is covariant ($i\R\ket{+}$).
The decomposition of $\mf u$ induced by the vertical/covariant tangent space decomposition (see \cref{eq:pull_back_horizontal_space}) is equal to $\mf t\oplus \orth{\mf t}$ because the action of $\sigma(S)$ on $\mc H$ is transitive on $\sigma(S)\ket{+}$.
We observe a different behaviour at the point $\ket{0}$:
\begin{align}
    T_{\ket{0}} \mc H
    &= 
    i\R\ket{0}
    +\bigg[
        \big[
            i\R\ket{0}
            \oplus\DE{0}
        \big]
        \oplus \De{\R\ket{1}\oplus i\R\ket{1}}
    \bigg]\nonumber\\
    &= \De{\R\ket{1}\oplus i\R\ket{1}} \oplus \DE{0} \oplus i\R\ket{0}\oplus\DE{0}.
\end{align}
As we can see, the operator $i\mbb I$ generates a vertical tangent as well, so that we obtain the induced vertical subspace $\uperp_{\ket{0}}=\Span_{i\R}\DE{\mbb I, Z}\neq \mf t$ and its covariant complement $\mf u^\|_{\ket{0}}=\Span_{i\R}\DE{X, Y}$; see \cref{sec:equi_and_co:covariance_algebras} for details.

For the tensor power representation of $\mc{SU}(2)$ decomposed as in \cref{eq:su2_tensorpower_tangent_decomp} and the state $\ket{01}$ we obtain only two vertical state tangents
$\mf t\ket{01}=\Span_\R\DE{i\ket{B_+}, \ket{B_-})}$, with the Bell states $\ket{B_\pm}=\frac{1}{\sqrt{2}}(\ket{00}\pm\ket{11})$.
The tangent space decompositions read
\begin{widetext}
\begin{align}
    T_{\ket{01}}\mc H
    &=
    \underset{\utcless\ket{01}}{\underbrace{\Span_\R\DE{i\ket{01},i\ket{10}}}}
    +\bigg[\big[\underset{\mf z(\mf t)\ket{01}}{\underbrace{\DE{0}}}
    \oplus\underset{\tcless\ket{01}}{\underbrace{\Span_\R\DE{i\ket{B_+},\ket{B_-}}}}\big]
    \oplus \underset{\orth{\mf t\ket{01}}}{\underbrace{\Span_\R\DE{i\ket{01},\ket{10},i\ket{10},i\ket{B_-},\ket{B_+}}}}\bigg]\\
    &=\Span_\R\DE{\ket{10},i\ket{B_-}, \ket{B_+}}\ \oplus\ \Span_\R\DE{i\ket{01}, i\ket{10}}\ \oplus\ \DE{0}\ \oplus\ \Span_\R\DE{i\ket{B_+}, \ket{B_-}}.
\end{align}
\end{widetext}
As we can see there are two vertical and two equivariant tangent directions, the latter of which lie in the $5$-dimensional covariant tangent subspace.
The induced decomposition of $\mf u(4)$ leads to the vertical subspace 
\begin{align}
    \uperp_{\ket{01}}=\Span_{i\R}\DE{X_1+X_2, Y_1+Y_2},
\end{align}
which is not a subalgebra.
That is, for the given configuration of $S$, $\sigma$, the action $\act{\mc H}$ and the point $\ket{01}$, not even the induced \emph{vertical} generators form an algebra.
This is due to $i(Z_1+Z_2)\in\mf t^0$ producing the zero tangent.

The state dependence of the tangent space decomposition becomes particularly clear if we move from $\ket{01}$ to its anti-symmetrization $\ket{\phi}=\frac{1}{\sqrt{2}}(\ket{01}-\ket{10})$:
\begin{align}
    T_{\ket{\phi}}\mc H
    &=i\R\ket{\phi} + \bigg[ \underset{\mf t\ket{\phi}}{\underbrace{[\DE{0}\oplus\DE{0}]}}\oplus T_{\ket{\phi}}\mc H \bigg]\\
    &=\de{T_{\ket{\phi}}\mc H \setminus i\R\ket{\phi}} \oplus i\R\ket{\phi}\oplus\DE{0}\oplus\DE{0}.
\end{align}
That is, at the point $\ket{\phi}\in\mc H$, all tangents are covariant, one of which is in addition equivariant.
This is due to $\ket{\phi}$ being the singlet state on two qubits, which is a one-dimensional irreducible representation of $\mc{SU}(2)$.
As a consequence, the group action on $\ket{\phi}$ becomes trivial and the state is unaffected by symmetry transformations.

The decomposition of $\mf u(4)$ based on this tangent space decomposition (see \cref{eq:pull_back_horizontal_space}) reads $\mf u(4) = \uperp_{\ket{\phi}} \oplus\mf u^\|_{\ket{\phi}}= \DE{0}\oplus \mf u(4)$ because all symmetry generators produce the zero tangent ($\mf t=\mf t^0$).
Both induced subspaces, vertical and covariant, are algebras (the former being trivial) and in particular we have the \emph{covariant group} $\mf u(4)$ at $\ket{\phi}$.

\section{Application to quantum circuits}\label{sec:circuitry}
In this section we apply the abstract concepts previously discussed to parametrized quantum gates, parametrized quantum states, and PQC-based cost functions.
To this end, we will choose common symmetry actions at these three levels and focus on examples.

\subsection{Quantum gates}\label{sec:circuitry:gates}
We begin with a generic action of a symmetry group $S$ on the unitary group $\mc M=\mc U(d)$ via a $d$-dimensional matrix representation and matrix multiplication (from the right):
\begin{align}\label{eq:action_on_gates}
    \sigma : S \to \mc U(d); \qquad
    \forall s\in S: s\act{\sigma} U =& U\sigma(s)\\
    \forall x\in \mf s: x\act{\sigma} U =& U\sigma(x).
\end{align}
This is the action we also treated in \cref{sec:equi_and_co:tangent_decomp_U}, giving rise to the tangent space $T_U\mc U=U\mf u$.
The equivariant subspace of the tangent space is that induced by $\mf u^{\mf t}$ by definition.
In addition, due to the transitivity of $\act{\sigma}$, the vertical and covariant subspaces are compatible with the decomposition $\mf u=\mf t\oplus \orth{\mf t}$: 
\begin{align}
    V_p\mc M &= \mf s\act{\sigma}\mc U=\mc U\mf t\\
    H_p\mc M &= \orth{\mf s\act{\sigma}\mc U}=\orth{\mc U\mf t}=\mc U \orth{\mf t}.
\end{align}
As a consequence, the induced decomposition $\mf u = \mf u^\|_U \oplus \uperp_U$ is the same as that into $\mf t$ and $\orth{\mf t}$.
Overall, the tangent space thus decomposes like $\mf u$ itself:
\begin{align}
    T_U\mc U =
    \underset{\text{covariant }\de{\orth{U \mf t}}}{\underbrace{U \mf r \oplus \lefteqn{\overset{\text{equivariant } \de{U\mf u^{\mf t}}}{\overbrace{\phantom{U\utcless \oplus U\mf z(\mf t)}}}}U\utcless}}\oplus \underset{\text{vertical } (U\mf t)}{\underbrace{U\mf z(\mf t) \oplus U\tcless}}.
\end{align}

Now consider a parametrized quantum circuit $U$ mapping $p$ parameters $\btheta$ to a unitary $U(\btheta)\in\mc U(d)$.
Its derivative with respect to the $j$-th parameter reads (see \cref{sec:preliminaries:pqc})
\begin{align}
    \partial_j U(\btheta) = U(\btheta)\Omega_j(\btheta),
\end{align}
for some (right-)effective generator $\Omega_j(\btheta)\in \mf u(d)$.
Denote an orthonormal basis (ONB) with respect to the trace inner product for $\mf u^{\mf t}$ as $\DE{y_a}_a$, and an ONB for the complement $\mf r \oplus \tcless$ as $\DE{z_a}_a$.
\begin{highlight}
The equivariant derivative from \cref{eq:equivariant_derivative_abstract} for parametrized unitaries reads
\begin{align}
    E_jU(\btheta)
    &= U(\btheta) \mc T_{\sigma(S)}[\Omega_j(\btheta)]\\
    &= U(\btheta) y_a \frac{1}{d}\tr{y_a^\dagger \Omega_j(\btheta)}\\
    &= U(\btheta) \De{\Omega_j(\btheta) - z_a \frac{1}{d}\tr{z_a^\dagger \Omega_j(\btheta)}}\nonumber,
\end{align}
with the twirling operator $\mc T_{\sigma(S)}$ and the Einstein convention of summing over repeated indices.
\end{highlight}

Note that we may use these basis elements directly to construct projectors, again because the inner product is unitarily invariant, i.e.,~$\frac{1}{d}\tr{(U x)^\dagger Uy}=\frac{1}{d}\tr{x^\dagger y}$.

\begin{highlight}
The covariant derivative from \cref{eq:covariant_derivative_abstract} for parametrized unitaries becomes
\begin{align}
    D_j U(\btheta)
    &= U(\btheta) \mbb H_{U(\btheta)} \Omega_j(\btheta)\\
    &= U(\btheta) \De{\Omega_j(\btheta) - x_a \frac{1}{d}\tr{x_a^\dagger \Omega_j(\btheta)}}\nonumber,
\end{align}
for an ONB $\DE{x_a}_a$ of $\mf t$.
\end{highlight}
We remark that\footnote{We do not ``mean'' anything by denoting an index as superscript or subscript, but write $A_j^a$ to increase readability.} $A_j^a(\btheta)=\frac{1}{d}\tr{x_a^\dagger \Omega_j(\btheta)}$ is, from a gauge theory perspective, the \emph{vector potential} for this symmetry action.
The computational recipes for $D_jU$ and $E_jU$ are very similar, which underlines the fact that they are both applications of the projected derivative.

\begin{table}[]
    \centering
    \begin{tabular}{rcc}
        $U(\btheta)$ \hspace{4pt}& $E_j U(\btheta)$ & $D_j U(\btheta)$\\\midrule
        general\hspace{4pt} & $U(\btheta) \mc T_S[\Omega_j(\btheta)]$ & \hspace{4pt}$U(\btheta) \mbb H_{U(\btheta)} \Omega_j(\btheta)$ \hspace{4pt}\\
        equivariant\hspace{4pt} & $\partial_j U(\btheta)$ & $D_j U(\btheta)\in U(\btheta)\utcless$ \\
        covariant\hspace{4pt} & $E_j U(\btheta)\in U(\btheta)\utcless$ & $\partial_j U(\btheta)$
    \end{tabular}
    \caption{\textbf{Equivariant and covariant derivatives of a parametrized unitary $U(\btheta)$} with respect to the symmetry action in \cref{eq:action_on_gates}, and their properties for equivariant or covariant $U$.
    If the derivative and the subgroup to which $U$ maps are both equivariant or both covariant, we obtain the partial derivative.
    For the ``off-diagonal'' combinations in the table, the derivative lies in the subspace $U(\btheta)\utcless$ of $T_{U(\btheta)}\mc U$ that is both equivariant and covariant.}
    \label{tab:equi_and_co_U}
\end{table}

If we choose $U(\btheta)$ to lie in the equivariant or covariant subgroup\footnote{Provided that the latter exists; see \cref{sec:equi_and_co:covariance_groups} for details.}, the equivariant and covariant derivatives project the effective generator into the intersection of the corresponding subspace of $\mf u$.
That is, on one hand the covariant derivative of an equivariant unitary and the equivariant derivative of a covariant unitary will both lie in $U(\btheta)\utcless$.
On the other hand, the equivariant (covariant) derivative of an equivariant (covariant) circuit will lie in $U(\btheta) \mf u^{\mf t}$ $\de{U(\btheta) \orth{\mf t}}$ and match its partial derivative.
We summarize this interplay in \cref{tab:equi_and_co_U}.
If $U(\btheta)$ is equivariant \emph{and} covariant itself, we obtain $\partial_jU(\btheta)=D_jU(\btheta)=E_jU(\btheta)$.

\paragraph*{Examples.}
First, we extend the single-qubit examples from \cref{sec:equi_and_co:tangent_decomp_U}.
For this, consider the unitary
\begin{align}\label{eq:parametrized_gate_ex}
    U(\btheta)&=\exp(i\theta_3)R_X(\theta_2)R_Y(\theta_1).
\end{align}
While the global phase component usually would not be of interest within a quantum circuit, we here explicitly include it for illustrative purposes.
Its right-effective generators are (see \cref{sec:ex_calc:1q_gate} for details)
\begin{align}
    \Omega_1 = -\frac{i}{2} Y,\quad
    \Omega_2(\btheta) = -\frac{i}{2} (c_1 X + s_1 Z),\quad
    \Omega_3 = i\mbb I,
\end{align}
with $c_1=\cos(\theta_1)$ and $s_1=\sin(\theta_1)$.
The symmetry representation $\mf t=i\R Z$ then leads to the covariant and equivariant derivatives
%
\begin{alignat}{4}\label{eq:circuitry_gates_ex1_0}
    D_1U(\btheta) &= -\frac{i}{2} U(\btheta) Y,\quad
    &&E_1U(\btheta) &&=&& 0\\\label{eq:circuitry_gates_ex1_1}
    D_2U(\btheta) &= -\frac{ic_1}{2} U(\btheta) X,\quad
    &&E_2U(\btheta) &&=&& -\frac{is_1}{2} U(\btheta) Z\\\label{eq:circuitry_gates_ex1_2}
    D_3U(\btheta) &= iU(\btheta),\quad
    &&E_3U(\btheta) &&=&& iU(\btheta).
\end{alignat}
That is, for the first two parameters we have $\partial_jU(\btheta)=D_jU(\btheta)+E_jU(\btheta)$ and for the global phase we observe $\partial_3U(\btheta)=D_3U(\btheta)=E_3U(\btheta)$.
If we switch to the representation $\mf t=i\R\mbb I$, we instead obtain, for $j=1,2$:
\begin{align} 
    \Omega_j\in\utcless \Rightarrow D_jU(\btheta)=E_jU(\btheta)=\partial_jU(\btheta).
\end{align}
The global phase derivatives are $D_3U(\btheta)=0$ (symmetry aligns with gate generator) and $E_3U(\btheta)=\partial_3U(\btheta)$ (global phases are an Abelian group).

\subsection{Quantum states}\label{sec:circuitry:states}
We now move to the action of the unitary group on quantum states, as anticipated in \cref{sec:equi_and_co:tangent_decomp_H}.
We will describe parametrized states via a unitary circuit $U(\btheta)$, applied to some initial state $\ket{\psi_0}$, i.e.,~$\ket{\psi(\btheta)}=U(\btheta)\ket{\psi_0}$.
We will consider two actions of the symmetry group $S$ on states, both based on a matrix representation $\sigma$.
They differ by the order of the parametrized unitary and the symmetry transformation\footnote{We reuse the notation $\act{L}$, as it is formally the same as the left action on unitaries.}:
\begin{alignat}{3}\label{eq:actions_on_states}
    s\act{L}\ket{\psi(\btheta)} &= \sigma(s)\ket{\psi(\btheta)} &&=&& \sigma(s)U(\btheta)\ket{\psi_0}\\
    s\act{\btheta}\ket{\psi(\btheta)} &= U(\btheta)\sigma(s) U^\dagger(\btheta)\ket{\psi(\btheta)} &&=&& U(\btheta)\sigma(s)\ket{\psi_0}.
\end{alignat}
The motivation for these different actions stems from applications for PQC cost functions:
if the symmetry enters via some (data-based) state preparation, $\act{\btheta}$ is the suitable action, if we assume the state before preparation is made invariant; for symmetries determined by the measurement observable, $\act{L}$ will be the right choice.
The symmetry generators $\mf s$ act accordingly by left multiplication at the same position.
Note that $\act{L}=\act{\btheta}$ for equivariant unitaries.

Throughout this work we mostly describe quantum states via state vectors, but of course the symmetry actions readily generalize to density matrices (which we use in \cref{sec:gauge_theory}, for example).
For this, we simply replace multiplication from the left by conjugation for symmetry transformations in $S$, and by the commutator for generators in $\mf s$.

As discussed in \cref{sec:equi_and_co:tangent_decomp_H}, the action on quantum states does not necessarily allow for a decomposition of the tangent space that is compatible with the decomposition $\mf u=\mf t\oplus\orth{\mf t}$.
In addition, any decomposition of $T_{\ket{\psi(\btheta)}}\mc H$ will depend on the group action we chose, because the inner product (see \cref{eq:inner_product_H}) does:
\begin{align}
    x\act{L} \ket{\psi(\btheta)} \perp y\act{L} \ket{\psi(\btheta)}&\Leftrightarrow \dE{\psi(\btheta)|x^\dagger y|\psi(\btheta)}=0\label{eq:state_action_left}\\
    x\act{\btheta} \ket{\psi(\btheta)} \perp y\act{\btheta} \ket{\psi(\btheta)}&\Leftrightarrow \dE{\psi_0|x^\dagger y|\psi_0}=0\label{eq:state_action_right}.
\end{align}
As a consequence, the decompositions in \cref{eq:state_space_decomp_nonorth} and \cref{eq:state_space_decomp_orth} depend on $\btheta$ for $\act{L}$, but not for $\act{\btheta}$.
The redundancies in $\mf u\act{}\ket{\psi(\btheta)}$ compared to $T_{\ket{\psi(\btheta)}}\mc H$ do not only disrupt the orthogonality between $\mf t\ket{\psi(\btheta)}$ and $\orth{\mf t}\ket{\psi(\btheta)}$, but orthonormality more generally.

In order to write down the projectors onto subspaces of $T_{\ket{\psi(\btheta)}}\mc H$ explicitly, we require ONBs for these subspaces, which we can construct using the respective Gram matrices.
Specifically, we denote an ONB of the equivariant subspace $\mf u^{\mf t}\act{}\ket{\psi(\btheta)}$ as
\begin{align}\label{eq:orthonormalize_ut_psi}
    \ket{U_a} &= \sqrt{\tilde{G}^+}_{ab}\ket{V_b}\,,\\
    \ket{V_a} &= y_a\act{}\ket{\psi(\btheta)}\,,
    \quad
    \tilde{G}_{ab} = \dE{V_a|V_b}\label{eq:orthonormalize_ut_psi_1}\,,
\end{align}
where $\DE{y_a}_a$ is a basis for the commutant algebra.

\begin{highlight}
The equivariant derivative of parametrized quantum states is
\begin{align}
    E_j\ket{\psi(\btheta)}
    &= \ket{U_a}\dE{U_a|\partial_j\psi(\btheta)}\\
    &= \ket{V_a}\tilde{G}^+_{ab} \dE{V_b|\partial_j\psi(\btheta)},
\end{align}
where $\tilde{G}^+$ denotes the pseudo-inverse of $\tilde{G}$.
\end{highlight}
\noindent In \cref{sec:circuitry:cost_functions} we discuss how to compute the Gram matrix and overlaps of the form $\dE{V_a |\partial_j\psi(\btheta)}$ in practice.

Similarly, for the symmetry generators $\mf t$ with a basis $\DE{x_a}_a$, we set up an ONB for the vertical subspace via
\begin{align}\label{eq:orthonormalize_t_psi}
    \ket{T_a} &= \sqrt{G^+}_{ab}\ket{X_b}\,,\\
    \ket{X_a} &= x_a\act{}\ket{\psi(\btheta)}\,,
    \quad
    G_{ab} = \dE{X_a|X_b}\,.
\end{align}
\begin{highlight}
The covariant derivative of parametrized quantum states reads
\begin{align}
    D_j\ket{\psi(\btheta)}
    &= \de{\mbb I - \ket{T_a}\dE{T_a|\,\cdot\,}} \ket{\partial_j\psi(\btheta)}\\
    &= \ket{\partial_j\psi(\btheta)} - \ket{X_a}G^+_{ab} \dE{X_b|\partial_j\psi(\btheta)}\label{eq:cov_deriv_states}.
\end{align}
The coefficients of the vertical component of the derivative again form the vector potential,
\begin{align}\label{eq:vector_potential_states}
    A^a_j(\btheta) = G^+_{ab}\dE{X_b|\partial_j\psi(\btheta)}.
\end{align}
\end{highlight}

\paragraph*{Examples.}
In addition to the examples for state tangent spaces in \cref{sec:equi_and_co:tangent_decomp_H}, we here continue those from the previous section.
We use the circuit from \cref{eq:parametrized_gate_ex} applied to the initial state $\ket{+}$, and the representation $\mf t = i\R Z$ together with the action $\act{\btheta}$ from \cref{eq:state_action_right}.
See \cref{sec:ex_calc:1q_state} for details.

The vertical subspace is the (real) span of $i\ket{-}$, which is a normalized state tangent vector already and does not depend on $\btheta$.
The equivariant tangent vectors are $\ket{V_{1,2}}=i\ket{\pm}$, which are orthonormal as well.
With these bases for $\mf t\ket{+}$ and $\mf u^{\mf t}\ket{+}$, the covariant and equivariant derivatives read
\begin{alignat}{4}
    D_1\ket{\psi(\btheta)} &= \partial_1 \ket{\psi(\btheta)},
    \quad
    &&E_1\ket{\psi(\btheta)} &&=&& 0,\\
    D_2\ket{\psi(\btheta)} &= -\frac{ic_1}{2} \ket{\psi(\btheta)},
    \quad
    &&E_2\ket{\psi(\btheta)} &&=&& \partial_2 \ket{\psi(\btheta)},\nonumber\\
    D_3\ket{\psi(\btheta)} &= \partial_3\ket{\psi(\btheta)} ,
    \quad
    &&E_3\ket{\psi(\btheta)} &&=&& \partial_3\ket{\psi(\btheta)}.\nonumber
\end{alignat}
We note that while most expressions resemble the modified derivatives from \cref{eq:circuitry_gates_ex1_0,eq:circuitry_gates_ex1_1,eq:circuitry_gates_ex1_2}, the equivariant derivative $E_2\ket{\psi(\btheta)}$ differs from $(E_2U(\btheta)\ket{+}$, because the Pauli $X$ component of $\Omega_2(\btheta)$ acts like a global phase at $\ket{+}$, which is an equivariant direction and hence is not being projected away.

If we move from $\act{\btheta}$ to $\act{L}$, the tangent space decompositions become parameter-dependent, and in particular the equivariant subspace does not always remain two-dimensional but has singularities with respect to $\btheta$.
This is reflected in the eigenvalues $1-s_1c_2$ and $1+s_1c_2$, one of which may vanish at a time.
The covariant derivatives $D_{1,2}\ket{\psi(\btheta)}$ switch roles, in some sense, as now the $R_X$ neighbors the symmetry action, and the left-effective generator of the $R_Y$ carries a component in the symmetry direction, which is being projected away in $D_1\ket{\psi(\btheta)}$.

Before moving on to cost functions, we prepare a second example, on two qubits.
We apply the circuit
\begin{align}
    U(\btheta)=CR_X^{(2, 1)} (\theta_3)R_Y^{(2)}(\theta_2) R_Y^{(1)}(\theta_1)
\end{align}
to some product initial state $\ket{\psi_0^{(1)}}\otimes\ket{\psi_0^{(2)}}$.
The symmetry group is $\mc{SU}(2)$, represented on the first qubit and acting after the circuit, i.e.,~via $\act{L}$.
In a detailed computation in \cref{sec:ex_calc:2q_state}, we find the vector potential
\begin{align}
    A_j^a(\btheta)
    = -\frac12 \de{\begin{array}{ccc}
        0 & A^0_1 & p_1^{(2)} \\
        p_0^{(2)}+c_3p_1^{(2)} & A^1_1 & 0 \\
        s_3 p_1^{(2)} & A^2_1 & 0 
    \end{array}},
\end{align}
where we abbreviated $p_\ell^{(2)}=\left|\bra{\ell}R_Y(\theta_2)\ket{\psi_0^{(2)}}\right|^2$.
As we can see, $\theta_1$ causes changes in the direction of the symmetry transformations $Y^{(1)}$ and $Z^{(1)}$, whereas $\theta_3$ only contributes in the direction of $X^{(1)}$.
The rotation $R_Y^{(2)}(\theta_2)$ has overlap with all three symmetry directions, depending on the parameter position $\btheta$.
In \cref{sec:entangling_app} we will use this example to construct a maximally entangling circuit.

\subsection{PQC cost functions}\label{sec:circuitry:cost_functions}
We now move to the common setup of symmetry-aware variational quantum algorithms and geometric quantum machine learning.
For this, we consider cost functions of the form 
\begin{align}\label{eq:circuitry:pqc_cost_function}
    C(\btheta) = \bra{\psi_0}U^\dagger (\btheta) M U(\btheta)\ket{\psi_0},
\end{align}
where $M$ is a Hermitian observable and $\ket{\psi_0}$ is an initial state, which could depend on encoded data.
The symmetry actions on cost functions we consider are $\act{L}$ and $\act{\btheta}$ at the state level from \cref{eq:actions_on_states}.
Correspondingly, the tangent space decompositions and projectors are the same as for the state derivatives, too.
However, there are two crucial differences between the state level and cost functions:
first, the cost function is a measurable quantity whereas the full state vector $\ket{\psi(\btheta)}$ is in general intractable for interesting use cases of quantum computers.
Accordingly, the equivariant and covariant derivatives are more interesting for cost functions than for state vectors in practice.
Second, the measurement observable has an impact on the equivariant and covariant derivatives of $C(\btheta)$, as we will see in the following.

The common partial derivatives of $C$ take the form
\begin{align}
    \partial_jC(\btheta)
    &= \bra{\psi_0} [U^\dagger(\btheta) M U(\btheta), 
    U^\dagger(\btheta)\partial_j U(\btheta)]\ket{\psi_0}\\
    &=\dE{\psi_0|[\widetilde{M}(\btheta),\Omega_j(\btheta)]|\psi_0},
\end{align}
where $\widetilde{M}(\btheta)=U^\dagger(\btheta) M U(\btheta)$ and $\Omega_j(\btheta)$ is the right-effective generator of the full unitary $U(\btheta)$.
For $\act{\btheta}$, the symmetry acts on the initial state $\ket{\psi_0}$ and we need to project the state tangents $U(\btheta)\Omega_j(\btheta)\ket{\psi_0}$ onto the corresponding subspace.
For $\act{L}$, we instead use the left-effective generators $\Omega^L_j$:
\begin{align}
    \partial_jC(\btheta)
    &= \bra{\psi(\btheta)} [M, (\partial_j U(\btheta))U^\dagger(\btheta)] \ket{\psi(\btheta)}\\
    &= \dE{\psi(\btheta)|[M,\Omega_j^L(\btheta)]|\psi(\btheta)},
\end{align}
and write the tangent vector of the parametrized state as $\Omega^L_j(\btheta)\ket{\psi(\btheta)}$.

\begin{table*}[]
    \centering
    \begin{tabular}{rcc}
         $C(\btheta)$ & $E_jC(\btheta)$ & $D_jC(\btheta)$ \\\midrule
         general &
         $\dE{\psi(\btheta)|M|U_a}\dE{U_a|\partial_j\psi(\btheta)}$ &
         $\partial_jC(\btheta)-\dE{\psi(\btheta)|M|T_a}\dE{T_a|\partial_j\psi(\btheta)}$ \\
         equivariant &
         $\partial_jC(\btheta)$ &
         $\partial_jC(\btheta)$ \\
         covariant &
         $\dE{\psi(\btheta)|M|U_a}\dE{U_a|T_b}\dE{T_b|\partial_j\psi(\btheta)}$ &
         $\partial_jC(\btheta)$ \\
    \end{tabular}
    \caption{\textbf{Equivariant and covariant derivatives of PQC-based cost functions} of the form in \cref{eq:circuitry:pqc_cost_function}.
    The structure of $E_jC$ and $D_jC$ is similar, but they differ in the tangent subspace to which the partial derivative of the parametrized quantum state is being projected.
    This is visible in the used subspace basis, given by $\DE{\ket{U_a}}_a$ (\cref{eq:orthonormalize_ut_psi}) for $E_jC$ and by $\DE{\ket{T_a}}_a$ (\cref{eq:orthonormalize_t_psi}) for $D_jC$.
    In addition, we need to take the complement for the covariant derivative $D_jC$, subtracting the projected part from $\partial_j C$.
    Equivariant PQC cost functions are those that use equivariant $U(\btheta)$ and $M$, i.e.,~$[U(\btheta), \mf t]=[M,\mf t]=\DE{0}$.
    Covariant PQC cost functions only exist in special cases or for particular values of $\btheta$; see \cref{sec:equi_and_co:covariance_groups} for details.
    }
    \label{tab:equi_and_co_C}
\end{table*}

\begin{highlight}
We compute the equivariant derivative of a PQC cost function for the symmetry actions $\act{L/\btheta}$ as
\begin{alignat}{2}
    E_jC(\btheta)
    &=
    \underbrace{2\dE{\psi(\btheta)|M|V_a}}
    &&\tilde{G}_{ab}^+ 
    \underbrace{\dE{V_b|\partial_j\psi(\btheta)}}\\
    &=
    \hspace{19pt} m^E_a(\btheta) 
    &&\tilde{G}_{ab}^+ 
    \hspace{15pt} \omega^E_{bj}(\btheta).
\end{alignat}
\end{highlight}
Here we introduced $m^{E}_a(\btheta)$, the changes of the cost function corresponding to movements in the equivariant tangent space, and $\omega^E_{bj}(\btheta)$, the overlaps between the equivariant tangents and the $j$th partial derivative of the parametrized quantum state.
For the two symmetry actions, they can be written as
\begin{align}
    m^E_a(\btheta) 
    &=\begin{cases}
        \dE{\psi_0|[\widetilde{M}(\btheta), y_a]|\psi_0} & \text{ for } \act{\btheta}\\
        \dE{\psi(\btheta)|[M, y_a]|\psi(\btheta)} & \text{ for } \act{L}
    \end{cases}\label{eq:sym_deriv_as_commutator}\\
    \omega^E_{bj}(\btheta)
    &= \begin{cases}
        \frac12\dE{\psi_0|\{y_b^\dagger, \Omega^R_j(\btheta)\}|\psi_0} & \text{ for } \act{\btheta}\\
        \frac12\dE{\psi(\btheta)|\{y_b^\dagger, \Omega^L_j(\btheta)\}|\psi(\btheta)}& \text{ for } \act{L}
    \end{cases}\,.
\end{align}
The Gram matrix $\tilde{G}$, defined in \cref{eq:orthonormalize_ut_psi_1}, differs between the two actions as well,
\begin{align}
    \tilde{G}_{ab}
    &= \begin{cases}
        \frac12\dE{\psi_0|\{y_a^\dagger, y_b\}|\psi_0} & \text{ for } \act{\btheta}\\
        \frac12\dE{\psi(\btheta)|\{y_a^\dagger, y_b\}|\psi(\btheta)}& \text{ for } \act{L}
    \end{cases}\,.
\end{align}

For equivariant circuits, the projector $\mbb P_{\mf u^{\mf t}}=\ket{V_a}\tilde{G}_{ab}^+\dE{V_b|\,.\,}$ does not have any effect and we obtain $E_jC=\partial_jC$.
In addition, for $\act{L}$, if $M$ is in the commutant of $\mf u^{\mf t}$, $m^E_a(\btheta)$ vanishes and thus does $E_jC=0$.
The commutant of $\mf u^{\mf t}$, which is the bicommutant of $\mf t$ contains $\mf t$ but may be larger in general.
For $\act{\btheta}$, $U(\btheta)$ needs to be in said bicommutant as well if we want to deduce $m^E_a(\btheta)=0$.

Due to the similarity between \cref{eq:orthonormalize_ut_psi} and \cref{eq:orthonormalize_t_psi}, the covariant derivative $D_jC$ takes a form similar to $E_jC$.
We simply exchange the bases $\DE{y_j}_j\to\DE{x_j}_j$ and the Gram matrices $\tilde{G}\to G$, or alternatively the bases $\DE{\ket{V_a}}_a\to\DE{\ket{X_a}}_a$ and denote the corresponding factors as $m_a(\btheta)$ and $\omega_{bj}(\btheta)$.

\begin{highlight}
The result is the covariant derivative of PQC cost functions,
\begin{align}
    D_jC(\btheta) = \partial_j C(\btheta) - m_a(\btheta) G_{ab}^+ \omega_{bj}(\btheta).
\end{align}
\end{highlight}
We call the term $\vec{m}(\btheta)$ the \emph{symmetry derivative}, because it quantifies the changes in $C(\btheta)$ caused by infinitesimal symmetry transformations.
The Gram matrix $G$ characterizes the mutual linear dependence of the symmetry tangent vectors and $\omega(\btheta)$ measures the changes in symmetry directions caused by varying the parameters $\btheta$ of the PQC.
$G$ and $\omega$ combine into the vector potential $A_j^a(\btheta)=G_{ab}^+ \omega_{bj}(\btheta)$.

For covariant circuits\footnote{See \cref{sec:equi_and_co:covariance_groups,sec:equi_and_co:connect} for details on covariant groups.}, the projection $\mbb H$ is redundant and we have $D_jC(\btheta)=\partial_j C(\btheta)$, which might only hold for particular parameter positions. 
For $\act{L}$, equivariant observables with $[M,\mf t]=0$ imply $m_a=0$, so that $D_jC=\partial_j C$.
For $\act{\btheta}$, $m_a=0$ additionally requires $U(\btheta)$ to be equivariant.
In other words, if the cost function is symmetry-invariant with respect to a specific group action, the vertical part of the derivative $m_aG_{ab}^+\omega_{bj}$ vanishes because $m_a$ does.
It is important to note that the converse need not be true: a vanishing vertical projection may be due to $\omega_{bj}(\btheta)$ vanishing, i.e.,~due to parameter changes not generating symmetry transformations.
We summarize the equivariant and covariant derivatives of PQC cost functions in \cref{tab:equi_and_co_C}, for general circuits and special cases.

\subsubsection{Computing \texorpdfstring{$E_jC$}{EjC} and \texorpdfstring{$D_jC$}{DjC}}
With PQC-based cost functions we consider quantities that can be measured in practice.
Correspondingly, we here will discuss how to obtain the equivariant and covariant derivatives on a quantum computer.
Throughout this section, we present statements concerning the equivariant derivative in parentheses.

First, we need to compute entries $m^{(E)}_a(\btheta)=\dE{\psi(\btheta)|M|Z_a}$ for $\ket{Z_a}=\ket{X_a}$ ($\ket{Z_a}=\ket{V_a}$).
Its entries take the form of derivatives with respect to inserted gates generated by the symmetry generators (the commutant generators):
\begin{align}
    m^{(E)}_a(\btheta) &= \partial_t \bra{\psi_a(\btheta,t)}M\ket{\psi_a(\btheta,t)}\big|_{t=0}\\
    \ket{\psi(\btheta, t)} &=
    \begin{cases}
        U(\btheta)\exp(z_at)\ket{\psi_0} & \text{ for } \act{\btheta}\\
        \exp(z_at)U(\btheta)\ket{\psi_0} & \text{ for } \act{L}\\
    \end{cases}\,,
\end{align}
with $z_a=x_a$ ($z_a=y_a$).
As we assumed $\DE{x_a}_a$ ($\DE{y_a}_a$) to be an ONB of $\mf t$ ($\mf u^{\mf t}$), we require at most $|\DE{z_a}_a|$ such derivatives.
There is a multitude of methods to compute them, including finite differences, parameter-shift rules~\cite{mitarai2018quantum,Vidal_Theis_18,wierichs2022general,kyriienko2021generalized}, and Hadamard tests~\cite{Guerreschi_Smelyanski_17,li2017efficient,romero2018strategies}, which require that the gate $\exp(z_at)$, and sometimes the generator $z_a$, can be implemented on hardware.
Some of these methods become particularly simple for the required rotation angle $t=0$.

Next, the Gram matrix $G$ ($\tilde{G}$) is required, which has $|\DE{z_a}_a|^2$ entries of the form 
\begin{align}
    \dE{Z_a|Z_b}&=
    \begin{cases}
        \frac{1}{2}\bra{\psi_0} \DE{z_a,z_b}\ket{\psi_0} &\text{ for } \act{\btheta}\\
        \frac{1}{2}\bra{\psi(\btheta)} \DE{z_a,z_b}\ket{\psi(\btheta)} &\text{ for } \act{L}
    \end{cases}\,.
\end{align}
Depending on the structure of $\mf t$ ($\mf u^{\mf t}$) and the initial state, we may compute these overlaps classically.
If this is prohibitively expensive, we may compute the Hermitian anticommutators $\DE{z_a, z_b}$ classically and evaluate their expectation values.
Should the anticommutators be intractable classically, we can instead adapt the technique for $\omega_{bj}^{(E)}$ described below to measuring $\dE{Z_a|Z_b}$.
Recall that the Gram matrix need not have full rank.
This means that evaluating and diagonalizing it first will reveal potential savings in computing $m_a^{(E)}$ and $\omega_{bj}^{(E)}$ by moving to the non-singular subspace of $G$ ($\tilde{G}$) first.
Note, though, that this may make the individual terms more expensive to compute, which has to be weighed against the reduction in dimension.

The third and last quantity we need to measure consists of the overlaps between the state derivatives and the vertical (commutant) tangent directions:
\begin{align}\label{eq:compute_overlaps_equi_and_co}
    \omega^{(E)}_{bj}(\btheta)
    &= \begin{cases}
        \frac12\dE{\psi_0|\{z_b^\dagger, \Omega^R_j(\btheta)\}|\psi_0} & \text{ for } \act{\btheta}\\
        \frac12\dE{\psi(\btheta)|\{z_b^\dagger, \Omega^L_j(\btheta)\}|\psi(\btheta)}& \text{ for } \act{L}
    \end{cases}\,.
\end{align}
They can be evaluated with a modified Hadamard test, which we detail in \cref{sec:hadamard_test_F_c}.
For some scenarios, simpler approaches may be possible.
For example, we may compute 
\begin{align}
    \widetilde{z_a}^j(\btheta)
    &=
    \begin{cases}
        U_{[:j]}(\btheta)(-iz_a)U^\dagger_{[:j]}(\btheta) &\text{ for }\act{\btheta}\\
        U^\dagger_{[j:]}(\btheta)(-iz_a)U_{[j:]}(\btheta) &\text{ for }\act{L}
    \end{cases}\,,
\end{align}
by evolving the symmetry generator with a part of the circuit on a classical computer.
Then we may obtain $\omega^{(E)}_{bj}$ via
\begin{align}
    \omega_{bj}^{(E)}(\btheta)
    &=
    i \partial_j \bra{\psi_0}U_{[:j]}^\dagger(\btheta)\widetilde{z_b}^j(\btheta)U_{[:j]}(\btheta)\ket{\psi_0}.
\end{align}
Overall, this makes the covariant (equivariant) derivative computable with circuits similar to the original PQC, but a considerable
amount of such auxiliary circuits is required.
The effort scales with the dimension of the symmetry (commutant) algebra, which can in the worst case become exponentially large, making the computation inefficient.

\subsubsection{Example: entangling circuit from covariance}\label{sec:entangling_app}
Here we briefly show an example of how to use the symmetry-aware approach to PQCs in an application.
Our optimization setup could be obtained without considering symmetries, but we hope that the perspective of this work will offer additional design strategies for more complex tasks.

Suppose we wanted to find a circuit on two qubits that entangles them maximally, starting from a product state $\ket{\psi_0}=\ket{\psi_0^{(1)}}\otimes\ket{\psi_0^{(2)}}$, and assume that we have repeated access to $\psi_0$.
We will find the maximally entangling circuit by constructing an ansatz $U(\btheta)$ and optimizing its parameters.
Note that a maximally entangled state, reduced to one of the qubits, results in the maximally mixed single-qubit state $\rho_{\mathrm{mm}}=\frac12\mbb I$, and that only maximally entangled states have this property\footnote{If we restrict ourselves to pure states on the full system, as prepared by unitary circuits.}.
In addition, note that $\rho_\mathrm{mm}$ is the only single-qubit state that commutes with all Pauli operators.
In other words, the map $\rho_\mathrm{mm}:\mathrm{Herm}\to\R$ via expectation values is invariant under $\mc{SU}(2)$ tranformations on either qubit, acting via $\act{L}$ on $\ket{\psi(\btheta)}$.

For this, we use as cost function the norm of the symmetry derivatives $m^{(q)}_{a\nu}(\btheta)=2\dE{\psi(\btheta)|M^{(q)}_\nu|V_a}$ for Pauli basis elements $M^{(q)}_\nu\in\DE{\mbb I, X, Y, Z}$ acting on the $q$th qubit.
With $6$ symmetry generators ($3$ per qubit, we do not consider global phase symmetries) and $8$ observables, we need to measure $48$ derivatives.
Before specifying the ansatz, let us point out some redundancies that reduce this number.
The symmetry derivatives can be written as a commutator expectation value between $M_\nu^{(q)}$ and $x_a$, similar to \cref{eq:sym_deriv_as_commutator}.
This commutator vanishes for $x_a$ and $M^{(q)}_\nu$ acting on different qubits, but also for $M_\nu^{(q)}\propto \mbb I$ or $M_\nu^{(q)}\propto x_a$.
In addition, we note that the condition to produce a maximally mixed state on one of the two qubits is sufficient, so that we may skip the group action and observables on, say, the second qubit altogether.
Thus, we are left with $3$ derivatives, given by the Pauli expectation values on the first qubit, so that
\begin{align}\label{eq:entangling_app:cost_function}
    C(\btheta) = \sum_{M\in \DE{X, Y, Z}} \bra{\psi(\btheta)}M^{(1)}\ket{\psi(\btheta)}^2.
\end{align}
As mentioned before, one could have constructed this cost function without considering symmetry transformations, but the above approach may generalize to cases that are not tractable manually.
\begin{figure}
    \centering
    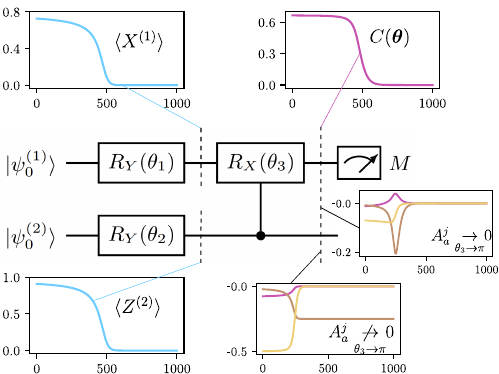
    \caption{\textbf{Parametrized quantum state used to prepare a maximally entangled state from some arbitrary product input state.}
    The measured observables $M$ lead to symmetry derivatives $m_{a\nu}$, which we use to define a cost function $C(\btheta)$, \cref{eq:entangling_app:cost_function}.
    Minimizing $C$ (\emph{top right inset}, iterations on all x-axes) yields parameters $\btheta^\ast$ that prepare the maximally entangled state on either qubit.
    This is achieved by maximizing the effect of the $CR_X$ gate, requiring to set $\braket{X^{(1)}}$ on the first qubit and $\braket{Z^{(2)}}$ on the second qubit to zero (\emph{left insets}), and $\theta_3$ to $\pi$.
    Some entries of the vector potential $A$ vanish at $\theta_3=\pi$, demonstrating that a vanishing covariant derivative need not imply cost function invariance (\emph{center right inset}).
    Other entries may or may not vanish, depending on $\theta_{1, 2}$ (\emph{bottom right inset}).
    }
    \label{fig:entangling_app}
\end{figure}
We choose the circuit ansatz
\begin{align}
    U(\btheta) 
    &= CR_X^{(2,1)}(\theta_3)R_Y^{(2)}(\theta_2)R_Y^{(1)}(\theta_1),
\end{align}
and apply it to some randomly initialized product state $\ket{\psi_0^{(1)}}\otimes\ket{\psi_0^{(2)}}$; see \cref{fig:entangling_app}.
Then we minimize $C(\btheta)$ via gradient descent and find the expected parameters:
$\theta_{1,2}$ are used to adjust the product state such that the $CR_X$ gate has a maximal effect, which is witnessed by the vanishing expectation values $\braket{X^{(1)}}$ and $\braket{Z^{(2)}}$ before the entangling gate.
$\theta_3$ is being set to $\pi$, making it equivalent to a maximally entangling $\mathrm{CNOT}$ gate up to a local phase gate.
We provide more details in \cref{sec:ex_calc:2q_cost}.

In this application, we used the symmetry derivatives $m_{a\nu}(\btheta)$ directly, rather than the covariant derivative $D_jC(\btheta)$.
This is necessary because the vector potential $A_j^a(\btheta)$ may vanish for specific parameter positions, suppressing corresponding contributions of the symmetry derivative to $D_jC(\btheta)$.
As a consequence, minimizing the distance $\|\partial_jC(\btheta)-D_jC(\btheta)\|=\|m_aA_j^a\|$ between the covariant and partial derivatives is insufficient to guarantee a symmetry-invariant cost function, forcing us to use the symmetry derivative.

\subsubsection{Commuting-generator circuits}
We briefly discuss a special class of PQCs that offer savings when computing (symmetric) derivatives: commuting-generator circuits.
The generators of these circuits satisfy $[H_j, H_k]=0\ \forall j,k$ so that the gates commute as well, $[U_j,U_k]=0$.
Instantaneous quantum polynomial (IQP) circuits, which have been widely studied~\cite{bremner2016average,ni2012commuting,lee2021towards}, fall into this category of circuits.
In addition, commuting-generator circuits have recently been investigated in \cite{bowles2023backpropagation} and shown to admit fast gradient computation schemes in some cases.
We will show that they exhibit a particularly simple structure of the covariant and equivariant derivatives, and thus also of the covariant quantum natural gradient we introduce in \cref{sec:qng}.

The commutativity simplifies the left and right-effective generators to $\Omega_j^{R/L}=-iH_j$.
This reduces the overlaps between gate derivatives and symmetry (commutant) directions from \cref{eq:compute_overlaps_equi_and_co} to
\begin{align}
    \omega_{bj}^{(E)}(\btheta) = 
    \begin{cases}
        \frac12 \bra{\psi_0}\DE{z_b,-i H_j}\ket{\psi_0} & \text{ for } \act{\btheta}\\
        \frac12 \bra{\psi(\btheta)}\DE{z_b,-iH_j}\ket{\psi(\btheta)} & \text{ for } \act{L}\\
    \end{cases}\,,
\end{align}
which usually makes them accessible with simpler methods than the Hadamard test from \cref{sec:hadamard_test_F_c}.
Instead---assuming that $H_j$ and $z_b$ can be handled classically---we may compute the anticommutator directly and evaluate its expectation value with a direct measurement in the state $\ket{\psi_0}$ or $\ket{\psi(\btheta)}$.
In particular, for $\act{\btheta}$ the overlaps $\omega_{bj}$ do no longer depend on $\btheta$.

Interestingly, the covariant (equivariant) derivative reduces to the partial derivative if the gate generators \emph{anticommute} with the symmetry (commutant) generators $z_b$.
This provides a new way to obtain inherently covariant derivatives, without making use of equivariant circuits and without putting any constraint on the observable $M$.
In fact, even the partial derivative of $\ket{\psi(\btheta)}$ itself is covariant (equivariant) in this scenario.

Commuting-generator circuits are of particular interest when the commutativity structure between the gate generators $\DE{H_j}_j$ and the observable $M$ is known.
By including the symmetry of interest in this structure, we may understand the covariant derivative of these specialized fast-gradients circuits, and compute them at similar cost as function evaluations.

\subsection{Commentary on applications design}
As mentioned before, in this work we focus on theoretical tools to incorporate symmetries in variational quantum computing, and finding suitable applications for the various approaches we present is left for future work.
Still, we here comment briefly on applications design.

The covariant derivative promises to project out parameter updates that are not of interest for an application, effectively reducing the dimension of the parameter subspace traversed during optimization.
We note that it only provides new optimization paths if the cost function itself does not incorporate the symmetry yet: an equivariant Hamiltonian will have vanishing symmetry derivatives
$m_a(\btheta)$,
making the partial derivative covariant.

Similarly, the circuit ansatz will lead to an already covariant partial derivative $\partial_jf(\btheta)$ if the corresponding gate does not generate movement in any symmetry direction at the point of the symmetry action, annihilating
$\omega_{bj}(\btheta)$.
As a consequence, the covariant derivative offers an alternative for incorporating the symmetry into the VQA whenever the cost function itself can not be made symmetry-aware.
It replaces reducing the parameter count during the ansatz design stage by symmetry-informed training, and promises to move in the right directions within a bigger parameter space.
In contrast to the predominant approach in the literature, this makes the impact of symmetries implicit.

Similar to the covariant derivative, the equivariant derivative allows to incorporate symmetries at the stage of training, e.g.,~whenever modifying the circuit ansatz is not an option.
This may enable the VQA to find symmetry-aware circuits in ans\"atze that are not equivariant in general, e.g.,~because they are tailored to hardware requirements.
For this, we note that measuring the equivariant derivative can be achieved via decompositions of the equivariant generators into linear combinations.
This is a consequence of the linearization we perform when moving from the global to the local perspective.
It removes the need to implement the full generator on hardware, which is a potential roadblock for the conventional GQML approach.

The used initial state will play an important role for equivariant training, as it defines the symmetry sector within which the constructed circuits lie.
For equivariant PQCs, the partial derivative reproduces the equivariant derivative by design, because projecting onto the equivariant subspace is defined at the level of the unitary algebra and equivariant gates form a group.

\section{Covariant quantum natural gradient}\label{sec:qng}
In this section we investigate the relationship between the covariant derivative of PQCs and the quantum natural gradient.
We start with a brief introduction and then establish a connection between the metric that underlies QNG---the quantum Fisher information---and the covariant derivative of a particular symmetry group.
This connection also allows us to generalize the natural gradient to arbitrary continuous symmetry groups.

\subsection{Quantum natural gradient}
The QNG of PQCs exploits the fact that global phase factors of quantum states do not have a physical impact.
It uses the geometry of projective Hilbert space, the state space that results from this redundancy, to obtain a more meaningful update direction for optimization tasks.
For a PQC with a cost function $C$, the QNG is given by
\begin{align}\label{eq:natural_gradient}
    \overline{\nabla} C(\btheta) &= \frac12 F(\btheta)^+ \cdot \nabla C(\btheta)\\
    F_{jk}(\btheta) &= \real{\braket{\partial_j\psi(\btheta)|\partial_k\psi(\btheta)}}\\
    &-\braket{\partial_j\psi(\btheta)|\psi(\btheta)}\braket{\psi(\btheta)|\partial_k\psi(\btheta)}\nonumber,
\end{align}
where $F$ is the Fubini-Study metric or quantum Fisher information, or the real part of the quantum geometric tensor\footnote{Up to rescaling: $F$ is the Fubini-Study metric, $4F$ is the quantum Fisher information, and $F=\real{G}$ where $G$ is the quantum geometric tensor~\cite{facchi2010classical}. As we comment below, the generic choice of the metric on the K\"ahler space $T_\psi\mc H$ is $2F$, just between the two.}, and $F^+$ is its pseudo-inverse.
It captures the geometry of the projective Hilbert space and as such, it does not depend on the observable that gives rise to the cost function $C$.
Note that for the update rule of the QNG optimization algorithm, the prefactor $\frac12$ usually is absorbed in the learning rate\footnote{This is also the approach commonly taken by derivations that focus on optimization, e.g.,~in~\cite[Sec.~2.2]{stokes2020quantum}.}.
Alternatively, it can be absorbed in $F^+$ by using the generic metric $2F$ on the K\"ahler space instead of $F$.

There is a multitude of approaches to QNG in the literature so that we restrict ourselves to a brief selection in the following.
In the context of PQCs, the goal to simulate imaginary-time evolution led to update schemes based on the quantum Fisher information~\cite{li2017efficient,mcardle2019variational}.
The relation to the so-called natural gradient from classical optimization theory was established in~\cite{stokes2020quantum} and analyses of the cost and performance of QNG include~\cite{wierichs2020avoiding,van2021measurement,yamamoto2019natural}.
Improvements as well as approximations to QNG were developed in~\cite{stokes2020quantum,koczor2022quantum,gacon2021simultaneous,van2021measurement} and in~\cite{gacon2023variational} a dual formulation is used to avoid the explicit computation of the metric.
A generalization of QNG via a projected derivative based on learning data has been proposed in~\cite{haug2023generalization}.
Finally, a more mathematical approach is taken in~\cite{hackl2020geometry}, which contains a thorough review of the geometry of projective Hilbert space (Chapter 3) and derives a number of variational principles in general (Chapter 4) and for PQCs in particular (Chapter 5).

\subsection{QNG and the covariant derivative}
As discussed in~\cite[Chapter 3]{hackl2020geometry}, the rescaled Fubini-Study metric $2F$ is the canonical metric on the tangent space of projective Hilbert space when viewing it as a K\"ahler space.
In the coordinates provided by a PQC, the matrix $F$ should match the Gram matrix of the covariant state derivatives induced by the global phase symmetry that leads to projective Hilbert space in the first place~\cite{requist2023quantum}.
In the following, we confirm this explicitly.

We already discussed the symmetry group of global phases $S=\mc U(1)\triangleleft\mc U(d)$ with its Lie algebra represented as $\mf t=i\R\mbb I$ in examples throughout the previous sections.
On the state level, this symmetry leads to the covariant derivative (\cref{eq:cov_deriv_states})
\begin{align}
    D_j \ket{\psi(\btheta)}
    &= \ket{\partial_j\psi(\btheta)} + A_j \ket{\psi(\btheta)}\\
    A_j
    &=-i\imag{\bra{\psi_0}\Omega_j(\btheta)\ket{\psi_0}}=-\braket{\psi(\btheta)|\partial_j\psi(\btheta)},
\end{align}
where we used $\braket{\psi(\btheta)|\partial_j\psi(\btheta)}\in i\R$ because $\Omega_j^\dagger=-\Omega_j$.
The Gram matrix of these covariant derivatives, using the real inner product $\dE{\,\cdot\,,\,\cdot\,}$, then is
\begin{align}\label{eq:fubini_from_cov_deriv}
    \dE{D_j\psi(\btheta)|D_k\psi(\btheta)}
    &= \real{\braket{\partial_j\psi(\btheta)|\partial_k\psi(\btheta)}+A_j^\ast A_k}\nonumber\\
    &=F_{jk},
\end{align}
i.e.,~it matches the Fubini-Study metric as expected.

The covariant derivatives in \cref{eq:fubini_from_cov_deriv} tempt us to generalize the Fubini-Study metric and the natural gradient to the homogeneous space (the space of states up to symmetries) corresponding to an arbitrary symmetry group $S$.
\begin{highlight}
The result is the \emph{covariant quantum natural gradient (CQNG)}, defined as
\begin{align}\label{eq:s_natural_gradient}
    \overline{\nabla}^S C(\btheta) = \frac12 F^S(\btheta)^+\cdot \nabla^S C(\btheta).
\end{align}
\end{highlight}
Here we denoted the vector of covariant derivatives as $(\nabla^S C(\btheta))_j=D_j C(\btheta)$ and made use of the Gram matrix $F^S$ of the covariant derivatives,
\begin{align}
    F^S_{jk}
    &=\dE{\partial_j\psi|\mbb H_\psi|\partial_k\psi}\\
    &=\dE{\partial_j\psi|\partial_k\psi}-\dE{\partial_j\psi|T_a}\dE{T_a|\partial_k\psi}.
\end{align}
The pseudo-inverse in \cref{eq:s_natural_gradient} is required because $F^S$---like the Fubini-Study metric---might be singular in the coordinates given by $\ket{\psi(\btheta)}$.
This is not a problem but merely indicates redundancies in the parametrization at the position $\btheta$ and requires us to remove the corresponding kernel of $F^S$ from the computation.

The principle underlying the (quantum) natural gradient is to approximate imaginary time evolution, which monotonically decreases the energy of the evolution Hamiltonian, within a given state parametrization.
This is formalized with the \emph{McLachlan minimal error (or variational) principle}~\cite{mclachlan1964variational,hackl2020geometry}, with respect to the real inner product on $T_\psi\mc H$.
In \cref{sec:qng_comp} we generalize this to the homogeneous space defined by the symmetry $S$, corresponding to an inner product that is agnostic to the vertical subspace of $T_\psi\mc H$.
We then derive the gradient flow of the variational parameters and find the CQNG in \cref{eq:s_natural_gradient} above.

The Gram matrix $F^S$ can be obtained on a quantum computer using Hadamard test-based constructions.
For the first term, $\dE{\partial_j\psi|\partial_k\psi}$, which is part of the Fubini-Study metric as well, this is discussed, e.g.,~in \cite{mcardle2019variational,li2017efficient,Guerreschi_Smelyanski_17}.
The second term is a product of the (rectangular) matrix $\dE{\partial_j\psi|X_a}$ with its own transpose.
We discussed its computation together with recipes to obtain $D_j C(\btheta)$ in \cref{sec:circuitry:cost_functions}.

Looking back to QNG, the CQNG with respect to global phases, we note that in \cref{eq:natural_gradient} the covariant derivative $\nabla^{\mc U(1)} C(\btheta)$ has been replaced by the partial derivatives $\nabla C(\btheta)$.
This is due to the fact that any PQC is equivariant with respect to global phases, so that $\nabla^{\mc U(1)} C(\btheta)=\nabla C(\btheta)$ (see \cref{sec:circuitry:cost_functions}).
Simultaneously, the equivariance with respect to global phases implies that $\partial_j C(\btheta)=E_j C(\btheta)$.
This points to an alternative generalization of QNG not to CQNG but to an \emph{equivariant quantum natural gradient}, which would be defined in analogy to \cref{eq:s_natural_gradient}.

The CQNG respects the metric of the parameter landscape that results from optimizing over the homogeneous space rather than the full Hilbert space.
For QNG, this homogeneous space is the projective Hilbert space, which provides a particularly suitable stage to describe quantum states.
However, in applications with additional symmetries, $\overline{\nabla}^S C(\btheta)$ takes those into account and can be expected to allow for more efficient optimization strategies.
As already discussed e.g.,~in~\cite{hackl2020geometry,stokes2020quantum} and references therein, advanced optimization algorithms can be adapted to make use of natural gradients and exploit their focus on relevant parameter directions.

\section{Physical gauge theory}\label{sec:gauge_theory}
In this section we want to explicitly connect the covariant derivative from \cref{sec:equi_and_co} to the covariant derivatives in field theory, as it is used for high energy particle physics, for example.
From a constructive point of view, this can be seen as a demonstration that the covariant derivative based on tangent space decompositions indeed transforms like a covariant object.
As for standard physical gauge theories, we will consider local symmetry transformations, with the variational parameters of PQCs taking the place of spacetime coordinates.
This allows to capture all (continuous) degrees of freedom that are redundant for the mathematical description of the PQC state due to its symmetries.
For a single two-level system, or qubit, the gauge theoretical perspective on time evolution and its relationship to Hilbert space geometry was discussed in \cite{bruno2011gauge}.

We want to point out two differences to the typical setup in physical gauge theory:
first, PQCs directly implement the full description of the parameterized quantum state, including symmetry transformations.
As a consequence, the vector potential is just as tangible as any quantity commonly measured in QML and VQAs, and we will obtain an explicit expression for $A_j^a$ in \cref{eq:vector_potential}.
Second, in our work we rarely care about the quanta arising from the conserved quantities that are related to the symmetry transformations.
In contrast they are of central importance in particle physics, as these quanta are the interaction particles.

We will closely follow \cite[Ch.~15]{peskin_schroeder} in this section.
We start with a parametrized quantum state, given by its density matrix $\rho(\btheta)=U(\btheta)\rho_0 U^\dagger(\btheta)$.
In addition, we consider the adjoint action $\act{}$ of the symmetry group $S$ on density matrices induced by a suitable linear representation $\sigma$, 
$s\act{}\rho(\btheta)=\sigma(s)\rho(\btheta) \sigma(s)^\dagger$.
\begin{convention}
We will not denote the representation $\sigma$ in the following but imply it whenever writing group elements next to density matrices.
\end{convention}
The symmetry transformation we consider may itself depend on the variational parameters $\btheta$, so that the density matrix transforms like
\begin{align}
    \rho(\btheta)\longrightarrow s(\btheta)\act{}\rho(\btheta) = s(\btheta)\rho(\btheta)s^\dagger(\btheta).
\end{align}

If we now want to define a derivative with respect to the parameters $\btheta$ as an infinitesimal difference, we need to compare $\rho(\btheta+\epsilon \vec{n})$ to $\rho(\btheta)$ for some direction vector $\vec{n}$ and thus need to make them transform in the same way under $s(\btheta)$.
For this, we introduce the \emph{comparator} $\Gamma(\vec{\phi}, \btheta)$ that connects $\rho(\btheta)$ between any two parameter positions $\btheta$, $\vec{\phi}$.
We restrict it to be a unitary channel and require it to transform as
\begin{align}
    \Gamma(\vec{\phi}, \btheta)\longrightarrow s(\vec{\phi}) \Gamma(\vec{\phi}, \btheta) s^{-1}(\btheta),
\end{align}
so that $\Gamma(\vec{\phi}, \btheta)\act{}\rho(\btheta)$ transforms like $\rho(\vec{\phi})$.
We may restrict $\Gamma$ further to lie in $S$ and are still guaranteed that it allows for the transformation behaviour above, because $S$ is a group.
In addition, we fix the comparator between $\btheta$ and itself to be the identity (channel), $\Gamma(\btheta, \btheta)=\mbb I$.

For small differences between the parameter positions, we obtain the expansion
\begin{align}
    \Gamma(\btheta+\varepsilon \vec{n}, \btheta)\act{}\rho(\btheta)
    &=\rho(\btheta) + \varepsilon n_j A_j^a [x_a, \rho(\btheta)] + \mc O (\varepsilon^2)\nonumber,
\end{align}
where $\{x_a\}_a$ is a basis for $\mf t$, the represented Lie algebra of $S$, and $A$ consists of the expansion coefficients of the first-order contribution of $\Gamma$, and is called the \emph{vector potential}.
As before, we used the Einstein summation convention.

As planned, the comparator allows us to define the covariant derivative as the limit of a directional finite difference:
\begin{align}\label{eq:cov_deriv_gauge_def}
    n_j \Delta_j \rho(\btheta) = \lim_{\varepsilon\to 0}\frac{1}{\varepsilon} \left[\rho(\btheta+\varepsilon \vec{n}) - \Gamma(\btheta+\varepsilon\vec{n}, \btheta)\act{}\rho(\btheta)\right].
\end{align}
Below we will show that $\Delta_j$ is equivalent to $D_j$ from \cref{eq:covariant_derivative_abstract}.
The expansion of $\Gamma$ above and the Taylor expansion of $\rho(\btheta+\varepsilon\vec{n})$ lets us rewrite $\Delta_j$ as
\begin{align}
    n_j \Delta_j \rho(\btheta) &= 
    \lim_{\varepsilon\to 0}\Large[n_j \partial_j \rho(\btheta)-n_j A_j^a [x_a, \rho(\btheta)] + \mc O(\varepsilon)\Large]\nonumber \\
    \Rightarrow \Delta_j \rho(\btheta) &= \partial_j\rho(\btheta) - A_j^a [x_a,\rho(\btheta)].
\end{align}

This modified derivative is termed covariant because it transforms like $\rho(\btheta)$ itself.
To show this, we compute the transformation behaviour of the vector potential from that of $\Gamma$ (see \cref{sec:gauge_theory_comp}):
\begin{align}
    [A_j^a x_a, \circ] \longrightarrow \left[s(\btheta)\left(\Lambda_j(\btheta) + A_j^a x_a\right)s^\dagger(\btheta), \circ \right ],
\end{align}
with the effective generator $\Lambda_j(\btheta) = s^\dagger(\btheta)\partial_j s(\btheta)$ of $s(\btheta)$.
The covariant derivative then can be shown to transform covariantly:
\begin{align}
    \Delta_j\rho(\btheta)\longrightarrow s(\btheta)\act{} \Delta_j \rho(\btheta).
\end{align}
Finally, we need to show that the covariant derivative introduced here is indeed the same as that introduced in \cref{sec:equi_and_co:covariance_algebras}.
For this, it is sufficient to show that the vector potential defined through the latter transforms correctly, as this will lead to the correct transformation behaviour of the covariant derivative.
In \cref{sec:gauge_theory_comp} we match the vector potential
\begin{align}\label{eq:vector_potential}
    A_j^a= G^+_{ab}\dE{\psi(\btheta)|x_b^\dagger \partial_j|\psi(\btheta)}
\end{align}
from \cref{eq:vector_potential_states} with the one above.
We then show that the (super-)operator $A_j^a [x_a, \circ]$ indeed transforms correctly.

\begin{highlight}
We thus identified the covariant derivatives with each other, i.e.,~$\Delta_j = D_j$.
\end{highlight}

\section{Conclusion}\label{sec:conclusion}
In this work we investigated the impact of symmetries on parametrized quantum circuits (PQCs) beyond the conventional approach of equivariant circuit ans\"atze.
For this, we focused on symmetry-aware derivatives, in particular the equivariant derivative, the covariant derivative, and the covariant quantum natural gradient.
Geometrically, our approach realizes symmetries locally, while the design of circuit ans\"atze can be understood as a global approach.

We started with the notion of a projected derivative which allows us to constrain the directions of change to some subspace.
This also constrains changes in the space of PQC parameters, e.g.,~during training.
The equivariant derivative restricts the common derivative to directions that are unaffected by symmetry transformations, moving the design principle of equivariant ans\"atze to the level of training.
Thus, if a PQC is equivariant already, the equivariant derivative reduces to the partial derivative.
Instead, the covariant derivative projects out symmetry directions, i.e.,~directions of change that would be caused by symmetry transformations.
This differs from the equivariant derivative in two important aspects.

First, the conserved subspaces differ, and neither is contained in the other; on one hand, there are symmetry directions that are left unchanged by other symmetry transformations (equivariant but not covariant).
On the other hand, there are non-symmetry directions that are modified by symmetry transformations (covariant but not equivariant).
Second, the equivariant subspace has a particularly nice structure---it constitutes a subalgebra.
This makes the notion of equivariant ans\"atze well-defined in the first place.
The covariant subspace does not have this structure in general, so that covariant PQC ans\"atze can only be defined in special cases.
More precisely, there is no notion of a covariant subgroup of the unitary group in general.

In addition to the general framework above, we discussed the implications at the level of parametrized unitaries, parametrized quantum states, and PQC cost functions.
Symmetries are defined at the group level, as are parametrized unitaries, which makes the analysis of the latter particularly convenient.
For quantum states and cost functions, however, the redundancy in the description via group actions complicates the projected derivatives, because the decompositions of the state tangent space are state-dependent.
For the covariant derivative we can not rely on a global reference of covariant quantum gates, so that this state dependence is especially important.

After describing the subtleties introduced by this, we explained how to obtain the equivariant and covariant derivatives on a quantum computer, using circuits similar to the original PQC.
In addition we demonstrated with a toy problem that our understanding of symmetries acting on PQCs allows to find quantum circuits with desired properties, using a symmetry-aware optimization task.

Next, we investigated the quantum natural gradient (QNG) and its relation to the covariant derivative.
While the relationship of QNG to global phase symmetries and the geometry of Hilbert space was well-known, we explicitly made the connection to the covariant derivative introduced above and showed that it gives rise to the quantum Fisher information.
This allowed us to generalize the quantum Fisher information and QNG to arbitrary continuous symmetry groups.
QNG has proven useful for gradient-based optimization of PQC cost functions in the past, and our extension allows to take the modified geometry of the quotient between Hilbert space and symmetry transformations into account.
This reduced the QNG to the degrees of freedom that are essential for the physical system of interest.

The covariant derivative plays an important role in gauge theory, which famously powered the development of modern particle physics throughout the last century and until today.
We showed how our definition of the covariant derivative compares to the traditional approach in physical gauge theory and showed their equivalence.
While this is not surprising from a mathematical point of view, it enables the transfer of established tools and concepts from physical gauge theory to geometric quantum machine learning.

Equivariant circuits are commonly used to construct QML models that incorporate symmetries in the learning task at hand in the form of priors that constrain the learning process favourably.
The equivariant derivative can replace this design tactic whenever modifying the PQC ansatz itself is not feasible.
Moreover, a symmetry group could also be defined through transformations that hurt the model performance.
This makes the covariant derivative a useful tool, as it explicitly excludes symmetry transformations from the training and allows the optimization to traverse the parameter space along a desired subspace.
As covariant quantum gates in general do not form a subgroup, it is not possible to realize this constraint globally, making the covariant derivative the only choice for this task.
Our new perspective on symmetries in parametrized quantum circuits expands the toolbox of geometric QML and may prove useful to shine new light on existing problems, QML models and optimization techniques.

\section{Acknowledgements}
DW is thankful for many useful discussions on the subject with Roeland Wiersema, David Wakeham, and Kor-binian Kottmann. ML was supported  by the U.S. Department of Energy (DOE), Office of Science, Office of Advanced Scientific Computing Research, under the Accelerated Research in Quantum Computing (ARQC) program as well as by the  Center for Nonlinear Studies at Los Alamos National Laboratory (LANL).  MC acknowledges support by the Laboratory Directed Research and Development (LDRD) program of LANL under project numbers 20230049DR and  20230527ECR. This work was also supported by LANL ASC Beyond Moore’s Law project.

\newpage
\bibliographystyle{quantum}
\bibliography{bib}

\appendix
\onecolumngrid


\section{Covariant derivative from quantum circuit perspective}\label{sec:circuitry_app}

In this appendix, we reproduce the results about the covariant derivative from \cref{sec:circuitry} in a less abstract approach.
Instead of mathematical abstractions we build on intuition about unitary operations and quantum states, and we treat notions from differential geometry from the perspective of quantum circuits.
Similar to the structure of \cref{sec:circuitry}, we discuss the covariant derivative on three ``levels'', namely for quantum gates (\cref{sec:circuitry_app:gates}), quantum states (\cref{sec:circuitry_app:states}), and PQC-based cost functions (\cref{sec:circuitry_app:cost_functions}).

\subsection{Quantum gates}\label{sec:circuitry_app:gates}

\subsubsection{Not a section about representations}\label{sec:circuitry_app:notrep}
Before we dive into the unitary group and its tangent spaces, we briefly need to mention group and algebra representations, which we will not keep track of explicitly.
A proper introduction can be found in~\cite{hall2015lie_1,fulton2004representation_1}.

A representation $\sigma$ of a group $S$ on a vector space sends group elements to invertible matrices with the correct shape for the vector space.
In addition, $\sigma$ is compatible with multiplication in $S$ and matrix products, so that $\sigma(g)\sigma(g') = \sigma(g g')$.
Representations are an essential tool for many applications and for understanding the deeper structure behind groups and the ways they can interact with other spaces.
\begin{convention}
    However, in this appendix we will ignore the step of representations, and simply denote the \emph{represented} groups.    
\end{convention}
This means that we do not think about abstract groups but only about unitary matrices that interact with matrix-vector and matrix-matrix products.
A representation of an algebra is defined similarly, with the requirement that $\sigma([x, y])=[\sigma(x), \sigma(y)]$.
We do not use those either but only talk about already represented (Lie) algebras.

\subsubsection{The unitary group and its tangent spaces}
The $d$-dimensional unitary matrices $\mc U(d)\equiv\mc U$ form a group under matrix multiplication.
This means that quantum gates, described by unitary matrices, can be composed into new gates and that for each gate $U$ there is a gate $U^{-1}$ that reverts it, given by its adjoint $U^{-1}=U^\dagger$ (i.e.,~$U^\dagger U=\mbb I\ \forall U\in\mc U$).
Any unitary can be expressed as the exponential
$U=\exp(-iH)$ of some skew-Hermitian matrix $-iH$, and any such exponential is unitary.
This equivalence is obvious for some standard gates, like single-qubit rotations 
\begin{align}
    R_P(\theta)=\exp\de{-i\frac{\theta}{2}P}
\end{align}
about some Pauli operator $P\in\DE{X, Y, Z}$, but also holds for gates like
\begin{align}
    \mathrm{CNOT}=\exp\de{-i \frac{\pi}{4} (Z-\mbb I)\otimes(\mbb I-X)}.
\end{align}
This equivalence between unitaries and skew-Hermitian matrices---or generators---will be a very useful perspective.
For example, it allows us to determine the (real) dimension of $\mc U(d)$ and its subgroups:
skew-Hermitian matrices in $d$ dimensions have $d^2$ real degrees of freedom so that $\dim(\mc U(d))=d^2$.
More importantly, the generator perspective makes it convenient to investigate the directions in which a given unitary can be modified.
For this, consider two unitaries $U,U'\in\mc U$ that are infinitesimally close together.
Because $\mc U(d)$ is a group, and each group element is the exponential of some skew-Hermitian matrix, we may write
\begin{align}\label{eq:close_by_unitary_expansion}
    U'=U\exp(-i\varepsilon H)=U-i\varepsilon UH+\mc{O}(\varepsilon^2),
\end{align}
for some finite $H$ and an infinitesimal $\varepsilon$.
This allows us to express the tangent space $T_U\mc U$ of $\mc U$ at $U$, i.e.,~the space of directions in which $U$ can change, by skew-Hermitian matrices.
In particular, we may fix an orthonormal basis (ONB) $\mc B$ for skew-Hermitian matrices with respect to the trace inner product
\begin{align}\label{eq:HS_inner_product}
    (x, y)\mapsto \frac{1}{d}\tr{x^\dagger y}\ \text{ and write }\ T_U\mc U = \Span_\R\DE{Ux | x\in\mc B}.
\end{align}
At the identity matrix we simply have $\mf u = T_{\mbb I}\mc U=\Span_\R \mc B$, the space of skew-Hermitian matrices.
Besides the vector space operations, $\mf u$ comes with an additional ``interaction'' of matrices, given by the matrix commutator.
We can check that the commutator of skew-Hermitian matrices is a skew-Hermitian matrix again\footnote{$[x, y]^\dagger = [y^\dagger, x^\dagger] = - [x, y].$}
and therefore that $\mf u$ is closed under the commutator.
This\footnote{Together with the alternativity condition $[x, x]=0$ for all matrices $x$ and the Jacobi identity.} makes $\mf u(d)$ a Lie algebra, more specifically it is \emph{the} Lie algebra associated to the unitary group $\mc U(d)$.
The above analysis shows us that it directly corresponds to the directions in which we can unitarily modify the identity operation $\mbb I$.

The tangent space at any other point $U$ in $\mc U$ essentially looks like $\mf u$, as these spaces are isomorphic.
In addition, multiplying skew-Hermitians $x, y$ with the unitary $U$ from the left does not change their inner product, so that the set $U\mc B$ again is an ONB, now for $T_U\mc U=U\mf u$.
In fact, we see that there is a simple way of relating tangents at different unitaries $U, V$ with each other, using multiplication from the left by the unitary that connects $U$ and $V$. 

In the context of quantum gates on an $N$-qubit system, the Pauli basis $\mc B=-i\DE{\mbb I, X, Y, Z}^{\otimes N}$ is a common choice.
We can modify any gate $U$ by multiplying a rotation gate\footnote{Including the rotation gate about $\mbb I$, a global phase gate.} to it from the right, effectively executing it before $U$ itself.
The analysis above tells us that for small rotation angles this will yield independent modifications of $U$ for each Pauli word, corresponding to $d^2=4^N$ directions.

\subsubsection{Symmetry groups and vertical directions}\label{sec:circuitry_app:gates:symmetry}
Now let us consider a connected symmetry Lie group $S$ that is a unitary subgroup\footnote{In this section we will not be precise about the mathematical requirements on $S$, which include compactness and that it is at least one-dimensional.}.
In analogy to the full unitary group, we are interested in the directions in which a symmetry transformation can modify a unitary.
At the identity, which is contained in any subgroup, these directions form a subspace, and even a subalgebra, of $\mf u$, which we denote as $\mf t$\footnote{While $\mf s$ might seem the more natural name here, we use $\mf t$ for consistency with the other parts of the manuscript.}.
It is the Lie algebra of $S$, and thus contains exactly those skew-Hermitian matrices that yield elements of $S$ when exponentiated, i.e.,~$S=\exp(t)$.
It has the same dimension as $S$ itself.

Away from the identity, the symmetry directions are given by the \emph{vertical subspace}
\begin{align}
    V_U\mc U = \DE{Ux | x\in \mf t} = U\mf t.
\end{align}
We want to highlight two important differences between the pairs $(\mc U, \mf u)$ and $(S, \mf t)$:
first, we consider the way we represent the group elements.
For the unitary group $\mc U(d)$, we simply chose to represent each abstract group element by ``its matrix''.
A representation, as mentioned in \cref{sec:circuitry_app:notrep}, maps elements from an abstract group to $d$-dimensional matrices, which leads to redundancy and requires us to fix this map.
For example, the unitary group $\mc U(2)$ composed of single-qubit gates is a subgroup of all two-qubit gates $\mc U(4)$, but there is no generic way to embed the former into the latter.
Does $\mc U(2)$ act on the first or on the second qubit?
Or on both qubits simultaneously, by copying the gate?
Or only on the subspace spanned by $\DE{\ket{01}, \ket{10}}$?
We therefore need to fix $S$, the subgroup of unitary matrices to which the abstract group is mapped.
We do this implicitly in the following.

The second difference concerns the order of the unitary $U$ and the skew-Hermitian matrices that define the directions of change.
For the full unitary group, we could have written $U'=\exp(-iJ)U$ instead of $U'=U\exp(-iH)$ in \cref{eq:close_by_unitary_expansion}, which leads to a different Lie algebra element $-iJ\neq -iH$, in general, because $H$ and $U$ need not commute.
Still, the tangent space would remain the same under this change, because we can write 
\begin{align}
    \exp(-iJ)=U\exp(-iH)U^\dagger = \exp(-iUHU^\dagger),
\end{align}
and $-iUHU^\dagger$ again is a skew-Hermitian matrix.
For the subgroup $S$, however, the space $V_U\mc U=U\mf t$ as defined above and the alternative definition $V'_U\mc U=\mf t U$ will \emph{not} be equivalent, unless $S$ satisfies additional assumptions\footnote{Namely that it be a \emph{normal} subgroup.}.
This is not an issue but simply means that after choosing a definition of the tangent space we need to comply with this choice in the following.
From a more abstract point of view, we are fixing an \emph{action of $S$ on $\mc U(d)$} (see \cref{sec:equi_and_co:equivariance_groups}).

The symmetry group defines the vertical subspace, but what about the remaining directions?
We already introduced the trace inner product for skew-Hermitian matrices and noted that it can be applied to translated matrices $Ux$ as well.
Thus it fixes the orthogonal complement to the vertical subspace, the \emph{covariant, or horizontal, subspace}
\begin{align}
    H_U\mc U
    &= \DE{Ux|x\in\mf u,\ \tr{x^\dagger U^\dagger y}=0\ \forall y\in U\mf t}\\
    &= U\DE{x \in \mf u| \tr{x^\dagger y}=0\ \forall y\in\mf t}\\
    &= U\orth{\mf t}.
\end{align}
As we can see from $\orth{U\mf t}=H_U\mc U=U\orth{\mf t}$, the decomposition of $T_U\mc U$ into covariant and vertical components is compatible with the translation by group elements.

With the decomposition of $T_U\mc U$ come projectors $\mbb V_U$ and $\mbb H_U=\mbb I-\mbb V_U$ onto the vertical and covariant subspace, respectively.
Using an ONB $\mc B_{\mf t}$ for $\mf t$ with respect to the trace inner product, the vertical projector is given by
\begin{align}
    \mbb V_U = \sum_{y\in\mc B_{\mf t}} Uy\,\frac{1}{d}\tr{y^\dagger U^\dagger\ \cdot\ }=\frac{1}{d}\sum_{y\in U\mc B_{\mf t}} \tr{y^\dagger\ \cdot\ } y,
\end{align}
which maps a direction $Ux\in U\mf u$ to a new direction that is a superposition of only the vertical directions $Uy\in V_U\mc U$, with coefficients $\frac{1}{d}\tr{y^\dagger x}\in \R$.

The projector onto the covariant space is simply given by $\mbb H_{U}=\mbb I-\mbb V_{U}$.
Using the projectors we can decompose any direction $Ux$ into its vertical and covariant part.
This idea is at the heart of the covariant derivative, which aims at \emph{only ever moving in the covariant direction}.

\subsubsection{The covariant derivative on \texorpdfstring{$\mc U$}{U}}\label{sec:circuitry_app:gates:cov_deriv}
We now want to put the space of directions into a more concrete context by considering a parametrized unitary $U(\btheta)$ as in \cref{sec:preliminaries:pqc} which depends on $p$ real-valued parameters $\btheta$.
Such a parametrization could be given by a sequence of rotation gates, a pulse-level description of a quantum system, or simply by a full parametrization of a unitary gate as discussed e.g.,~in \cite{wiersema2023here}.
Any parametrization comes with tangents via its partial derivatives (see \cref{sec:preliminaries:pqc}):
\begin{align}
    \partial_j U(\btheta)
    =U(\btheta)\Omega^R_j(\btheta)
    \in T_{U(\btheta)}\mc U=U(\btheta)\mf u.
\end{align}
$\Omega^R_j(\btheta)\in\mf u$ is the \emph{right-effective generator} of $U$ with respect to $\theta_j$ at the position $\btheta$.
Intuitively, the effective generator captures the direction of the change to $U(\btheta)$ that will be caused by increasing the parameter $\theta_j$ infinitesimally.
Above we discussed the alternative convention of writing tangents as $\mf u(d) U$ instead of $U\mf u(d)$.
In this convention the derivative is given by $\Omega^L_j(\btheta) U(\btheta)$ with the \emph{left-effective generator} $\Omega^L_j(\btheta)=U(\btheta)\Omega^R_j(\btheta)U^\dagger(\btheta)$.
Unless mentioned otherwise, we mean the right-effective generator when skipping the label $R/L$.

For a circuit consisting of an individual rotation gate, for example, the two effective generators are the same, and equal to the generator of the gate itself:
\begin{align}
    U(\btheta)=\exp\de{-\frac{i}{2}\theta_1 H}\quad\Rightarrow\quad 
    \partial_1 U(\btheta)
    = U(\btheta)\Omega_1
    = \Omega_1 U(\btheta),\quad \Omega_1=-\frac{i}{2}H.
\end{align}
However, for more complex parametrizations---and in fact already for a sequence of simple rotation gates---the effective generator(s) will depend on the parameter position $\btheta$, and $\Omega^R_j(\btheta)\neq\Omega^L_j(\btheta)$ in general.

We now may define the \emph{covariant derivative} of a parametrized unitary $U(\btheta)$ with respect to its parameter $\theta_j$.
It is the projection of the partial derivative onto the covariant subspace at $U(\btheta)$:
\begin{align}
    D_j U(\btheta)
    &= \mbb H_{U(\btheta)} \partial_j U(\btheta)\\
    &= \partial_j U(\btheta) - \sum_{y\in\mc B_{\mf t}} U(\btheta) y \frac{1}{d}\tr{y^\dagger U^\dagger(\btheta) \partial_j U(\btheta)}\\
    &= U(\btheta)\de{\Omega_j(\btheta)-\sum_{y\in\mc B_{\mf t}} y \frac{1}{d} \tr{y^\dagger \Omega_j(\btheta)}}.
\end{align}
As we can see from the last line, we may alternatively understand the covariant derivative first projecting $\Omega_j$ onto the covariant subspace of $\mf u=T_{\mbb I}\mc U$ and left translating it afterwards, i.e.,~$D_j U(\btheta)=U(\btheta)\mbb H_{\mbb I}\Omega_j(\btheta)$.
This is due to the compatibility between left translation and the definition of the vertical subspaces, which guarantees that $\mbb H_U U=U\mbb H_{\mbb I}$.

\subsubsection{Example: \texorpdfstring{$\mc U(1)<\mc U(2)$}{U(1)<U(2)}}\label{sec:circuitry_app:gates:example}
To understand the covariant derivative better, we discuss a simple example.
We set $d=2$ and look at the symmetry group $\mc U(1)$, which we represent as subgroup of $\mc U(2)$ via $S=\DE{\exp(-i\alpha Z)| \alpha \in [0, 2\pi)}$ so that $\mf t=-i\R Z$.
The vertical direction within $T_U\mc U(2)$ then is $iUZ$, and the space of covariant directions is spanned by $\DE{iU, iUX, iUY}$.

In addition, we pick the parametrized unitary
\begin{align}\label{eq:example_circuit_gates}
    U(\btheta)&=R_X(\theta_2)R_Y(\theta_1)
    =\exp\de{-i\frac{\theta_2}{2}X}\exp\de{-i\frac{\theta_1}{2}Y},
\end{align}
which has the effective generators
\begin{align}
    \Omega_1 (\btheta) &= -\frac{i}{2} Y \\
    \Omega_2 (\btheta) &= -\frac{i}{2} R_Y(-\theta_1)XR_Y(\theta_1)
    =-\frac{i}{2}\de{\cos(\theta_1)X+\sin(\theta_1)Z}.
\end{align}
Note that $R_X$ and $R_Y$ seem to point into purely covariant directions.
However, due to $[X,Y]\propto Z$, the generator $\Omega_2(\btheta)$ contains a contribution corresponding to the vertical direction.
Thus, following the partial derivatives would lead to a movement in \emph{both} the vertical and covariant subspaces.
Applying the definition for the covariant derivative above, we obtain
\begin{align}
    D_1 U(\btheta) &= -i\frac{1}{2} U(\btheta) Y \\
    D_2 U(\btheta) &= U(\btheta)\De{\Omega_2(\btheta)- (-iZ)\frac{1}{2}\tr{iZ \Omega_2(\btheta)}}
    =-i\frac{\cos(\theta_1)}{2}U(\btheta) X.
\end{align}
The vertical component of $\partial_2 U(\btheta)$ has been removed successfully and the covariant derivatives lie in the covariant subspace $H_{U(\btheta)}\mc U(2)$.

It is instructive to note that this statement only concerns the direction of change and does not imply anything about the position $U(\btheta)$ in the unitary group itself; in particular, for the parameters $\btheta^\ast=(\pi,\pi)^T$ the point $U\de{\btheta^\ast}=-iZ$ itself is an element of the symmetry group, despite the fact that the partial derivative is covariant at this point, $\partial_j U(\btheta^\ast)=D_j U(\btheta^\ast)$.

\subsection{Quantum states}\label{sec:circuitry_app:states}
\subsubsection{Changing states with quantum gates---Tangents of the state space}
In the previous section we were concerned with unitaries and the possible directions in which we can change them, i.e.,~the tangent space of $\mc U$.
In this section we turn to the Hilbert space $\mc H$ and discuss similar concepts for parametrized quantum states.
The core idea is the same as in \cref{sec:circuitry_app:gates}, but there are a few differences that we want to discuss in detail.
We don't consider parametrized states in their own right but in particular look at those which are created by PQCs, which is motivated by the use of PQCs in practice, together with the symmetries that appear in applications.
This impacts our definition of the tangent space of $\mc H$ and requires us to take special care of redundancies.

Consider a quantum state $\ket{\psi}\in\mc H$.
We are interested in different ways to modify this state using a unitary gate.
More precisely, we do not look for new quantum states we may move $\ket{\psi}$ \emph{to}, but for \emph{infinitesimal} changes corresponding to directions to move in, i.e.,~tangent vectors.
Similar to \cref{eq:close_by_unitary_expansion}, we may use the expansion of the exponential to write a state $\ket{\psi'}$ close by $\ket{\psi}$ as
\begin{align}\label{eq:close_by_state_expansion}
    \ket{\psi'}=\exp(-i\varepsilon H)\ket{\psi}=\ket{\psi}-i\varepsilon H\ket{\psi} +\mc{O}(\varepsilon^2),
\end{align}
from which we read off the tangent $\-iH\ket{\psi}$.
This means that we can express the tangent space of $\mc H$ that corresponds to unitary evolutions using the tangent space of the unitaries, $T_{\mbb I}\mc U=\mf u$:
\begin{align}
    T_{\psi}\mc H = \Span_\R \DE{x\ket{\psi}|x\in\mc B} = \mf u \ket{\psi},
\end{align}
where $\mc B$ again is some basis for the space of skew-Hermitian matrices.
The span is over $\R$ because directions of the form $ix\ket{\psi}$---and therefore non-real multiples of $x\ket{\psi}$---are not necessarily related to unitary operations.

Intuitively, the set $\mc B\ket{\psi}$ is overcomplete, because there are many gates that modify any given state $\ket{\psi}$ in the same way\footnote{For example, all Hermitian matrices for which $\ket{\psi}$ is an eigenstate will lead to a direction proportional to $i\ket{\psi}$.}.
This intuition is confirmed by counting dimensions:
as discussed before, $\mf u$ is $d^2$-dimensional.
In contrast, a tangent vector at any state $\ket{\psi}$ only has $d$ complex, i.e.,~$2d$ real, degrees of freedom.
Among those, there are the directions $i\ket{\psi}$, which generates global phase transformations, and $\ket{\psi}$, which generates rescalings of the state vector and is not unitary, thus not being part of $T_\psi\mc H$ above.

We obtain $\dim(T_\psi\mc H)=2d-1$ real degrees of freedom for the tangent space of $\mc H$ corresponding to unitary evolutions.
This confirms that $\mc B\ket{\psi}$ is highly overcomplete to span $T_\psi\mc H$, namely close to $d/2$-fold.
As indicated in the beginning of the section, we will need to keep this redundancy in mind, but it will nonetheless be useful to represent the tangent space $T_\psi\mc H$ via $\mf u$.

On the real vector space $T_{\psi}\mc H$ we use the inner product
\begin{align}
    \dE{\,\cdot\,|\,\cdot\,}
    : T_{\psi}\mc H \times T_{\psi}\mc H \to \R
    : (\ket{X}, \ket{Y})\mapsto \dE{X|Y} = \real{\braket{X|Y}},
\end{align}
where $\braket{X|Y}$ is the usual complex-valued inner product on $\mc H$ (and $T_\psi\mc H$).
The latter induces a metric, leading to our real-valued inner product above, which is also called \emph{quantum geometric tensor}.
It also carries a symplectic form on $\mc H$, given by the imaginary part of $\braket{X|Y}$, which we will not use explicitly at this point~\cite{hackl2020geometry}.

If two directions $\ket{X}$ and $\ket{Y}$ are given by skew-Hermitian matrices $x$ and $y$ applied to $\ket{\psi}$, the following equality will occasionally be useful\footnote{There is a certain ambiguity in the notation $\dE{X|O|Y}$ for non-Hermitian operators $O$. We always mean $\real{\bra{X}(O\ket{Y})}$, i.e.,~any operators between the states are read as denoted, not as part of the bra.}:
\begin{align}\label{eq:real_to_anticom}
    \dE{X|Y}=\real{\bra{\psi}x^\dagger y\ket{\psi}}
    = -\frac12 \bra{\psi}\DE{x,y}\ket{\psi}.
\end{align}
As we can see, the anticommutator of the generators quantifies the overlap between the directions they generate.
This is contrast to expectation value derivatives, which are expressed as commutators between the observable and a generator.
It would be interesting to understand the role of the anticommutator in this scenario from a physical perspective.

\subsubsection{Vertical changes of quantum states}
Like for unitary gates, a symmetry group $S<\mc U$ defines a tangent subspace of $\mc H$.
Given a basis $\mc B_{\mf t}$ for $\mf t$, the tangent space of $S$ at $\mbb I$, the vertical directions at $\ket{\psi}$ are given by
\begin{align}
    V_{\psi}\mc H = \mf t \ket{\psi}= \Span_\R\DE{x\ket{\psi}|x\in\mc B_{\mf t}},
\end{align}
and we may characterize the covariant subspace using the inner product $\dE{\,\cdot\,|\,\cdot\,}$, i.e.,
\begin{align}
    H_{\psi}\mc H
    = \DE{x\ket{\psi}\in T_\psi\mc H|\dE{\psi|x^\dagger y|\psi}=0\,\forall y\in\mc B_{\mf t}}.
\end{align}
Similarly to $\mc B\ket{\psi}$, the set of directions $\mc B_{\mf t} \ket{\psi}$ may be redundant to span $V_{\psi}\mc H$, and the latter might have fewer dimensions than $\mf t$ if some generators $x$ create linearly dependent tangent vectors.
However, for sufficiently small symmetry groups there might be no redundancy at all and we \emph{can} have $\dim V_{\psi}\mc H=|\mc B_{\mf t}|$.
An example for this would be the tensor power representation of $\mc{SU}(2)$, together with a suitable choice of the point $\ket{\psi}$.
In this case, $|\mc B_{\mf t}|=3<2^{N+1}-1$ for any number of qubits $N$, so that the set $\mc B_{\mf t}\ket{\psi}$ need not have redundancies.
Also see \cref{sec:equi_and_co:tangent_decomp_U,sec:equi_and_co:tangent_decomp_H} for more details on this example.

Note that the degree of redundancy will depend not only on $S$, but also on the specific state $\ket{\psi}$, so that moving through state space can change $\dim V_\psi\mc H$.
Intuitively, this means that the decomposition of $T_\psi\mc H$ is not compatible with moving through state space by left translation with unitaries if we define the vertical space at each point in $\mc H$ locally.
Such a local definition of the tangent space decomposition results from the symmetry action $\act{L}$ in \cref{eq:state_action_left}.
Instead, we may fix a point $\ket{\psi_0}$ at which the above definition holds, and \emph{define} the vertical subspace at other points via left translation:
\begin{align}
    V_{U\ket{\psi_0}}\mc H\coloneqq U V_\psi \mc H.
\end{align}
This corresponds to the symmetry action $\act{\btheta}$ in \cref{eq:state_action_right}.

To quantify the redundancy of the set $\mc B_{\mf t}\ket{\psi}$, we introduce its Gram matrix with respect to $\dE{\,\cdot\,|\,\cdot\,}$:
\begin{align}
    G_{ab}(\psi)=G_{ab} 
    &= \dE{\psi | x_a^\dagger x_b |\psi}= -\frac12 \bra{\psi} \DE{x_a, x_b}\ket{\psi},
\end{align}
where we enumerated the basis elements $\DE{x_a}_a$ and made use of \cref{eq:real_to_anticom} to arrive at the second expression.
Gram matrices are positive semi-definite and vanishing eigenvalues indicate linear dependencies between the vectors $\mc B_{\mf t}\ket{\psi}$.
Correspondingly, the rank of $G$ equals the dimension of $V_{\psi}\mc H$.
Accordingly, we may use $G$ to create an ONB of $V_{\psi}\mc H$ with respect to $\dE{\,\cdot\,|\,\cdot\,}$.
We define
\begin{align}
    \ket{T_a(\psi)}=\ket{T_a} = \de{\sqrt{G^+}}_{ab}x_b\ket{\psi}
    \quad \de{\ \Rightarrow \dE{T_a|T_b}=\delta_{ab},\ a\leq \mathrm{rank}\ G\ },
\end{align}
where $G^+$ denotes the pseudoinverse\footnote{Due to $G$ being positive semi-definite, $\sqrt{G^+}$ can be defined on the level of eigenvalues: zero is mapped to zero and all other values are mapped to the unique positive root of their inverse.} of $G$ and we make use of the Einstein convention, i.e.,~we imply summation over the repeated index $b$.
On the level of operators, this defines generators $t_a=\de{\sqrt{G^+}}_{ab}x_b$ that are orthonormal under $(t_a, t_b)\mapsto \dE{\psi|t_a^\dagger t_b|\psi}=\delta_{ab}$ for $a\leq\mathrm{rank}\ G$.

Note that we might obtain one or more vanishing vectors $\ket{X_a}$ (operators $t_a$), which we implicitly exclude from the basis in the following.
The ONB allows us to write out the projectors onto the vertical and covariant subspaces as
\begin{align}\label{eq:projectors_states}
    \mbb V_{\psi} = \ket{T_a}\dE{T_a|\:\cdot\:}, \quad \mbb H_\psi=\mbb I-\ket{T_a}\dE{T_a|\:\cdot\:}.
\end{align}

\subsubsection{The covariant derivative on \texorpdfstring{$\mc H$}{H}}\label{sec:circuitry_app:states:cov_deriv}
We now turn to the covariant derivative of quantum states that are prepared by PQCs as introduced in \cref{sec:preliminaries:pqc}.
As discussed before, there are multiple actions of the symmetry group on such states, and each will define a covariant derivative.
Here we focus on the action $\act{\btheta}$ that applies the symmetry group element at a fixed reference state $\ket{\psi_0}$.
At a given state $\ket{\psi(\btheta)}=U(\btheta)\ket{\psi_0}$, we write the tangent space as\footnote{The space $T_{\psi(\btheta)}\mc H$ itself does not depend on the symmetry action. However, we rewrite its elements for easy comparison with the subspaces below.}
\begin{align}
    T_{\psi(\btheta)}\mc H
    =\Span_\R\DE{U(\btheta)x\ket{\psi_0} | x\in\mc B}
    = U(\btheta)\mf u\ket{\psi_0},
\end{align}
which is split into a vertical and a covariant subspace via
\begin{alignat}{3}\label{eq:vert_hor_states}
    V_{\psi(\btheta)}\mc H
    &= U(\btheta)\mf t\ket{\psi_0}
    &&=&&\Span_\R\DE{U(\btheta)x_a\ket{\psi_0}|x_a\in \mc B_{\mf t}}\\
    H_{\psi(\btheta)}\mc H
    &= \orth{\de{U(\btheta)\mf t\ket{\psi_0}}}
    &&=&& \DE{U(\btheta)y\ket{\psi_0}|y\in\mf u, \dE{\psi_0 | x_a^\dagger y|\psi_0}=0\,\forall x_a\in\mc B_{\mf t}}.
\end{alignat}
It will be important that the orthogonal decomposition is not simply inherited from the decomposition of $\mf u$ into $\mf t$ and $\orth{\mf t}$.
A notable feature of this symmetry action is that the induced decomposition of $T_{\psi(\btheta)}\mc H$ is compatible with unitary transformations.
Moving from $\ket{\psi(\btheta)}$ to $\ket{\psi'}=W\ket{\psi(\btheta)}$ simply moves the vertical and covariant spaces to $WV_{\psi(\btheta)}\mc H$ and $WH_{\psi(\btheta)}\mc H$.
As a consequence, the corresponding Gram matrix
\begin{align}\label{eq:circuitry_app:gram_right_action}
    G_{ab} = \dE{\psi_0|x_a^\dagger x_b|\psi_0}
\end{align}
no longer depends on $\btheta$ but only on the reference state $\ket{\psi_0}$ of the symmetry action.
This will be a useful feature when computing covariant derivatives in practice.
It also fixes the dimension of the vertical and covariant subspaces as long as we keep $\ket{\psi_0}$ fixed.
The ONB of $V_{\psi(\btheta)}\mc H$ is given by
\begin{align}\label{eq:states:onb}
    \ket{T_a(\btheta)} = (\sqrt{G^+})_{ab} U(\btheta) x_b\ket{\psi_0} = U(\btheta) \ket{T_a},
\end{align}
i.e.,~it is the ONB defined at $\ket{\psi_0}$ by the constant Gram matrix and transported by left translation with $U(\btheta)$.

We finally can introduce the covariant derivative of parametrized quantum states.
As with the unitary group, we simply chain the partial derivative with $\mbb I-\mbb V=\mbb H$:
\begin{align}
    D_j \ket{\psi(\btheta)}
    &= \ket{\partial_j \psi(\btheta)} - \mbb V_{\psi(\btheta)} \ket{\partial_j \psi(\btheta)} \label{eq:states:cov_deriv0}\\
    &= \ket{\partial_j\psi(\btheta)} - \ket{T_a(\btheta)}\dE{T_a(\btheta)|\partial_j\psi(\btheta)}\label{eq:states:cov_deriv1}\\
    &= \ket{\partial_j\psi(\btheta)} \label{eq:states:cov_deriv2}- U(\btheta) x_a\ket{\psi_0} (G^+)_{ab} \dE{\psi_0|x_b^\dagger \Omega_j(\btheta)|\psi_0}.
\end{align}
Here we filled in the effective generator $\Omega_j(\btheta)$ of the parametrized unitary from \cref{sec:circuitry_app:gates}, i.e.,~$\Omega_j(\btheta)=U^\dagger(\btheta)\partial_j U(\btheta)$.

\subsubsection{Example: \texorpdfstring{$\mc U(1)<\mc U(2)$}{U(1)<U(2)} and \texorpdfstring{$\mc U(1)\triangleleft \mc U(2)$}{U(1)<|U(2)}}
As a first example, consider the circuit from \cref{eq:example_circuit_gates} applied to the initial state $\ket{+}$, specifically $\ket{\psi(\btheta)}=R_X(\theta_2)R_Y(\theta_1)\ket{+}$.
Recall its effective generators $\Omega_1(\btheta)=-\frac{i}{2}Y$ and $\Omega_2(\btheta)=-\frac{i}{2}(\cos(\theta_1)X+\sin(\theta_1)Z)$.
In addition, we consider two versions of the symmetry group $\mc U(1)$:
\begin{alignat}{3}
    S_1(S)
    &= \DE{\exp(i\alpha Z)|\alpha\in[0,2\pi)}
    &&\quad \Rightarrow\quad \DE{x_a}_a
    &&=\DE{iZ}
    \\
    S_2(S)
    &= \DE{e^{i\varphi}|\varphi\in[0,2\pi)}
    &&\quad \Rightarrow\quad \DE{x_a}_a
    &&=\DE{i\mbb I}.
\end{alignat}
Here $S_1$ is the group we used in \cref{sec:circuitry_app:gates} and $S_2$ represents $\mc U(1)$ as the normal subgroup of $\mc U(2)$, namely the group of global phases.
The Gram matrix is $G=(1)$ for both versions, so that the covariant derivatives are
\begin{alignat}{3}
    D_1^{S_1}\ket{\psi(\btheta)}
    &= -\frac{1}{2}U(\btheta)\ket{-}
    &&\qquad D_2^{S_1}\ket{\psi(\btheta)} 
    &&= -\frac{i}{2}\cos(\theta_1)\ket{\psi(\btheta)} \\
    D_1^{S_2}\ket{\psi(\btheta)}
    &= -\frac{1}{2}U(\btheta)\ket{-}
    &&\qquad D_2^{S_2}\ket{\psi(\btheta)} 
    &&=-\frac{i\sin(\theta_1)}{2}U(\btheta)\ket{-}.
\end{alignat}
The covariant derivatives with respect to $\theta_2$ differ from the usual partial derivative.
For $S_1$, the change between $D_2^{S_1}\ket{\psi(\btheta)}$ and $\partial_2\ket{\psi(\btheta)}$ is consistent with that of the derivatives of $U(\btheta)$ in \cref{sec:circuitry_app:gates}, that is $D^{S_1}_2\ket{\psi(\btheta)}=\de{D_2U(\btheta)} \ket{+}$.
For $S_2$, there would not be any modification at the unitary level because $\tr{-i\mbb I\Omega_1(\btheta)}=0$, but we find $D_2^{S_2}\ket{\psi(\btheta)}$ to differ from $\partial_2\ket{\psi(\btheta)}$.
This is because the inner products on $T\mc U(d)$ and $T_{\ket{+}}\mc H$ are not equivalent.
The covariant derivatives with respect to $\theta_1$ do not differ from $\partial_1\ket{\psi(\btheta)}$, which is in line with the covariant derivative on the unitary level.
However, we note that this need not be the case: changing $\ket{\psi_0}$ from $\ket{+}$ to the eigenstate $\ket{i}$ of $Y$ leads to $D^{S_2}_1\ket{\psi(\btheta)}=0$, because the $R_Y$ gate acts like a global phase gate in this case.

\subsection{PQC-based cost functions}\label{sec:circuitry_app:cost_functions}

In this section we apply the covariant derivative to cost functions based on parametrized quantum circuits.
Such cost functions are used for quantum machine learning tasks and variational quantum algorithms, for example.
More specifically, applications in which symmetry actions on the cost function are of interest include classification tasks with data symmetry and the variational quantum eigensolver for Hamiltonians with symmetries.
We will discuss the covariant derivative for those two applications in the following.

We start with the task of classifying data points subject to some equivalence relation that is based on a symmetry group as also discussed in \cref{sec:preliminaries:symmetries}.
Assume that this task is addressed with an encoding unitary $W$ applied to an initial state $\ket{\phi}$, followed by a sequence of trainable gates and the measurement of some observable $M$, leading to the cost function
\begin{align}
    C_{\vec{x}}(\btheta) &= \bra{\psi_{\vec{x}}(\btheta)} M \ket{\psi_{\vec{x}}(\btheta)},\\
    \ket{\psi_{\vec{x}}(\btheta)} &= U(\btheta)\ket{\psi_0(\vec{x})}=\prod_{j=p}^1 U_j(\theta_j)\ \ W(\vec{x})\ket{\phi}.
\end{align}
At this point we do not put any constraints on the unitary $U(\btheta)$ or the Hermitian measurement observable $M$.
We consider each gate $U_j$ to depend on a scalar parameter $\theta_j$ but the setting can be generalized to multi-parameter gates.
For convenience we do not denote the dependence on $\vec{x}$ when it is clear from the context.

We further assume that $W$ be equivariant\footnote{See \cref{sec:equi_and_co:equivariance_groups} for details on the notion of equivariance.} w.r.t.~$S$ and that $\ket{\phi}$ be invariant under $S$\footnote{Technically, we may just assume that the associated density matrix $\ket{\phi}\!\!\bra{\phi}$ is invariant.}.
Under these assumptions, which capture a common setup, encodings of two data points $\vec{x}$, $\vec{x'}$ related by a symmetry transformation $s\in S$ satisfy $W(\vec{x'})\ket{0}=sW(\vec{x})s^\dagger\ket{0}$.
From this, we already see that the symmetry action $\act{\btheta}$, which inserts $s\in S$ between the parametrized unitary $U(\btheta)$ and the encoded state vector $\ket{\psi_0(\vec{x})}$, is a suitable choice.
Accordingly, we will work with the (right-)effective generators $\Omega_{j}(\btheta)=U^\dagger(\btheta)\partial_j U(\btheta)$ (see \cref{sec:circuitry_app:gates}).

The covariant derivative of $C$ with respect to the parameter $\theta_j$ is then given by
\begin{align}
    D_j C(\btheta)
    &= 2\real{\bra{\psi(\btheta)}M D_j\ket{\psi(\btheta)}}\\
    &= \partial_j C(\btheta) - 2\dE{\psi(\btheta)|M|T_a}\dE{T_a|\partial_j\psi(\btheta)}.
\end{align}
For practical purposes, we will make use of the Gram matrix and the basis elements $\mc B_{\mf t}=\{x_a\}$ from the previous section:
\begin{align}\label{eq:cov_deriv_coms0}
    D_j C(\btheta)
    &= \partial_j C(\btheta)- 2\dE{\psi_0| \widetilde{M}(\btheta) x_a |\psi_0}(G^+)_{ab}\dE{\psi_0|x_b^\dagger \Omega_j(\btheta) |\psi_0}\\
    &= \partial_j C(\btheta)+\frac12 \bra{\psi_0}[\widetilde{M}(\btheta),x_a]\ket{\psi_0}\label{eq:cov_deriv_coms1} (G^+)_{ab}\bra{\psi_0}\DE{x_b, \Omega_j(\btheta)}\ket{\psi_0}\\
    &\eqqcolon\partial_j C(\btheta) + m_a\ (G^+)_{ab}\ \omega_{bj},
\end{align}
where we denoted $\widetilde{M}(\btheta)=U^\dagger(\btheta) M U(\btheta)$ and used the Gram matrix entries from \cref{eq:circuitry_app:gram_right_action}.
In the last line we abbreviated the expectation values of the (anti)commutators with $\widetilde{M}$ ($\frac12 \Omega_j$) as $m_a$ ($\omega_{b, j}$).

In the second scenario, we do not have an encoding layer but a fixed initial state $\ket{\psi_0}$.
The symmetry instead enters via the observable $M$, which usually encodes the problem we want to solve.
More specifically, we consider symmetries to be operators that commute with $M$, so that 
\begin{align}
    C(\btheta)
    =\bra{\psi(\btheta)}M\ket{\psi(\btheta)}
    = \bra{\psi(\btheta)}s^\dagger sM\ket{\psi(\btheta)}
    = \bra{\psi(\btheta)}s^\dagger M s\ket{\psi(\btheta)}
    = \bra{\psi'(\btheta)}M\ket{\psi'(\btheta)}
    =C'(\btheta),
\end{align}
i.e.,~the cost function is invariant under the symmetry acting on $\ket{\psi(\btheta)}$ via multiplication from the left.
This is exactly the symmetry action $\act{L}$ in \cref{eq:state_action_left}.
The covariant derivative for this action becomes
\begin{align}
    D_j C(\btheta)
    &= \partial_j C(\btheta)- 2\dE{\psi(\btheta)| M x_a |\psi(\btheta)}(G^+)_{ab}\dE{\psi(\btheta)|x_b^\dagger \Omega^L_j(\btheta) |\psi(\btheta)}\\
    &= \partial_j C(\btheta)+\frac12 \bra{\psi(\btheta)}[M,x_b]\ket{\psi(\btheta)}(G^+)_{ab}\bra{\psi(\btheta)}\DE{x_b, \Omega_j(\btheta)}\ket{\psi(\btheta)}.
\end{align}
It differs from the expression in \cref{eq:cov_deriv_coms0} by moving from $\ket{\psi_0}$ to $\ket{\psi(\btheta)}$, replacing $\widetilde{M}(\btheta)$ by $M$, and exchanging $\Omega_j(\btheta)$ for the \emph{left}-effective generator $\Omega^L_j(\btheta)$.
In particular, the used Gram matrix becomes $G_{ab} = \dE{\psi(\btheta)|x^\dagger_a x_b|\psi(\btheta)}$.

We discuss how to compute the quantities in the covariant derivatives for these two symmetry actions in \cref{sec:circuitry:cost_functions}.

\section{Useful facts about Lie groups and Lie algebras}\label{sec:lie_lemmas}

In this section we collect a few small lemmas used throughout the main text, as well as a detailed description of the Lie algebra decomposition for \cref{sec:equi_and_co:connect} in \cref{sec:algebra_4_split_derivation}.
We tailor these statements to our needs but they can otherwise be seen as common knowledge in Lie theory or geometric topology.

First, we simplify our lives by clarifying the following:

\begin{lemma}[Commutativity with algebra and group coincide]\label{lemma:group_and_algebra_commutants}
    Consider a connected group $S$, represented as $\sigma(S)<\mc U$, and its Lie algebra $\mf s$, represented as $\mf t= \sigma(\mf s)$.
    The commutant algebra $\mf u^{\mf t}$ of unitary algebra elements that commute with $\mf t$ is equal to the commutant algebra $\mf u^{\sigma(S)}$ of unitary algebra elements that commute with $\sigma(S)$.
\end{lemma}

\emph{Proof.} For any $x\in\mf u^{\mf t}$ and $s\in S$, we have 
\begin{align}
    \sigma(s) x \sigma(s)^\dagger
    = \exp(t \sigma(y)) x \exp(-t \sigma(y))
    = \exp(t\,\mathrm{ad}_{\sigma(y)})(x)
    = x \ \Rightarrow\ x\in \mf u^{\sigma(S)}.
\end{align}
Here we first used that $\sigma(s)=\exp(t\sigma(y))$ for some $y\in \mf s$ because $\sigma(S)$ is connected, and then $\mathrm{Ad}_{\exp(x)}=\exp(\mathrm{ad}_x)$.
Conversely, take $x\in \mf u^{\sigma(S)}$ and $y\in\mf t$, so that
\begin{align}
    [y, x]
    = \frac{\mathrm d}{\mathrm d t} \exp(ty) x \exp(-ty) \big|_{t=0}
    = \frac{\mathrm d}{\mathrm d t} x \big|_{t=0}
    = 0 \ \Rightarrow\ x\in\mf u^{\mf t}.
\end{align}
Here we simply used that $\exp(t y)\in \sigma(S)$. \hfill $\square$

We therefore do not have to specify whether we mean $\mf u^{\sigma(S)}$ or $\mf u^{\mf t}$ when we talk about \emph{the} commutant algebra.
Next, we consider the relation between Lie group and Lie algebra commutants:

\begin{lemma}[Lie algebra of commutant is commutant algebra]\label{lemma:algebra_of_commutant}
    Consider a group $S$ represented as subgroup $\sigma(S)<\mc U$ and its Lie algebra $\mf s$, represented as $\mf t = \sigma(\mf s)$.
    The Lie algebra of the commutant group $\mc U^{\sigma(S)}$ is equal to the commutant algebra $\mf u^{\sigma(S)}$.
\end{lemma}

\emph{Proof.} The commutant group is given by $\mc U^{\sigma(S)}=\DE{U\in \mc U| \sigma(s) U\sigma(s)^\dagger=U\ \forall s \in S}$.
Accordingly, any element $x$ in the Lie algebra of $\mc U^{\sigma(S)}$ must satisfy $\sigma(s)\exp(tx)\sigma(s)^\dagger=\exp(tx)$ for any $t\in\R$.
This implies 
\begin{align}
    \sigma(s) x\sigma(s)^\dagger
    =\frac{\mathrm{d}}{\mathrm{d} t}\sigma(s)\exp(tx)\sigma(s)^\dagger\big|_{t=0}
    =\frac{\mathrm{d}}{\mathrm{d} t}\exp(tx)\big|_{t=0}=x,
\end{align}
so that $[\sigma(s), x]=0\ \forall s\in S$, i.e.,~$x\in \mf u^{\sigma(S)}$.
Conversely, consider $x\in\mf u^{\sigma(S)}$, so that $\mathrm{Ad}_{\sigma(s)}(x)=x$ for all $s\in S$.
Then, for all $s\in S, t\in\R$
\begin{align}
    \sigma(s) \exp(tx) \sigma(s)^\dagger
    = \sum_{n=0}^\infty \sigma(s)\frac{(tx)^n}{n!} \sigma(s)^\dagger
    = \sum_{n=0}^\infty \frac{(t \sigma(s) x\sigma(s)^\dagger )^n}{n!}
    = \exp(tx),
\end{align}
i.e.,~$x$ generates group elements that commute with $\sigma(S)$, and thus lies in the Lie algebra of $\mc U^{\sigma(S)}$. \hfill$\square$

The above lemma states that $T_{\mbb I} \mc U^{\sigma(S)}=\mf u^{\sigma(S)}$.
If $S$ is connected, \cref{lemma:group_and_algebra_commutants} additionally implies
\begin{align}
    T_{\mbb I} \mc U^{\sigma(S)}=\mf u^{\sigma(S)} = \mf u^{\mf t}.
\end{align}
\cref{lemma:algebra_of_commutant} asserts that the local approach we take in \cref{sec:equi_and_co:equivariance_algebras} is compatible with the notion of equivariant circuits.
It also is the reason why equivariant generators from $\mf u^{\sigma(S)}$, created via twirling for example, lead to equivariant circuits within $\mc U^{\sigma(S)}$.

Finally, we state a small lemma about the induced decomposition discussed in \cref{sec:induced_split} for the case of a free group action.
A more general version can be found in~\cite[Lemma 3.1.14]{ziller2013group}.
\begin{lemma}[Free group actions have an injective differential w.r.t. the group argument]\label{lemma:free_action}
    Consider a free action $\act{}$ of the unitary group $\mc U$ on a manifold $\mc M$.
    Then the map
    \begin{align}
        \act{}p: \mf u \to T_p\mc M: x\mapsto \frac{\mathrm{d}}{\mathrm{d}t}(\exp(xt)\act{}p)\big|_{t=0}
    \end{align}
    is injective.
\end{lemma}
\emph{Proof.}
Assume the converse, so that there are $x,y\in\mf u$ with $x\neq y$ but $x\act{}p=y\act{}p$.
Then
\begin{align}
    0 = x\act{}p - y\act{}p = (x - y)\act{}p = \frac{\mathrm{d}}{\mathrm{d}t}(\exp((x-y)t)\act{}p)\big|_{t=0},
\end{align}
where we used that $\act{}p$ is linear.
Flows on $\mc M$ are identified uniquely by their starting point and the derivative at that point.
As the flow $p(t)=p\ \forall t$ satisfies has the starting point $p$ and the same derivative as $\exp((x-y)t)\act{}p$ at $t=0$, we find that $\exp((x-y)t)$ stabilizes $p$, which is a contradiction to $\act{}$ being a free group action.
\hfill $\square$

A consequence of this lemma is that the map $\act{}p$ is bijective on its image in $T_p\mc M$.
In particular, the decomposition of $\mf u$ described in \cref{sec:induced_split} that is induced by the decomposition $T_p\mc M=V_p\mc M\oplus H_p\mc M$ in \cref{def:equi_and_co:vertical_tangents} matches the generic split into $\mf t$ and $\orth{\mf t}$.
In addition $\mf t^0=\DE{0}$.

\subsection{Details on Lie algebra decomposition}\label{sec:algebra_4_split_derivation}
Here we derive the decomposition of the unitary Lie algebra into three subalgebras and a remainder vector subspace. 
We start with a small lemma:
\begin{lemma}[The commutant of an algebra is an algebra]\label{lemma:commutant_algebra}
    Consider a Lie algebra $\mf g\subset\mf u$ and a subalgebra $\mf h\subseteq\mf g$. The commutant of $\mf h$ in $\mf g$, $\mf g^{\mf h}=\DE{x\in\mf g|[x,z]=0\ \forall z\in\mf h}$ is a subalgebra of $\mf g$.
\end{lemma}
\emph{Proof.} Due to bilinearity of the commutator, we have
\begin{align}
    [\alpha x+y, z]=0\ \forall x, y\in\mf g^{\mf h}, \alpha\in\R, z\in\mf h,
\end{align}
making $\mf g^{\mf h}$ a vector subspace of $\mf g$.
In addition, $\mf g^{\mf h}$ is closed under the Lie bracket because
\begin{align}\label{eq:commutant_is_closed}
    \forall x,y\in\mf g^{\mf h},z\in\mf h: [[x,y], z] = [x, [y,z]]-[y, [x,z]] = 0\ \Rightarrow\ [x,y]\in\mf g^{\mf h},
\end{align}
where we used the Jacobi identity. \hfill $\square$

We apply this lemma to two algebra pairs:
first, to $\mf u^{\mf t}\subset \mf u$ where $\mf t$ is a represented symmetry algebra, $\mf t = \sigma(\mf s)$.
Second, to the center $\mf z(\mf s)=\mf s^{\mf s}$ of $\mf s$, which is the commutant of $\mf s$ within itself. 
Both $\mf u^{\mf t}$ and $\mf z(\mf s)$ are subalgebras due to \cref{lemma:commutant_algebra}.
We remark that $\mf z(\mf s)$ and an algebra representation of $\mf s$ interact nicely:
\begin{align}\label{eq:centers_coincide}
    \de{\sigma([x, y]) = 0 \Leftrightarrow [\sigma(x), \sigma(y)] = 0}\quad \Rightarrow\quad \sigma(\mf z(\mf s)) = \mf z(\sigma(\mf s))=\mf z(\mf t),
\end{align}
where we only used the fact that $\sigma$ is an algebra homomorphism, i.e.,~it preserves the Lie bracket.
In the following, we will implicitly make use of this identity.
We will also make use of the fact that $\mf u^{\mf t} \cap \mf t=\mf z(\mf t)$ by definition, and that $\mf z(\mf t)\subseteq\mf z\de{\mf u^{\mf t}}$.

Next, we consider the trace inner product $(x, y)\mapsto \frac{1}{d}\tr{x^\dagger y}$ on $\mf u(d)$.
Any vector subspace $\mf h\subset\mf g$ of an algebra $\mf g\subset\mf u(d)$ has a well-defined orthogonal complement within $\mf g$,
\begin{align}
    \orth{\mf h} = \DE{x\in \mf g\,\big|\,\frac{1}{d}\tr{z^\dagger x}=0\ \forall z\in \mf h},
\end{align}
which is a vector space due to linearity of the trace.
The following lemma is concerned with the special case where $\mf h$ in addition lies within the center of $\mf g$.
\begin{lemma}[Trace inner product and centers]\label{lemma:tip_centers}
    Consider a unitary algebra $\mf g$ and a subspace $\mf h\subset\mf z(\mf g)$ of its center.
    The orthogonal complement $\orth{\mf h}$ within $\mf g$ with respect to the trace inner product is an algebra.
\end{lemma}
\emph{Proof.} As mentioned above, the orthogonal complement $\orth{\mf h}$ is a vector subspace of $\mf g$.
Furthermore, it is closed under the Lie bracket because $\mf g$ is and because
\begin{align}
    \forall x, y\in\orth{\mf h}, z\in\mf h:\quad
    \tr{z^\dagger[x, y]}
    =\tr{[z^\dagger,x] y}
    =0
    \ \Rightarrow\ [x,y]\in \orth{\mf h}.
\end{align}
Here we used the cyclicity of the trace and the commutativity condition between $z$ and $x$ due to $z\in \mf h\subset \mf z(\mf g)$.
This shows that $\orth{\mf h}$ is an algebra, and a subalgebra of $\mf g$ in particular.\hfill $\square$

We apply this lemma to two cases:
first, let $\mf g=\mf u ^{\mf t}$ for a represented algebra $\mf t=\sigma(\mf s)$ as before, and take $\mf h =\mf z(\mf t)$.
As remarked above, $\mf z(\mf t)\subset \mf z(\mf u^{\mf t})=\mf z(\mf g)$.
The lemma thus tells us that the space $\utcless\coloneqq \orth{\mf z (\mf t)}$ is an algebra, where the orthogonal complement is taken within $\mf g=\mf u^{\mf t}$.
Second, let $\mf g = \mf t$ and take the same subspace as before, $\mf h =\mf z(\mf t)$.
We find that $\tcless\coloneqq \mf z(\mf t)$ is an algebra, where now the orthogonal complement is taken within $\mf g=\mf t$.

Overall, we showed with \cref{lemma:commutant_algebra,lemma:tip_centers} that $\mf z(\mf t)$, $\utcless$ and $\tcless$ are subalgebras.
In addition, $\utcless$ and $\tcless$ are orthogonal to $\mf z(\mf t)$ by construction.
It remains to check that $\utcless$ and $\tcless$ are orthogonal.
\begin{lemma}
    If $S$ is compact, the orthogonal complement $\tcless$ of $\mf z(\mf t)$ within $\mf t$ and the orthogonal complement $\utcless$ of $\mf z(\mf t)$ within $\mf u^{\mf t}$ are mutually orthogonal.
\end{lemma}
\emph{Proof.} 
$S$ being compact implies that $\sigma(S)$ is compact, so that its Lie algebra $\mf t$ is reductive~\cite[Cor.4.25]{knapp2002lie}.
Concretely, this means that $\mf t$ decomposes into a direct sum between its center and a semisimple algebra.
As we performed exactly this decomposition manually above, we find that $\tcless$ is semisimple, and thus $\tcless=[\tcless,\tcless]$.
That is, we can write any element $x\in\tcless$ as the commutator of two other elements $x_1, x_2\in\tcless$.
Now, consider $y\in \utcless$ and $x\in\tcless$.
Then
\begin{align}
    \tr{x^\dagger y} = - \tr{[x_1, x_2] y} = - \tr{x_1 [x_2, y]} = 0,
\end{align}
where we first used $x^\dagger=-x$ and wrote $x=[x_1, x_2]$.
Then we applied the cyclicity of the trace together with $[x_2, y]=0$, which holds because $x_2\in\tcless\subset\mf t$ and $y\in\utcless\subset\mf u^{\mf t}$, similar to the proof of \cref{lemma:tip_centers}.\hfill $\square$

We thus decomposed $\mf u^{\mf t} \cup \mf t$ into three mutually orthogonal subalgebras.
Denoting the orthogonal complement of $\mf u^{\mf t} \cup \mf t$ in $\mf u$ as $\mf r$, we complete the decomposition as
\begin{align}
     \mf u = \mf r \oplus \utcless\oplus \mf z(\mf t)\oplus \tcless,
\end{align}
with three Lie subalgebras of $\mf u$ and the remainder vector space $\mf r$.
The latter is not an algebra in general, as we will also see in examples in the main text.

\section{Subtleties of covariance, equivariance, and invariance}\label{sec:subtleties_variances}
In this section, we discuss two perspectives covariance, equivariance, and invariance that showcase sources for confusion between these concepts and possible reasons for the inconsistent use of these names between research communities.
We begin with a matrix-level condition that turns out to imply invariance and no invariance, depending on the context.
Then we discuss why the covariant derivative may be seen as an equivariant map.

\subsection{Invariance or just equivariance? Context matters}\label{sec:invariance_or_not}

\begin{figure}
    \centering
    \includegraphics[width=0.95\textwidth]{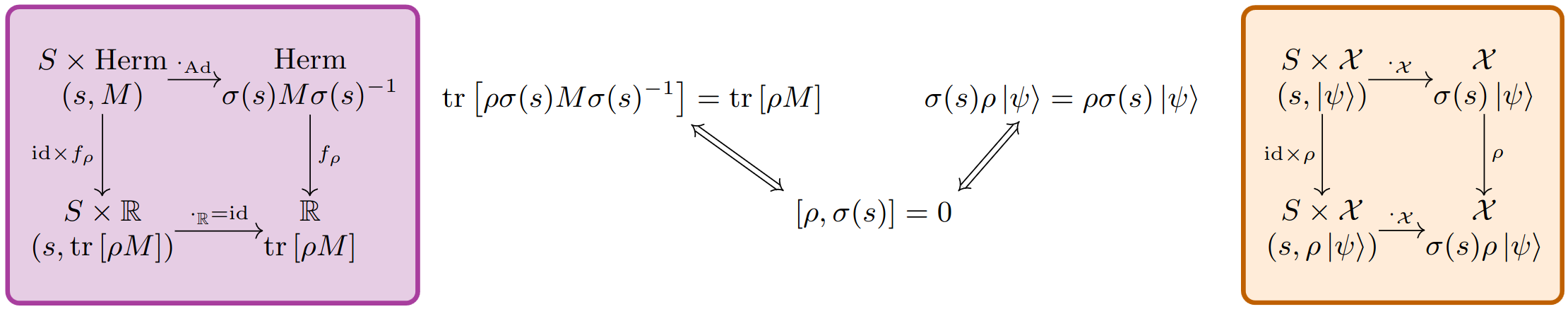}
    \caption{\textbf{A density matrix $\rho$ as an equivariant map.} (\emph{left}) $\rho$ as a map from the space of observables $\mathrm{Herm}$ to $\R$, using the trace inner product.
    The commuting diagram implies that $\rho$ is invariant with respect to the adjoint symmetry action $\act{\mathrm{Ad}}$ on $\mathrm{Herm}$ and the trivial representation on $\R$.
    The invariance is a special case of equivariance and implies $[\rho, \sigma(s)]=0$.
    (\emph{right}) $\rho$ interpreted as a map from $\mc X=\C^d$ to itself.
    The commuting diagram leads to the same condition $[\rho, \sigma(s)]=0$ on $\rho$, but only implies equivariance, not invariance, with respect to the left action $\act{\mc X}$ on $\mc X$.
    }
    \label{fig:equivariance_rho}
\end{figure}
Here we discuss how different interpretations of the same physical object, together with different symmetry actions, affect whether or not the object is invariant.
As an example, consider a density matrix $\rho$ that maps observables to real numbers via the trace inner product:
\begin{align}
    f_\rho : \mathrm{Herm} \to \R : M \mapsto \tr{\rho M}.
\end{align}
Choose the adjoint action associated to a matrix representation $\sigma$ on the observables and the trivial representation on $\mbb R$.
$f_\rho$ being invariant implies
\begin{align}
    \tr{\rho (\sigma(s) M \sigma(s)^{-1})}
    &= f_\rho(s\act{\mathrm{Ad}} M)
    = s\act{\mbb R} f_\rho(M)
    = \tr{\rho M}\quad\forall s\in S, M\in \mathrm{Herm}\nonumber\\
    \Rightarrow 
    [\rho, \sigma(s)]&=0\ \forall s\in S\nonumber.
\end{align}
If we interpret $\rho$ as a map acting on $\mc X=\mbb C^d$ instead\footnote{Not on $\mc H$, because $\rho$ need not map to a valid quantum state.}, together with the a representation $\act{\mc X}$ of $S$ on that space, the matrix relation $[\rho, \sigma(s)]=0$ leads to 
\begin{align}
    \rho(s\act{\mc X}\ket{\psi})= \rho \sigma(s)\ket{\psi}=\sigma(s)\rho \ket{\psi} =s\act{\mc X}\rho(\ket{\psi}),
\end{align}
which implies equivariance, but not invariance, of $\rho$ with respect to $\sigma$ as representation on $\mc X=\mc Y$.
This highlights the importance of the group actions/representations for statements about equivariance or invariance.

Finally, note that a PQC together with an observable is commonly called \emph{equivariant} if the resulting cost function is \emph{invariant} with respect to one of two symmetry actions:
for data-based QML applications, the symmetry is motivated by the input data and the invariance condition reads $C(\btheta,s\act{\text{data}}\vec{x})=C(\btheta, \vec{x})$.
This is usually guaranteed by making the data encoding and parameterized unitaries equivariant, and by using initial states and observables that commute with the symmetry representation on quantum states.

\subsection{The covariant derivative as an equivariant map}\label{sec:covariant_derivative_is_equivariant}
Throughout this work, we stressed that equivariance and covariance as defined in \cref{sec:equi_and_co} are different concepts.
There is a good reason for using the notion of equivariance for the covariant derivative, though.
We will discuss this statement for the second example in \cref{sec:equi_and_co:covariance_algebras}, i.e.,~the manifold $\mc M=\mc H$, the symmetry action $\act{\mc H}$ associated to a representation $\sigma$, and $f(\btheta)=\ket{\psi(\btheta)}$.
This is just for illustration purposes but holds more generally.
The covariant derivative $D_j$ maps the function $\ket{\psi}:\R^p\to \mc H$ to a new function
\begin{align}
    D_j \ket{\psi}:\R^p \to T\mc H : \btheta \mapsto \mbb H_{\psi(\btheta)} \partial_j \ket{\psi(\btheta)}.
\end{align}
That is, it takes a parametrized quantum state and returns a parametrized state tangent.
$D_j$ is a linear map because $\mbb H$ and $\partial_j$ both are linear.

In contrast to the other parts of the manuscript, we will here denote the Lie algebra representation as $\mathrm{d}\sigma$, as it is the differential of $\sigma$ after all.
Similary, we denote the action of the algebra on $\mc H$ induced by $\act{\mc H}$ with its own symbol $\odot_{\mc H}$.
Finally, $\act{\mc H}$ lets us define an action of $S$ on the tangent spaces $T\mc H$, which is denoted as $\act{T\mc H}$.
It is the differential of $\act{\mc H}$ if we fix the acting group element. 
In short, we can act with $S$ on $\mc H$ via $\act{\mc H}$, with $\mf s$ on $\mc H$ via $\odot_{\mc H}$, and with $S$ on $T\mc H$ via $\act{T\mc H}$.
All these actions are expressed as matrix-vector products.

For a local symmetry transformation $s(\btheta)$, we compute
\begin{align}
    D_j (s(\btheta) \act{\mc H} \ket{\psi(\btheta)})
    &= \mbb H_{\psi(\btheta)} \partial_j \sigma(s(\btheta)) \ket{\psi(\btheta)}
    = \mbb H_{\psi(\btheta)} \de{\partial_j \sigma(s(\btheta))}\ket{\psi(\btheta)} + \mbb H_{\psi(\btheta)} \sigma(s(\btheta))\ket{\partial_j\psi(\btheta)}.
\end{align}
Here we first used the definition of $\act{\mc H}$ via $\sigma$ and the matrix-vector product and then the product rule.
For the first term, we use the definition of the differential $\mathrm{d}\sigma$ and that of $\odot_{\mc H}$ via $\mathrm{d}\sigma$ and the matrix-vector product:
\begin{align}
    \mbb H_{\psi(\btheta)} \de{\partial_j \sigma(s(\btheta))}\ket{\psi(\btheta)}
    = \mbb H_{\psi(\btheta)}  \mathrm{d}\sigma(\partial_j s(\btheta))\ket{\psi(\btheta)}
    = \mbb H_{\psi(\btheta)}\ \ \partial_j s(\btheta)\odot_{\mc H}\ket{\psi(\btheta)}.
\end{align}
Clearly, $\partial_j s(\btheta)$ is an element of the tangent space $T_{s(\btheta)}S=s(\btheta)\mf s$.
As the operator $\mbb H_{\psi(\btheta)}$ projects out symmetry tangents at $\ket{\psi(\btheta)}$, we find that the first term vanishes.

For the second term, we use the action on $T\mc H$ and that it commutes with the projector.
This is because it is a differential and thus linear.
We find
\begin{align}
    \mbb H_{\psi(\btheta)} \sigma(s(\btheta))\ket{\partial_j\psi(\btheta)}
    = \mbb H_{\psi(\btheta)}\ s(\btheta)\act{T\mc H}\ket{\partial_j\psi(\btheta)}
    = s(\btheta)\act{T\mc H}\mbb H_{\psi(\btheta)}\ket{\partial_j\psi(\btheta)}
    = s(\btheta)\act{T\mc H} D_j\ket{\psi(\btheta)}.
\end{align}
Combining all steps, we have 
\begin{align}
    D_j\de{s(\btheta)\act{\mc H}\ket{\psi(\btheta)}}=s(\btheta)\act{T\mc H} D_j\ket{\psi(\btheta)},
\end{align}
showing that $D_j$ is equivariant in the sense of \cref{def:equivariance} with respect to the group actions $\act{\mc H}$ and $\act{T\mc H}$.
As a commutative graph, we may write
\begin{center}
    \includegraphics[width=0.4\textwidth]{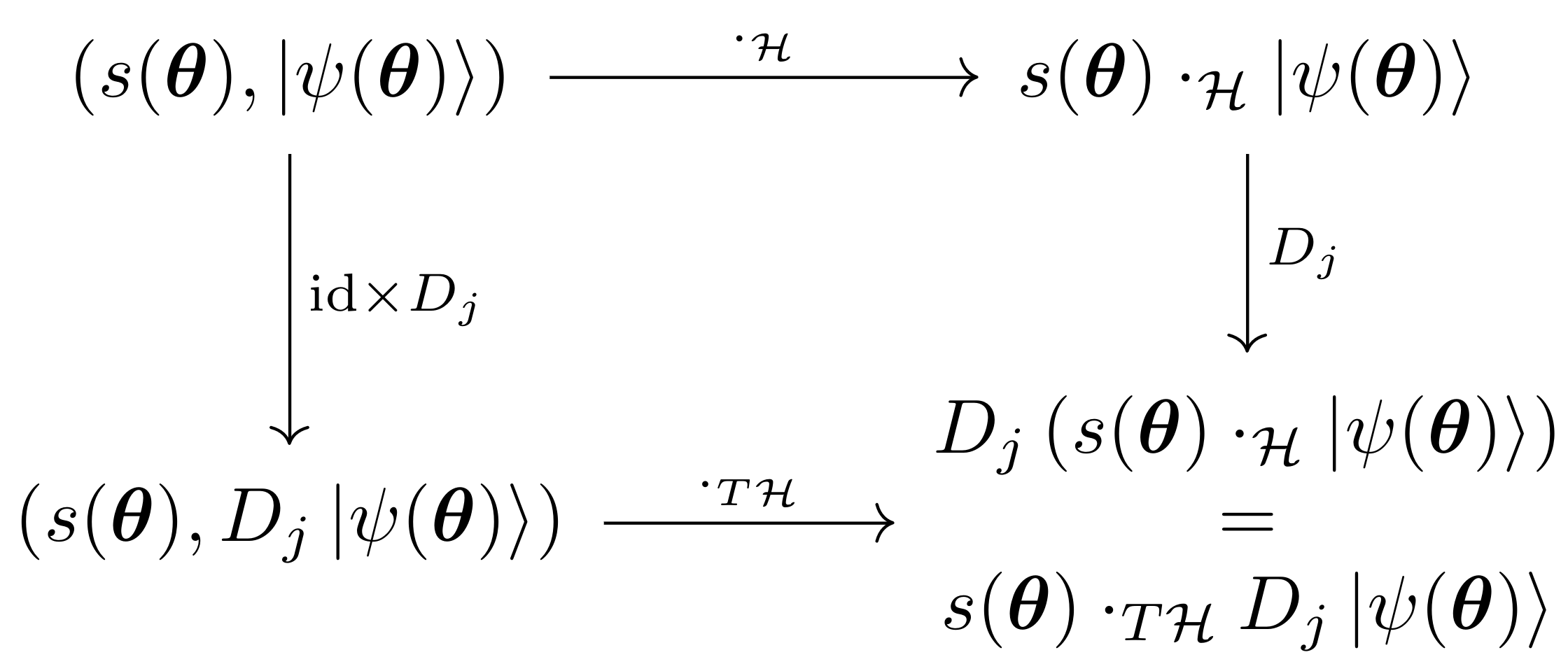}
\end{center}
%
%
This equivariance property is behind the statement that $D_j$ be covariant (by its transformation behaviour) in the context of physical gauge theory, offering a lot of potential for confusion.
We discuss this ``proof of covariance'' in more detail in \cref{sec:gauge_theory,sec:gauge_theory_comp}.

\section{Computations for examples}\label{app:example_calculations}
In this section we perform a few basic calculations for the examples in \cref{sec:circuitry}.

\subsection{Single-qubit example for quantum gates}\label{sec:ex_calc:1q_gate}
In \cref{sec:circuitry:gates} we look at the unitary $U(\btheta)=\exp(i\theta_3)R_X(\theta_2)R_Y(\theta_1)$.
We compute its partial derivatives and right-effective generators
\begin{alignat}{3}
    \partial_1U(\btheta) &= \exp(i\theta_3)R_X(\theta_2)R_Y(\theta_1)\frac{-i}{2}Y &&\Rightarrow \Omega_1 &&= -\frac{i}{2}Y\\
    \partial_2U(\btheta) &= \exp(i\theta_3)R_X(\theta_2)\frac{-i}{2}X R_Y(\theta_1) && && \\
    &=\exp(i\theta_3)R_X(\theta_2)R_Y(\theta_1) \de{-\frac{ic_1}{2}X-\frac{is_1}{2}Z} &&\Rightarrow \Omega_2(\btheta) &&= -\frac{i}{2}(c_1 X+s_1Z)\\
    \partial_3U(\btheta) &= \exp(i\theta_3)iR_X(\theta_2)R_Y(\theta_1) &&\Rightarrow \Omega_3 &&= i \mbb I,
\end{alignat}
where we abbreviated $c_1=\cos(\theta_1)$ and $s_1=\sin(\theta_1)$.
As we consider the action of the symmetry group $\sigma(S)=\DE{\exp(i\alpha Z)|\alpha\in[0, 2\pi)}$ on the unitary from the right, we need to project the right-effective generators onto the subspace $\mf t$ ($\mf u^{\mf t}$) for the covariant (equivariant) derivative.
The covariant derivative thus removes the $Z$ component in $\Omega_2(\btheta)$ and does not alter $\Omega_{1,3}$.
The equivariant derivative only keeps symmetry generators $iZ$ and global phases $i\mbb I$, reducing $\Omega_1$ to $0$ and $\Omega_2(\btheta)$ to the second component $-\frac{is_1}{2}Z$ while leaving $\Omega_3$ intact.

If we move from $\mf t =i\R Z$ to $\mf t=i\R\mbb I$, the generators $\Omega_{1, 2}(\btheta)$ are covariant already, but $\Omega_3$ is projected to $0$.
As the symmetry group in this case is that of global phases, all gates are equivariant with respect to $\sigma(S)$.
This implies that the partial derivatives are the equivariant derivatives already.

In a state space example below we will also make use of the left-effective generators
\begin{align}\label{eq:single_qubit_unitary_left_gen}
    \Omega_1^L(\btheta) = -\frac{i}{2}R_X(\theta_2)YR_X(-\theta_2) = -\frac{i}{2}(c_2Y+s_2 Z)\quad \Omega_2^L = -\frac{i}{2}X\quad \Omega_3^L=i\mbb I.
\end{align}

\subsection{Single-qubit example for quantum states}\label{sec:ex_calc:1q_state}
In \cref{sec:circuitry:states} we look at the unitary $U(\btheta)=\exp(i\theta_3)R_X(\theta_2)R_Y(\theta_1)$ from the previous section, applied to the initial state $\ket{+}$.
The partial derivatives are
\begin{align}
    \partial_j U(\btheta)\ket{+}
    &= U(\btheta)\Omega_j(\btheta)\ket{+}
    = \begin{cases}
        -\frac12 U(\btheta)\ket{-} & j=1\\
        -\frac12 U(\btheta)\de{c_1i\ket{+}+s_1i\ket{-}} & j=2\\
        U(\btheta)i\ket{+} & j=3
    \end{cases}\,.
\end{align}
We first consider the symmetry group $\mc U(1)<\mc U(2)$ represented as Pauli $Z$ rotations and the symmetry action $\act{\btheta}$ on the initial state.
The vertical subspace, which will be projected away by $D_j$, then is $U(\btheta)\mf t\ket{+}=\R U(\btheta) i\ket{-}$ and the equivariant subspace, comprising the directions preserved by $E_j$, is $U(\btheta)\mf u^{\mf t}\ket{+}=U(\btheta) \Span_{\R}\DE{i\ket{+}, i\ket{-}}$.
We find the derivatives
\begin{alignat}{7}
    D_1\ket{\psi(\btheta)} &= \partial_1\ket{\psi(\btheta)},
    \quad
    &&D_2\ket{\psi(\btheta)} &&=&& -\frac{ic_1}2 U(\btheta)\ket{+},
    \quad
    &&D_3\ket{\psi(\btheta)} &&=&& \partial_3\ket{\psi(\btheta)},\\
    E_1\ket{\psi(\btheta)} &= 0,
    \quad
    &&E_2\ket{\psi(\btheta)} &&=&& \partial_2\ket{\psi(\btheta)},
    \quad
    &&E_3\ket{\psi(\btheta)} &&=&& \partial_3\ket{\psi(\btheta)}.
\end{alignat}
If we switch to the symmetry action $\act{L}$, the left-effective generators of $U(\btheta)$, which we computed in \cref{eq:single_qubit_unitary_left_gen}, are useful:
\begin{align}
    \partial_j U(\btheta)\ket{+}
    &= \Omega_j^L(\btheta)U(\btheta)\ket{+}
    = \begin{cases}
        -\frac{i}{2} (c_2 Y + s_2 Z)\ket{\psi(\btheta)} & j=1\\
        -\frac{i}{2} X\ket{\psi(\btheta)} & j=2\\
        i\ket{\psi(\btheta)} & j=3
    \end{cases}\,.
\end{align}
The vertical and equivariant subspaces now are $\mf t\ket{\psi(\btheta)}$ and $\mf u^{\mf t}\ket{\psi(\btheta)}$, so that we require inner products of the form $\dE{\psi(\btheta)|A^\dagger B|\psi(\btheta)}$.
In particular, we compute
\begin{alignat}{4}
    \dE{\psi(\btheta)|(-iZ) iX|\psi(\btheta)} &= \real{i\braket{Y}_{\psi(\btheta)}}=0
    \quad
    &&\dE{\psi(\btheta)|(-i\mbb I) iX|\psi(\btheta)} &&=&& \real{\braket{X}_{\psi(\btheta)}}=c_1\\
    \dE{\psi(\btheta)|(-iZ) iY|\psi(\btheta)} &= \real{-i\braket{X}_{\psi(\btheta)}}=0
    \quad
    &&\dE{\psi(\btheta)|(-i\mbb I) iY|\psi(\btheta)} &&=&& \real{\braket{Y}_{\psi(\btheta)}}=s_1s_2\\
    \dE{\psi(\btheta)|(-iZ) iZ|\psi(\btheta)} &= \|\psi(\btheta)\|^2=1
    \quad
    &&\dE{\psi(\btheta)|(-i\mbb I) iZ|\psi(\btheta)} &&=&& \real{\braket{Z}_{\psi(\btheta)}}=-s_1c_2\\
    &   
    \quad
    &&\dE{\psi(\btheta)|(-i\mbb I) i\mbb I|\psi(\btheta)} &&=&& \|\psi(\btheta)\|^2=1.
\end{alignat}
In particular, we find the Gram matrix for the equivariant subspace
\begin{align}
    \tilde{G} = \de{\begin{array}{cc}
        1 & -s_1c_2 \\
        -s_1c_2 & 1
    \end{array}} \quad\Rightarrow\quad \tilde{G}^+ 
    = \de{\begin{array}{cc}
        (1-s_1c_2)^+ + (1+s_1c_2)^+ & (1-s_1c_2)^+ - (1+s_1c_2)^+ \\
        (1-s_1c_2)^+ - (1+s_1c_2)^+ & (1-s_1c_2)^+ + (1+s_1c_2)^+
    \end{array}},
\end{align}
where the pseudo-inverse of an matrix entry is the inverse if the entry is nonzero and zero otherwise.
We find that for $s_1c_2\in \DE{\pm 1}$ we indeed have a singular Gram matrix with reduced rank one.
This makes sense because for the corresponding rotation angles $\ket{\psi(\btheta)}$ produces a computational basis state, so that $i\mbb I$ and $iZ$ both are generators of global phases.

Using these inner products and the Gram matrix, we can compute the covariant and equivariant derivatives:
\begin{alignat}{4}
    D_1\ket{\psi(\btheta)} &= -\frac{ic_2}{2}Y\ket{\psi(\btheta)},
    \quad
    &&E_1\ket{\psi(\btheta)} &&=&& \tilde{G}^+\cdot \de{-\frac{i}{2}}\de{\begin{array}{c}
         Z\ket{\psi(\btheta)} \\
         s_1s_2c_2(Y-Z)\ket{\psi(\btheta)} 
    \end{array}}\\
    D_2\ket{\psi(\btheta)} &= \partial_2\ket{\psi(\btheta)},
    \quad
    &&E_2\ket{\psi(\btheta)} &&=&& \tilde{G}^+\cdot \de{\begin{array}{c}
         0 \\
         -\frac{ic_1}{2}X\ket{\psi(\btheta)} 
    \end{array}}\\
    D_3\ket{\psi(\btheta)} &= (i\mbb I+is_1c_2 Z)\ket{\psi(\btheta)},
    \quad
    &&E_3\ket{\psi(\btheta)} &&=&& \tilde{G}^+\cdot \de{\begin{array}{c}
         -is_1c_2\ket{\psi(\btheta)} \\
         i\ket{\psi(\btheta)} 
    \end{array}}.
\end{alignat}
It becomes clear that the symmetry action $\act{L}$ enhances the parameter dependence.
In particular, the global phase generator $i\mbb I$ is getting mixed with the symmetry generator $iZ$ because $U(\btheta)$ leads to finite overlaps with computational basis states, for which these generators are equivalent.

\subsection{Two-qubit example for quantum states}\label{sec:ex_calc:2q_state}
As a second example on the level of quantum states, we consider a two-qubit product state\footnote{We will not denote tensor products explicitly but imply it by writing factors acting on different qubits next to each other, usually in parentheses.} $\ket{\psi_0^{(1)}}\ket{\psi_0^{(2)}}$ to which we apply the circuit
\begin{align}
    U(\btheta)
    &=
    CR_X^{(2,1)}(\theta_3)R_Y^{(2)}(\theta_2)R_Y^{(1)}(\theta_1).
\end{align}
The symmetry group is $\mc{SU}(2)$ acting on the first qubit via $\act{L}$.
This setup will be used in \cref{sec:entangling_app}, with details provided below in \cref{sec:ex_calc:2q_cost}.
The partial derivatives read
\begin{align}
    \ket{\partial_j\psi(\btheta)}
    &= \de{\begin{array}{r}
        CR_X^{(2,1)}(\theta_3)R_Y^{(2)}(\theta_2)\de{-\frac{i}{2}Y^{(1)}}R_Y^{(1)}(\theta_1)\ket{\psi_0} \\
        CR_X^{(2,1)}(\theta_3)\de{-\frac{i}{2}Y^{(2)}}R_Y^{(2)}(\theta_2)R_Y^{(1)}(\theta_1)\ket{\psi_0} \\
        \de{-\frac{i}{2} X P_1}CR_X^{(2,1)}(\theta_3)R_Y^{(2)}(\theta_2)R_Y^{(1)}(\theta_1)\ket{\psi_0}
    \end{array}}
    = \de{\begin{array}{r}
        -\frac{i}2 Y^{(1)} CR_X^{(2,1)}(-2\theta_3)\ket{\psi(\btheta)} \\
        -\frac{i}2 Y^{(2)} R_{XZ}^{(1,2)}(\theta_3)\ket{\psi(\btheta)} \\
        \de{-\frac{i}{2} X P_1}\ket{\psi(\btheta)} 
    \end{array}},
\end{align}
where we denoted $P_1=\ket{1}\!\!\bra{1}$ and $R_{XZ}^{(1, 2)}(\theta_3)=\exp\de{-\frac{i}{2}\theta_3 X Z}$.
The vertical directions are $ix_a\ket{\psi(\btheta)}$ for $x_a\in\DE{X^{(1)}, Y^{(1)}, Z^{(1)}}$, which are orthonormal with respect to $\dE{\,\cdot\,|\,\cdot\,}$ already.
We find the vector potential
\begin{align}
    A_j^a(\btheta)
    = \dE{X_a|\partial_j\psi(\btheta)}
    = -\frac12 \dE{\psi(\btheta)\bigg|\de{\begin{array}{lll}
        0 & \tilde{c}_3 XY + \tilde{s}_3\mbb I X & \mbb I P_1\\
        \mbb I P_0+c_3 \mbb I P_1& \tilde{c}_3YY & 0 \\
        s_3\mbb I P_1 & \tilde{c}_3 ZY & 0
    \end{array}}\bigg|\psi(\btheta)},
\end{align}
with $\tilde{c}_3=\cos(\theta_3/2)$ and $\tilde{s}_3=\sin(\theta_3/2)$ as well as $P_0=\ket{0}\!\!\bra{0}$.
We can compute these expectation values in more detail:
\begin{align}
    A_j^a(\btheta) = -\frac12 \de{\begin{array}{ccc}
        0 & \braket{X}_1\braket{Y}_2 & \braket{P_1}_2 \\
        \braket{P_0}_2+c_3\braket{P_1}_2 & \tilde{c}_3 \de{\tilde{c}_3 \braket{Y}_1 - \tilde{s}_3 \braket{Z}_1}\braket{Y}_2 & 0 \\
        s_3 \braket{P_1}_2 & \tilde{c}_3 \de{\tilde{c}_3 \braket{Z}_1 + \tilde{s}_3 \braket{Y}_1}\braket{Y}_2 & 0 
    \end{array}},
\end{align}
where $\braket{M}_q$ denotes the single-qubit expectation value $\bra{\psi_0^{(q)}}R_Y^\dagger(\theta_q) M R_Y(\theta_q)\ket{\psi_0^{(q)}}$ for $q=1,2$.
That is, these expectation values are with respect to the single-qubit states before the $CR_X$ gate.

At $\theta_3=\pi$, which is at the optimum of the optimization in \cref{sec:ex_calc:2q_cost}, we find $A$ to take a particularly simple form, because most terms in the second column and some in the first vanish:
\begin{align}
    A_j^a(\btheta)\big|_{\theta_3=\pi}
    =-\frac12 \de{\begin{array}{ccc}
        0 & \braket{X}_1\braket{Y}_2 & \braket{P_1}_2\\
        \braket{Z}_2 & 0 & 0 \\
        0 & 0 & 0
    \end{array}}.
\end{align}
We do not compute the covariant or equivariant derivative explicitly at this point but use this example for the application in the next section.

\subsection{Two-qubit cost function example}\label{sec:ex_calc:2q_cost}
Here we provide some details for the application in \cref{sec:entangling_app}, using the calculations from the previous section.
As discussed in the main text, we want to obtain a quantum state at which expectation values are invariant with respect to local $\mc{SU}(2)$ transformations on the first qubit.
For this, we use the ansatz from the previous section and minimize the cost function (see \cref{eq:entangling_app:cost_function})
\begin{align}
    C(\btheta) = \sum_{M\in \DE{X, Y, Z}} \bra{\psi(\btheta)}M^{(1)}\ket{\psi(\btheta)}^2,
\end{align}
which is equivalent to the (squared) norm of the symmetry derivatives vector.
As this cost function attains its minimum $0$ if and only if the reduced density matrix on the first qubit is the maximally mixed state, we require the angle of the entangling $CR_X$ gate to attain the value $\theta_3=\pi$, maximizing the entangling power of the gate.
Simultaneously, the single-qubit $R_Y$ rotations need to transform the input product state such that the effect of the $CR_X$ gate is maximized.
The corresponding angles $\theta_{1,2}$ depend on the input state, but it is sufficient to demand that the expectation values $\braket{X^{(1)}}_1$ and $\braket{Z^{(2)}}_2$ vanish.
In addition, we note that $\braket{Z^{(2)}}_2=\braket{P_1^{(2)}}_2-\braket{P_2^{(2)}}_2=0$, together with $\braket{\mbb I^{(2)}}_2=\braket{P_1^{(2)}}_2+\braket{P_2^{(2)}}_2=1$, implies $\braket{P_1^{(2)}}_2=\frac12$.
Overall, we conclude that the vector potential at the optimized parameters $\btheta^\ast$ will take the values
\begin{align}
    A_j^a(\btheta^\ast) = \de{\begin{array}{ccc} 0 & 0 & -\frac14 \\ 0 & 0 & 0 \\ 0 & 0 & 0 \end{array}}.
\end{align}
We set up a numerical experiment with the described ansatz and randomly sampled input states $\ket{\psi_0^{(1,2)}}$.
After optimizing the parameter with a simple gradient descent routine, the cost function indeed attains its minimum $0$, and $\theta_3^\ast=\pi$ as expected. 
The expectation values $\braket{X^{(1)}}_1$ and $\braket{Z^{(2)}}_2$ vanish as well, so that we obtain the above values for the vector potential.
The behaviour of these quantities during the optimization process is depicted in \cref{fig:entangling_app}.

This toy example demonstrates how to make use of our newly developed understanding of symmetries.
We constructed a cost function that allowed us to find a specific quantum state with a particular symmetry.
This application also reminds us that the covariant derivative consists of multiple symmetry-induced quantities:
the partial derivative is modified with the product of the (inverse of the) Gram matrix, the symmetry derivatives, and the overlap of PQC derivatives with vertical tangent vectors.
If we want to obtain an invariant cost function, the symmetry derivatives $m_{a\nu}$ should vanish.
Due to the product structure above, demanding that the second term in the covariant derivative be $0$ is insufficient to guarantee $m_{a\nu}=0$.

\section{Measuring Hermitian anticommutators}\label{sec:hadamard_test_F_c}

\begin{figure}[h!]
    \centering
    \begin{quantikz}
        \lstick{$\ket{+}$} & [5mm]\ctrl{1}& \qw & \meter{}\rstick[wires=2]{$\qquad B=Y\otimes H_j$}\\
        \lstick{$\ket{0}$} & \gate{z_b} \qwbundle{N} & \gate{U_{[:j]}} &\meter{}
    \end{quantikz}
    \caption{\textbf{Modification of a generalized Hadamard test with which $\omega_{bj}$ can be measured.}
    If $z_b$ itself is not unitary, a linear decomposition $z_b=\chi_\ell w_\ell$ into unitaries can be used by running the above circuit with $w_\ell$ instead of $z_b$, for each $\ell$, and contracting the measured results with the coefficients $\chi_\ell$.}
    \label{fig:hadamard-circuit-omega_c_mu}
\end{figure}
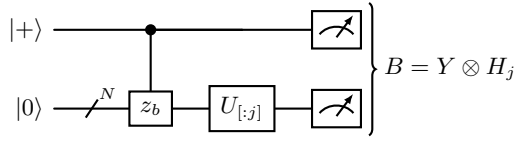

In this section we outline a technique to measure the anticommutator expectation values
\begin{align}
    \omega_{bj}^{(E)}
    = \frac12 \bra{\psi_0}\DE{z_b,\Omega^R_j(\btheta)}\ket{\psi_0}
    = \frac12 \bra{\psi_0}\DE{z_b\,\vec{,}\,U_{[:j]}^\dagger(\btheta)(-iH_j) U_{[:j]}(\btheta)}\ket{\psi_0}
\end{align}
on a qubit-based quantum computer, where $\Omega^R_j$ is the right-effective generator of a parametrized unitary $U(\btheta)$ and $z_b$ are orthonormal basis elements for $\mf t$ ($\mf u^{\mf t}$).
They are required to obtain the covariant (equivariant) derivative of PQC cost functions discussed in \cref{sec:circuitry:cost_functions,sec:circuitry_app:cost_functions}.
The technique is a modification of generalized Hadamard tests~\cite{Ortiz_Laflamme_01} similar to those presented in, e.g.,~\cite{Guerreschi_Smelyanski_17,li2017efficient,romero2018strategies,BravoPrieto2023variational}.

We consider a linear decomposition of $z_b$ into unitaries\footnote{If $z_b$ already is (proportional to) a unitary, we may use it directly, of course.}, $z_b=\chi_{\ell}w_{\ell}$ with $\chi_\ell\in\C$ and $w_\ell\in\mc U$, and the circuit on $N+1$ qubits presented in \cref{fig:hadamard-circuit-omega_c_mu}.
It prepares the state
\begin{align}
    \ket{\phi_\ell}
    = U_{[:j]}\,Cw_\ell \de{\ket{+}\otimes \ket{\psi_0}}
    = \frac{1}{\sqrt{2}}\De{\ket{0}\otimes U_{[:j]}\ket{\psi_0}+\ket{1}\otimes U_{[:j]}w_\ell\ket{\psi_0}},
\end{align}
so that measuring the observable $B = Y\otimes H_j$ leads to the expectation value
\begin{align}
    \braket{\phi_\ell|B|\phi_\ell}
    &= \frac12 \De{\braket{1|Y|0} \bra{\psi_0} w_\ell^\dagger U^\dagger_{[:j]} H_j U_{[:j]}\ket{\psi_0}
    +\braket{0|Y|1} \bra{\psi_0} U^\dagger_{[:j]} H_j U_{[:j]}  w_\ell \ket{\psi_0}}\\
    &= \frac12 \bra{\psi_0} \De{-w_\ell^\dagger \Omega_j(\btheta)+\Omega_j(\btheta)w_\ell}\ket{\psi_0}.
\end{align}
Contracting the measured expectation values for all $w_\ell$ with the coefficients $\chi_\ell$, we obtain
\begin{align}
    \chi_\ell \braket{\phi_\ell|B|\phi_\ell} = \frac12\bra{\psi_0} \De{-\chi_\ell w^\dagger_\ell \Omega_j(\btheta)+ \Omega_j(\btheta)\chi_\ell w_\ell}\ket{\psi_0} = \omega_{bj}.
\end{align}
It is noteworthy that the unitary constituents $w_\ell$ need to be unitary but not necessarily skew-Hermitian themselves, so that the results $\braket{\phi_\ell|B|\phi_\ell}$ individually may not be expectation values of anticommutators.

In cases where decompositions of $z_b$ are particularly hard to find, it is possible to exchange the roles of $z_b$ and $-iH_j$ above, requiring us to decompose $-iH_j$ into unitaries instead, and to execute the partial circuit $U_{[:j]}$ before applying the scheme in \cref{fig:hadamard-circuit-omega_c_mu}.
For equivariant circuits, we may move the $Cw_\ell$ gate through the partial circuit.
If possible, the transformed observable $(Cz_b^\dagger) B (Cz_b)$ can then be evaluated classically and measured in the state $U_{[:j]}(\btheta)\ket{\psi_0}$ without auxiliary qubits.

\section{Covariant quantum natural gradient and the McLachlan variational principle}\label{sec:qng_comp}

In this section we discuss the McLachlan variational principle~\cite{mclachlan1964variational} and apply it to the simulation of imaginary time evolution up to symmetries $S$ with a parametrized quantum circuit.
This derivation is based on Sec.~4.4 of~\cite{hackl2020geometry} and App.~I of~\cite{mcardle2019variational}.

As a preliminary, we start by defining a symmetry-aware metric on the tangent space $T_\psi\mc H$,
\begin{align}
    \Vert \ket{X} \Vert_S^2 \coloneqq \dE{X|\mbb H_\psi|X}.
\end{align}
It is based on the inner product $(\ket{X}, \ket{Y}) \mapsto \dE{X|\mbb H_\psi|Y}$, where $\mbb H_\psi$ is the projector onto the covariant subspace as defined in \cref{eq:projectors_states}.
Any contributions in the vertical subspace are being projected out, only measuring the relevant overlap in the covariant subspace.

The modified Schr\"odinger equation for imaginary time evolution reads
\begin{align}\label{eq:ite_schrodinger}
    \frac{\mathrm{d}}{\mathrm{d}t} \ket{\psi(t)}+(M-\braket{M}(t))\ket{\psi(t)}=0,
\end{align}
where the term $\braket{M}(t)\ket{\psi(t)}$ leads to a normalized solution given by $\ket{\psi(t)}=\sqrt{\braket{\psi_0|\exp(-2Mt)|\psi_0}}e^{-Mt}\ket{\psi_0}$.
This solution experiences damping of higher-energy contributions; denoting the eigenstates of $M$ that have finite overlap with $\ket{\psi_0}$ as $\DE{\phi_r}_r$ and their eigenvalues as $\DE{\lambda_r}_r$, we can compute
\begin{align}
    \left|\frac{\braket{\phi_r|\psi(t)}}{\braket{\phi_0|\psi(t)}}\right|
    = \left|\frac{e^{-\lambda_r t}\braket{\phi_r|\psi_0}}{e^{-\lambda_0 t}\braket{\phi_0|\psi_0}}\right|
    = e^{-(\lambda_r-\lambda_0)t} \left|\frac{\braket{\phi_r|\psi_0}}{\braket{\phi_0|\psi_0}}\right|
    \underset{t\to\infty}{\longrightarrow} 
    \begin{cases}
        0 & \lambda_r>\lambda_0 \\ 1 & \lambda_r=\lambda_0
    \end{cases}\,.
\end{align}
We therefore may hope that approximating this solution as closely as possible within the state space given by a parametrization $\ket{\psi(\btheta)}$ will allow us to minimize $\braket{M}(\btheta)$ towards $\lambda_0$.
For this, $\ket{\psi(\btheta)}$ replaces $\ket{\psi(t)}$ and the notion of time evolution is interpreted as the evolution of the variational parameters during training.
Correspondingly, we may write $\frac{\mathrm{d}}{\mathrm{d}t}\ket{\psi(\btheta)}=\dot{\theta}_j \ket{\partial_j\psi(\btheta)}$ with the usual shorthand notation $\dot\theta=\frac{\mathrm{d}\theta}{\mathrm{d}t}$.

An approximation using $\ket{\psi(\btheta)}$ will not satisfy \cref{eq:ite_schrodinger} in general, but minimizes the norm of the left hand side expression while restricting the evolution to the subspace that can be reached with the parametrization.
As we want to perform the minimization in a symmetry-aware manner, the relevant norm is $\Vert\,\cdot\,\Vert_S$ so that additionally the evolution may be restricted to the covariant subspace\footnote{We could choose not to project onto the covariant subspace while using the norm $\Vert\,\cdot\,\Vert_S$. The motivation to fix the vertical directions is to minimize optimization effort in vertical parameter directions, which will create symmetry transformations we do not care about.}.
The update directions that can be reached by the parametrization within $H_\psi\mc H$ are precisely given by the covariant derivatives $\DE{D_j\ket{\psi(\btheta)}}_j$.
We therefore can construct an orthonormal basis with respect to the metric $\dE{\,\cdot\,|\,\cdot\,}$ from these vectors, given by $\ket{v_j}=\sqrt{(F^S)^+}_{jk} D_k\ket{\psi(\btheta)}$ with $F^S_{jk}=\dE{D_j\psi(\btheta)|D_k\psi(\btheta)}$.
The best approximation to the imaginary time evolution up to symmetries then is given by the projection of $(\braket{M}(t)-M)\ket{\psi(\btheta)}$ onto this space.
We obtain
\begin{align}
    \dot{\theta}_\ell \partial_\ell\ket{\psi(\btheta)}
    &=\frac{\mathrm{d}}{\mathrm{d}t}\ket{\psi(\btheta)}
    = \ket{v_k}\dE{v_k|(\braket{M}(t)-M)|\psi(\btheta)}
    = \ket{v_k}\dE{v_k|M|\psi(\btheta)}\\
    \underset{\dE{v_j|\,\cdot\,}}{\Rightarrow} \qquad\dot{\theta}_\ell \dE{v_j|\partial_\ell\psi(\btheta)}
    &= \dE{v_j|M|\psi(\btheta)}
    =\sqrt{(F^S)^+}_{jk} \dE{D_k\psi(\btheta)|M|\psi(\btheta)}\\
    \underset{\sqrt{(F^S)^+}\cdot}{\Rightarrow} \hspace{2.45cm} \dot{\theta}_\ell
    &= \frac12 (F^S)^+_{\ell k}\  D_k C(\btheta).
\end{align}
In the first line we used that $\dE{v_k|\psi(\btheta)}=\sqrt{(F^S)^+}_{k\ell}\dE{\partial_\ell\psi(\btheta)|\mbb H_{\psi(\btheta)}|\psi(\btheta)}=0$ because $\ket{\psi}$ is not even included in the tangent space $T_\psi\mc H$, which we defined to only contain directions based on unitary evolution, let alone in the subspace onto which $\mbb H_\psi$ projects\footnote{In the setting where $S=\mc U(1)\triangleleft\mc U$, this fact can be derived from the normalization condition on $\ket{\psi(\btheta)}$ instead; see \cite{mcardle2019variational}.}.To obtain the last line, we used $\dE{v_j|\partial_\ell\psi(\btheta)}=\sqrt{(F^S)^+}_{jk}\dE{\partial_k\psi(\btheta)|\mbb H_\psi|\partial_\ell\psi(\btheta)}=\sqrt{F^S}_{j\ell}$.
With this, we derived the covariant quantum natural gradient shown in \cref{eq:s_natural_gradient}.
As discussed in the main text, for $S=\mc U(1)\triangleleft\mc U$ we obtain $F^S=F$ and $D_k C(\btheta)=\partial_k C(\btheta)$, so that we recover the conventional QNG.

\section{Physical gauge theory--details}\label{sec:gauge_theory_comp}
Here we provide details for the abbreviated derivation in \cref{sec:gauge_theory}.
As remarked before, we closely follow~\cite[Ch.~15]{peskin_schroeder}.
Recall that we have a symmetry group acting on density matrices via a matrix representation $\sigma$, which we do not denote explicitly, and conjugation.
From the expansion of the comparator
\begin{align}
    \Gamma(\btheta+\varepsilon \vec{n}, \btheta)\act{}\rho(\btheta)
    =\rho(\btheta) + \varepsilon n_j A_j^a [x_a, \rho(\btheta)] + \mc O (\varepsilon^2),
\end{align}
and the Taylor expansion of $\rho$,
\begin{align}
    \rho(\btheta+\varepsilon\vec{n})=\rho(\btheta)+\varepsilon n_j\partial_j\rho(\btheta)+\mc O(\varepsilon^2),
\end{align}
we compute the covariant derivative defined in \cref{eq:cov_deriv_gauge_def} as
\begin{align}
    n_j \Delta_j \rho(\btheta)
    &=
    \lim_{\varepsilon\to 0} \frac1\varepsilon\left[ \rho(\btheta)+\varepsilon n_j\partial_j\rho(\btheta)+\mc O(\varepsilon^2) - \rho(\btheta) - \varepsilon n_j A_j^a [x_a, \rho(\btheta)] - \mc O (\varepsilon^2)\right]\\
    &=n_j \left(\partial_j \rho(\btheta) - A_j^a [x_a,\rho(\btheta)]\right) + \underset{0}{\underbrace{\lim_{\varepsilon\to 0} \mc O(\varepsilon)}}.
\end{align}

In order to show covariance of $\Delta_j \rho(\btheta)$, we need to consider the transformation behaviour of $\Gamma$ for infinitesimal distances and the behaviour of $A_j^a$ that results from it:
\begin{align}
    \Gamma(\btheta+\varepsilon\vec{n},\btheta)\act{}\rho(\btheta)
    \longrightarrow&
    s(\btheta+\varepsilon\vec{n})\Gamma(\btheta+\varepsilon\vec{n},\btheta) s^{-1}(\btheta)\act{}[s(\btheta)\act{}\rho(\btheta)]\\
    =&s(\btheta+\varepsilon\vec{n})\act{}(\Gamma(\btheta+\varepsilon\vec{n},\btheta)\act{}\rho(\btheta))\\
    \Rightarrow 
    \rho(\btheta) + \varepsilon n_j A_j^a [x_a, \rho(\btheta)] + \mc O (\varepsilon^2)
    \longrightarrow &
    s(\btheta)\left(\mbb I +\varepsilon n_j \Lambda_j (\btheta)+\mc O(\varepsilon^2)\right)
    \act{}\left[\rho(\btheta) + \varepsilon n_j A_j^a [x_a, \rho(\btheta)] + \mc O (\varepsilon^2)\right]\\
    =& s(\btheta)\rho(\btheta)s^\dagger(\btheta) +\varepsilon n_j s(\btheta)[\Lambda_j(\btheta),\rho(\btheta)]s^\dagger(\btheta)\\
    &\phantom{s(\btheta)\rho(\btheta)s^\dagger(\btheta)}+ \varepsilon n_j A_j^a  s(\btheta) [x_a, \rho(\btheta)]s^\dagger(\btheta) +\mc O(\varepsilon^2)\\
    \Rightarrow
    A_j^a [x_a, \rho(\btheta)]
    \longrightarrow&
    s(\btheta)[\Lambda_j(\btheta), \rho(\btheta)]s^\dagger(\btheta) + A_j^a s(\btheta)[x_a, \rho(\btheta)]s^\dagger(\btheta)\label{eq:vector_potential_trafo_gauge}\\
    \Rightarrow
    [A_j^a x_a, \circ]
    \longrightarrow&
    \left[s(\btheta)\left(\Lambda_j(\btheta)+A_j^a x_a\right)s^\dagger(\btheta), \circ\right].
\end{align}
In this computation, we used the same notation for $S$ and $\mf s$ acting on $\rho$, via $\act{}$.
Further, we used $\partial_j s(\btheta) = s(\btheta)\Lambda_j(\btheta)$ with the effective generator $\Lambda_j(\btheta)\in\mf s$ of the symmetry transformation, so that $(\partial_j s(\btheta))\act{}\rho(\btheta) = s(\btheta)\act{}[\Lambda_j(\btheta), \rho(\btheta)]$.
Finally, note that in order to extract the transformation behaviour of $A_j^a[x_a, \circ]$ alone, we had to compensate for the transformation of $\rho(\btheta)$ itself in the second-to-last line.

We now can compute the transformation behaviour of $\Delta_j\rho(\btheta)$:
\begin{align}
    \Delta_j\rho(\btheta)
    \longrightarrow&
    \partial_j (s(\btheta)\act{}\rho(\btheta)) - \left[s(\btheta)\left(\Lambda_j(\btheta)+A_j^a x_a\right)s^\dagger(\btheta), s(\btheta)\rho(\btheta)s^\dagger(\btheta)\right]\\
    =& s(\btheta)[\Lambda_j(\btheta),\rho(\btheta)]s^\dagger(\btheta)+s(\btheta)\left(\partial_j\rho(\btheta)\right)s^\dagger(\btheta) - s(\btheta) [\Lambda_j(\btheta), \rho(\btheta)] s^\dagger(\btheta) - A_j^a s(\btheta)[x_a, \rho(\btheta)]s^\dagger(\btheta)\\
    =& s(\btheta) \left(\partial_j\rho(\btheta) - A_j^a[x_a, \rho(\btheta)] \right)s^\dagger(\btheta)\\
    =& s(\btheta)\act{} \Delta_j\rho (\btheta).
\end{align}
This shows that $\Delta_j\rho(\btheta)$ transforms like $\rho(\btheta)$, making it covariant in the physicists' sense.
Note that the $\Delta_j$ is defined via two objects that transform with $s(\btheta+\varepsilon\vec{n})$, in the limit of $\varepsilon\to 0$, and that the whole purpose of introducing the comparator $\Gamma$ was to make $\rho$ comparable at different parameter positions.
It is not surprising, then, that the result transforms with $s(\btheta)$.

To conclude this section we want to show equivalence of the covariant derivative introduced here and that of the main text (\cref{sec:circuitry:states}), defined via the projector onto the covariant subspace.
We compute the latter for the density matrix of a pure parametrized quantum state and extract the vector potential:
\begin{align}
    \Delta_j \rho(\btheta)
    &= 
    \de{\partial_j \ket{\psi(\btheta)}} \bra{\psi(\btheta)}
    + \ket{\psi(\btheta)} \partial_j \bra{\psi(\btheta)}
    - \de{\mbb V_{\psi} \ket{\psi(\btheta)}}\bra{\psi(\btheta)}
    - \ket{\psi(\btheta)}\De{\mbb V_\psi \ket{\psi(\btheta)}}^\dagger\\
    &= \partial_j \rho(\btheta) 
    - x_a \ket{\psi(\btheta)}\bra{\psi(\btheta)} G_{ab}^{+} 
    \dE{\psi(\btheta)|x_b^\dagger \partial_j|\psi(\btheta)}
    - \ket{\psi(\btheta)}\bra{\psi(\btheta)}x_a^\dagger G_{ab}^{+} 
    \dE{\partial_j \psi(\btheta)|x_b|\psi(\btheta)}\\
    &=\partial_j \rho(\btheta)
    -G_{ab}^{+} \dE{\psi(\btheta)|x_b^\dagger |\partial_j \psi(\btheta)} [x_a,\rho(\btheta)]\\
    \Rightarrow\ \ A_j^a &= G_{ab}^{+} \dE{\psi(\btheta)|x_b^\dagger |\partial_j \psi(\btheta)}.
\end{align}
To show that this is a valid vector potential for the gauge theoretical covariant derivative, we only need to certify its transformation behaviour.
To this end, we compute
\begin{align}
    A_j^a [x_a, \rho(\btheta)] 
    \longrightarrow&
    \de{\dE{\psi(\btheta)|s^\dagger(\btheta) x_c^\dagger x_d s(\btheta)| \psi(\btheta)}}^+_{ab}
    \dE{\psi(\btheta)|s^\dagger(\btheta)x_b^\dagger \partial_j s(\btheta)|\psi(\btheta)}[x_a, s(\btheta)\rho(\btheta)s^\dagger(\btheta)]\\
    =&
    \de{\dE{\psi(\btheta)|y_c^\dagger y_d | \psi(\btheta)}}^+_{ab} 
    \De{\dE{\psi(\btheta)|y_b^\dagger |\partial_j \psi(\btheta)}+\dE{\psi(\btheta)|y_b^\dagger \Lambda_j(\btheta) |\psi(\btheta)}}s(\btheta)[y_a, \rho(\btheta)]s^\dagger(\btheta)
    \label{eq:vector_potential_trafo}\\
    =&
    s(\btheta) \de{A_j^a[x_a, \rho(\btheta)] + \De{\Lambda_j(\btheta), \rho(\btheta)}} s^\dagger(\btheta)\label{eq:vector_potential_trafo_geom},
\end{align}
where we again used $\partial_j s(\btheta)=s(\btheta)\Lambda_j(\btheta)$ and denoted $y_a=s^\dagger(\btheta)x_as(\btheta)$.
Some comments are in order:
first, the operators $\DE{y_a}_a$ again form a basis for $\mf s$, because the trace inner product is invariant under unitary transformations and because the adjoint action of $s(\btheta)\in S$ on $\mf s$ maps to $\mf s$ again.
Second, in the last step we identified
\begin{align}
    \de{\dE{\psi(\btheta)|y_c^\dagger y_d | \psi(\btheta)}}^+_{ab} \dE{\psi(\btheta)|y_b^\dagger |\partial_j \psi(\btheta)} [y_a, \rho(\btheta)]
    = A_j^a [x_a, \rho(\btheta)],
\end{align}
which holds because our choice of basis for $\mf s$ may not impact the operator $A_j^a [x_a, \circ]$.
Last, for the second term in \cref{eq:vector_potential_trafo} we reverted the decomposition
\begin{align}
    \Lambda_j \ket{\psi(\btheta)}
    &= t_a \ket{\psi(\btheta)} \dE{\psi(\btheta)|t_a^\dagger\Lambda_j |\psi(\btheta)}
    =y_a \ket{\psi(\btheta)} \dE{\psi(\btheta)|y_b^\dagger\Lambda_j |\psi(\btheta)}\overline{G}^+_{ab},
\end{align}
with the ONB $t_a\ket{\psi(\btheta)}=\sqrt{\overline{G}^+}_{ab} y_b\ket{\psi(\btheta)}$, for the tangent space of $\mc H$ at $\ket{\psi(\btheta)}$, computed with the Gram matrix $\overline{G}_{ab}=\dE{\psi(\btheta)|y_a^\dagger y_b | \psi(\btheta)}$.

Comparing the result \cref{eq:vector_potential_trafo_geom} with \cref{eq:vector_potential_trafo_gauge}, we confirm that $A_j^a[x_a,\circ]$ transforms correctly.

\end{document}